\RequirePackage{fix-cm}
\documentclass[twocolumn,epjc3]{svjour3nohead}
\smartqed  % flush right qed marks, e.g. at end of proof
\RequirePackage{graphicx}
\RequirePackage{subfigure}
\RequirePackage{xspace}
\RequirePackage{verbatim}
 \RequirePackage[colorlinks,citecolor=blue,urlcolor=blue,linkcolor=blue]{hyperref}
\newcommand{\BLUE     }{{\sc BLUE}}
\newcommand{\ROOT     }{{\sc ROOT}}

\newcommand{\scen     }[1]{\ensuremath{\cal{#1}}\xspace}
\newcommand{\sA       }{\scen{A}}
\newcommand{\sB       }{\scen{B}}
\newcommand{\sC       }{\scen{C}}
\newcommand{\sD       }{\scen{D}}
\newcommand{\sE       }{\scen{E}}
\newcommand{\XZ    }[3]{\ensuremath{#1\,\pm #2\,\pm #3}\xspace}
\newcommand{\mt     }{\ensuremath{m_{\mathrm{top}}}\xspace}
\newcommand{\gev    }{\ensuremath{\mathrm{GeV}}\xspace}
\newcommand{\GeV    }{\ensuremath{\mathrm{GeV}}\xspace}
\newcommand{\Pdf    }[2]{\ensuremath{{\cal{P}}(#1, #2)}\xspace}
\newcommand{\Prob   }[2]{\ensuremath{P(#1, #2)}\xspace}
\newcommand{\chiq   }{\ensuremath{\chi^2}\xspace}
\newcommand{\chiqp  }{\ensuremath{{\chi^{\prime}}^{2}}\xspace}
\newcommand{\xii    }{\ensuremath{x_\mathrm{i}}\xspace}

\newcommand{\xj     }{\ensuremath{x_\mathrm{j}}\xspace}
\newcommand{\Xo     }{\ensuremath{X_\mathrm{1}}\xspace}
\newcommand{\Xt     }{\ensuremath{X_\mathrm{2}}\xspace}
\newcommand{\Xii    }{\ensuremath{X_\mathrm{i}}\xspace}

\newcommand{\xo     }{\ensuremath{x_\mathrm{1}}\xspace}

\newcommand{\xt     }{\ensuremath{x_\mathrm{2}}\xspace}

\newcommand{\xT     }{\ensuremath{x_\mathrm{T}}\xspace}

\newcommand{\so     }{\ensuremath{\sigma_{1}}\xspace}
\newcommand{\soq    }{\ensuremath{\sigma_{1}^{2}}\xspace}
\newcommand{\sok    }{\ensuremath{\sigma_{1k}}\xspace}
\newcommand{\sokq   }{\ensuremath{\sigma_{1k}^{2}}\xspace}

\newcommand{\sot    }{\ensuremath{\sigma_{12}}\xspace}

\newcommand{\st     }{\ensuremath{\sigma_{2}}\xspace}
\newcommand{\stq    }{\ensuremath{\sigma_{2}^{2}}\xspace}
\newcommand{\stk    }{\ensuremath{\sigma_{2k}}\xspace}
\newcommand{\stkq   }{\ensuremath{\sigma_{2k}^{2}}\xspace}

\newcommand{\stt    }{\ensuremath{\sigma_{22}}\xspace}

\newcommand{\ssta   }{\ensuremath{\sigma_\mathrm{Stat}}\xspace}

\newcommand{\ssys   }{\ensuremath{\sigma_\mathrm{Syst}}\xspace}

\newcommand{\ssyso  }{\ensuremath{\sigma_\mathrm{Syst1}}\xspace}
\newcommand{\ssyst  }{\ensuremath{\sigma_\mathrm{Syst2}}\xspace}
\newcommand{\si     }{\ensuremath{\sigma_{i}}\xspace}
\newcommand{\siq    }{\ensuremath{\sigma_{i}^{2}}\xspace}

\newcommand{\sik    }{\ensuremath{\sigma_{ik}}\xspace}
\newcommand{\sinv   }{\ensuremath{\sigma_{i0}}\xspace}
\newcommand{\sio    }{\ensuremath{\sigma_{i1}}\xspace}
\newcommand{\sit    }{\ensuremath{\sigma_{i2}}\xspace}

\newcommand{\sisys  }{\ensuremath{\sigma_\mathrm{i,Syst}}\xspace}
\newcommand{\sjk    }{\ensuremath{\sigma_{jk}}\xspace}
\newcommand{\sx     }{\ensuremath{\sigma_{x}}\xspace}
\newcommand{\sxq    }{\ensuremath{\sigma_{x}^{2}}\xspace}

\newcommand{\rhof   }{\ensuremath{\rho}\xspace}

\newcommand{\rhootk }{\ensuremath{\rho_{12k}}\xspace}
\newcommand{\rhooto }{\ensuremath{\rho_{121}}\xspace}
\newcommand{\rhootz }{\ensuremath{\rho_{120}}\xspace}
\newcommand{\rhoij  }{\ensuremath{\rho_{ij}}\xspace}
\newcommand{\rhoijk }{\ensuremath{\rho_{ijk}}\xspace}
\newcommand{\rhored }{\ensuremath{\rho_\mathrm{red}}\xspace}
\newcommand{\bet    }{\ensuremath{\beta}\xspace}

\newcommand{\fijk   }{\ensuremath{f_{ijk}}\xspace}
\newcommand{\fij    }{\ensuremath{f_{ij}}\xspace}
\newcommand{\fk     }{\ensuremath{f_{k}}\xspace}

\newcommand{\zotk   }{\ensuremath{z_{12k}}\xspace}
\newcommand{\zijk   }{\ensuremath{z_{ijk}}\xspace}
\newcommand{\pt     }{\ensuremath{p_\mathrm{t}}\xspace}
\newcommand{\ptmin  }{\ensuremath{p_\mathrm{t,min}}\xspace}
\newcommand{\rmin   }{\ensuremath{r_\mathrm{min}}\xspace}
\newcommand{\fout   }{\ensuremath{f_\mathrm{out}}\xspace}
 \newcounter{enumct}
 \newenvironment{Enumerate}{\begin{list}{\arabic{enumct}.}%
 {\usecounter{enumct}\setlength{\topsep}{0.2mm}%
 \setlength{\partopsep}{0.2mm}\setlength{\itemsep}{0.2mm}%
 \setlength{\parsep}{0.2mm}}}{\end{list}}
\journalname{Eur. Phys. J. C}
\begin{document}
\title{On the combination of correlated estimates of a physics observable}
\titlerunning{On the combination of correlated estimates of a physics observable}
\author{Richard Nisius\thanksref{e1,addr1}}
\authorrunning{Richard Nisius}
\thankstext{e1}{e-mail: Richard.Nisius@mpp.mpg.de}
\institute{Max-Planck-Institut f\"ur Physik (Werner-Heisenberg-Institut)
  F\"ohringer Ring 6, D-80805 M\"unchen, Germany \label{addr1}}
\date{Received: \today / Accepted: date}
% The correct dates will be entered by the editor
%
\maketitle
%
%------------------------------------------------------------------------------
%
\begin{abstract}
 The combination of a number of correlated estimates of a given observable is
 frequently performed using the Best Linear Unbiased Estimate (\BLUE) method.
 Most features of such a combination can already be seen by analysing the
 special case of a pair of estimates from two correlated estimators of the
 observable.
 Two important parameters of this combination are the weight of the less precise
 estimate and the ratio of uncertainties of the combined result and the more
 precise estimate. Derivatives of these quantities are derived with respect to
 the correlation and the ratio of uncertainties of the two estimates.

 The impact of using either absolute or relative uncertainties in the
 \BLUE\ combination is investigated on a number of examples including Peelle's
 Pertinent Puzzle.
 Using an example, a critical assessment is performed of suggested methods to
 deal with the fact that both the correlation and the ratio of uncertainties of
 a pair of estimates are typically only known with some uncertainty.
 Finally, a proposal is made to decide on the usefulness of a combination and to
 perform it. The proposal is based on possible improvements with respect to the
 most precise estimate by including additional estimates. This procedure can be
 applied to the general case of several observables.
\end{abstract}
%
%------------------------------------------------------------------------------
%
\section{Introduction}
\label{sec:introd}
 The combination of a number of correlated estimates of a single observable is
 discussed in Ref.~\cite{LYO-1988}. Here, the term estimate denotes a particular
 outcome (measurement) based on an estimator of the observable, which follows a
 probability density distribution~(pdf). The particular estimate obtained may be
 a likely or unlikely outcome given that distribution. Repeating the measurement
 numerous times with identical conditions, the estimates will follow the
 underlying multi-dimensional pdf of the estimators\footnote{In this paper, the
   discussion is restricted to Gaussian estimator pdfs.}.
 The analysis~\cite{LYO-1988} makes use of a \chiq\ minimisation to obtain the
 combined value expressed in the mathematically equivalent \BLUE\ language.

 Provided the estimators are unbiased, when applying this formalism the {\bf
   Best Linear Unbiased Estimate} of the observable is obtained with the
 following meaning:
 {\bf Best:} the combined result for the observable obtained this way has the
 smallest variance; {\bf Linear:} the result is a linear combination of the
 individual estimates; {\bf Unbiased Estimate:} when the procedure is repeated
 for a large number of cases consistent with the underlying multi-dimensional
 pdf, the mean of all combined results equals the true value of the observable.
 For a real situation, for which the estimates are obtained by experiments that
 cannot be repeated numerous times, when performing the combination one has to
 rely on this fact, although the combined value obtained from the particular
 estimates may be far away from the true value. This fact however should not be
 mistaken for a bias inherent to the method.

 The equations to solve the problem for the general case of $m$ estimates and
 $n$ observables with $m\ge n$ are given in Ref.~\cite{Val-0301}. They have been
 implemented in a software package~\cite{NIS-1301} that is embedded into the
 \ROOT\ analysis framework~\cite{BRU-9701}, but are not repeated here.
 However, the special case of two correlated estimates of the same observable is
 discussed in some detail. This is because already from this case the main
 features of the combination can easily be understood.

 This paper is organised as follows: the case of two estimators and the
 consequences of the conditional probability is explained in
 Section~\ref{sec:corest}.
 The equations for the combination of a pair of estimates are given in
 Section~\ref{sec:twoest}.
 This is followed by a discussion of the properties of the estimates to be
 combined in Section~\ref{sec:estpro}.
 The impact of assigning relative uncertainties is reviewed in
 Section~\ref{sec:relunc}.
 The concept of reduced correlations is outlined in Section~\ref{sec:redcor},
 and other methods, constructed to maximise the variance of the combined result,
 are discussed in Section~\ref{sec:maxvar}.
 Based on an example, the consequences of using these methods are discussed in
 Section~\ref{sec:example}.
 A detailed proposal on how to decide on a combination and how to perform
 investigations of its stability is given in Section~\ref{sec:decide}.
 Finally, conclusions are drawn in Section~\ref{sec:summar}.
%
%------------------------------------------------------------------------------
%
\section{Correlated estimators and conditional probabilities}
\label{sec:corest}
 Let \Xo\ and \Xt\ with variances \soq\ and \stq\ be two unbiased, but
 correlated Gaussian estimators of a true value \xT.
 They obey the two-dimensional pdf \Pdf{\Xo}{\Xt}, with identical mean values
 $\langle\Xo\rangle=\langle\Xt\rangle=\xT$ for the two estimators if calculated
 based on the entire pdf.
 With a correlation of the two estimators of \rhof\ the pdf reads:
%
%%%%%%%%%%%%%%%%%%%%%%%%%%%%%%%%%%%%%%%%%%%%%%%%%%%%%%%%%%%%%%%%%%%%%%%%%%%%%%%
\begin{eqnarray}
 \Pdf{\Xo}{\Xt} &=& \frac{1}{\sqrt{2\pi}\so}\frac{1}{\sqrt{2\pi}\st}
                    \frac{1}{\sqrt{1-\rho^2}}\cdot \label{eq:cond1} \\
                & & \exp{\left\{-
                    \frac{1}{2(1-\rho^2)}
                    \left(\frac{(\Xo-\xT)^2}{\soq}+\right.\right.} \nonumber \\
                & & \left.\left.
                    \frac{(\Xt-\xT)^2}{\stq} - 
                    \frac{2\rho(\Xo-\xT)(\Xt-\xT)}{\so\st}\right)\right\}
                    \nonumber
\end{eqnarray}
%%%%%%%%%%%%%%%%%%%%%%%%%%%%%%%%%%%%%%%%%%%%%%%%%%%%%%%%%%%%%%%%%%%%%%%%%%%%%%%
%
 The outcome of a pair of data analyses using these estimators will be two
 estimates denoted by \xo\ and \xt\ that will occur according to this
 pdf\footnote{Throughout this paper estimators, like \Xo, are denoted by upper
   case letters, whereas estimates, like \xo, are denoted by lower case
   letters.}. The estimates will have variances of \soq\ and \stq\ assigned, and
 their correlation is \rhof.
 Without loss of generality it is assumed that \Xo\ is as least as precise an
 estimator of \xT\ than \Xt\ is, such that $z\equiv \st/\so\geq 1$.

 In combinations of estimates of physics observables the typical situation is
 that one estimate, here \xo, is available, and the question arises what the
 improvement will be if also the information from another estimate, here \xt, is
 used, rather than determining \xT\ and its uncertainty solely based on \xo.
 Therefore, it is important to understand what is the likely outcome of
 \xt\ given the existence of \xo. This is most directly seen by analysing the
 conditional pdf for \Xt\ given $\Xo=\xo$ which reads:
%
%%%%%%%%%%%%%%%%%%%%%%%%%%%%%%%%%%%%%%%%%%%%%%%%%%%%%%%%%%%%%%%%%%%%%%%%%%%%%%%
\begin{eqnarray}
 \Pdf{\xo}{\Xt} &=&\frac{1}{\sqrt{2\pi}\so}
                   \exp{\left\{-\frac{1}{2}
                   \left(\frac{\xo-\xT}{\so}\right)^2\right\}}\cdot \\
                & &\frac{1}{\sqrt{2\pi}\st\sqrt{1-\rho^2}}\cdot\nonumber \\
                & &\exp{\left\{-\frac{1}{2}\left(
                   \frac{\Xt-\left[\xT+\rho z(\xo-\xT)\right]}
                   {\st\sqrt{1-\rho^2}}\right)^2\right\}} \nonumber\,.
\label{eq:cond2}
\end{eqnarray}
%%%%%%%%%%%%%%%%%%%%%%%%%%%%%%%%%%%%%%%%%%%%%%%%%%%%%%%%%%%%%%%%%%%%%%%%%%%%%%%
%
 A few facts are worth noticing, see also Refs.~\cite{LYO-8601,BAR-8901}, and a
 related discussion in Ref.~\cite{COX-0601}.
 Firstly, this conditional pdf for \Xt\ at a given fixed value of \xo\ is no
 longer centred at $\langle\Xt\rangle=\xT$ but at $\langle\Xt\rangle=\xT+\rho
 z(\xo-\xT)$.
 Although \Xt\ in itself is an unbiased estimator, given the existence of the
 estimate \xo\ and the correlation of the estimators, it is no longer
 distributed around the true value, except for the situation in which the value
 of the more precise estimate coincides with the true value, i.e.~$\xo=\xT$.
 This is a mere consequence of the correlation. As intuitively expected, in the
 case of positively correlated estimates, if one estimate is larger (smaller)
 than \xT\ the other also more likely will be larger (smaller). For negatively
 correlated estimates the situation is reversed.

 For $\rho>0$, and depending on whether $\rho z$ is larger (smaller) than unity,
 the mean $\langle\Xt\rangle$ is even further away from (closer to) the true
 value \xT\ than \xo is.
 Given that the distribution in \Xt\ is still symmetric around its mean, for
 $\rho>1/z$ in more than half of the cases in which $\xT< \xo$ also $\xT< \xo <
 \Xt$ is fulfilled.
 Secondly, the variance of \Xt\ no longer amounts to the initial value of
 \stq\ but it is reduced to $(1-\rho^2) \stq$ which vanishes for $\rhof = \pm
 1$, again a consequence of the correlation.
 Finally, for $\rhof=0$ the original values of the mean and width of the pdf for
 \Xt\ are recovered.
%
%%%%%%%%%%%%%%%%%%%%%%%%%%%%%%%%%%%%%%%%%%%%%%%%%%%%%%%%%%%%%%%%%%%%%%%%%%%%%%%
\begin{figure*}[tbp!]
\centering
\subfigure[$\rhof=   0$]{\label{fig:Peellzr}\includegraphics[width=0.329\textwidth]{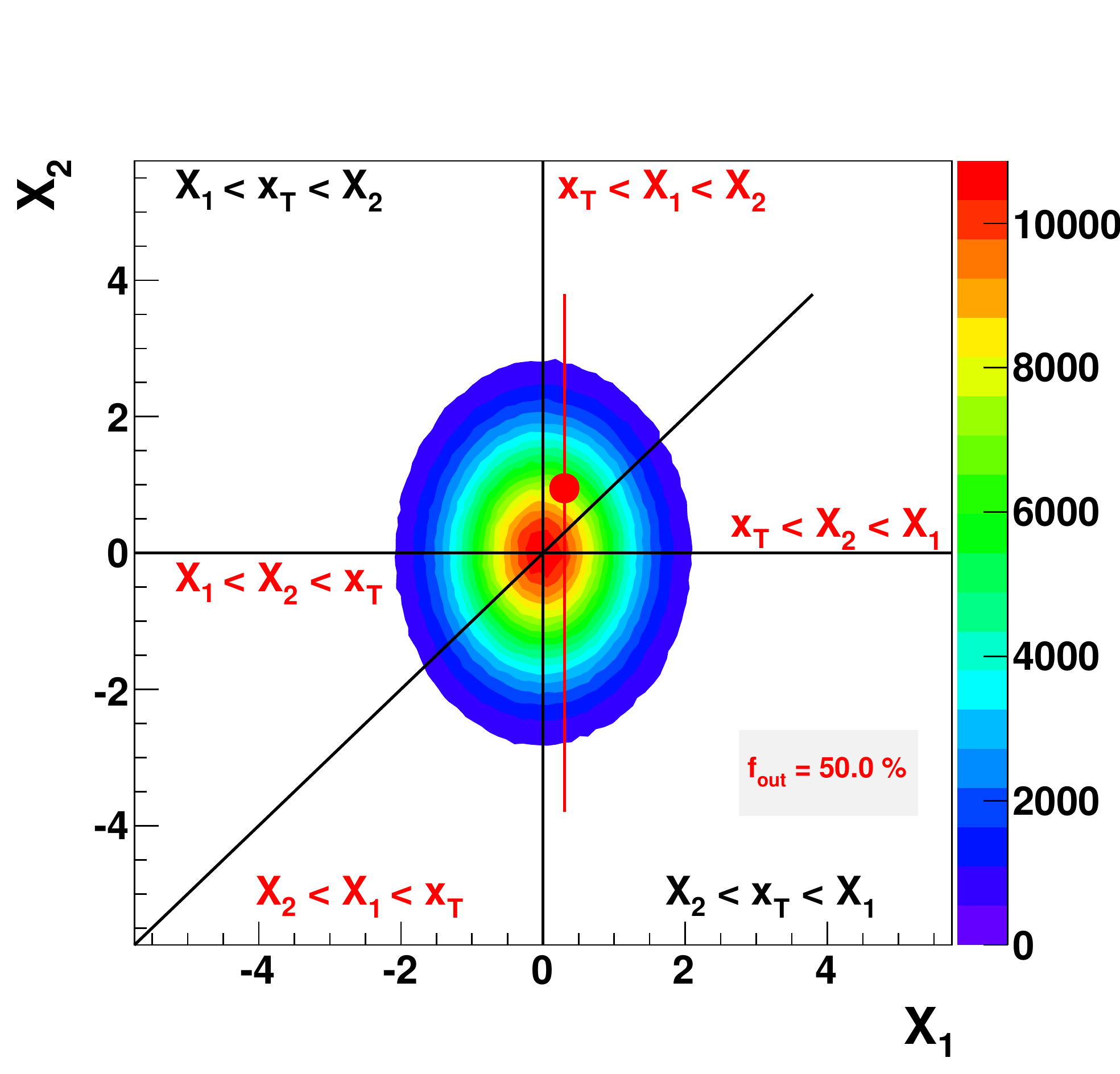}}
\subfigure[$\rhof= 0.9$]{\label{fig:Peellpr}\includegraphics[width=0.329\textwidth]{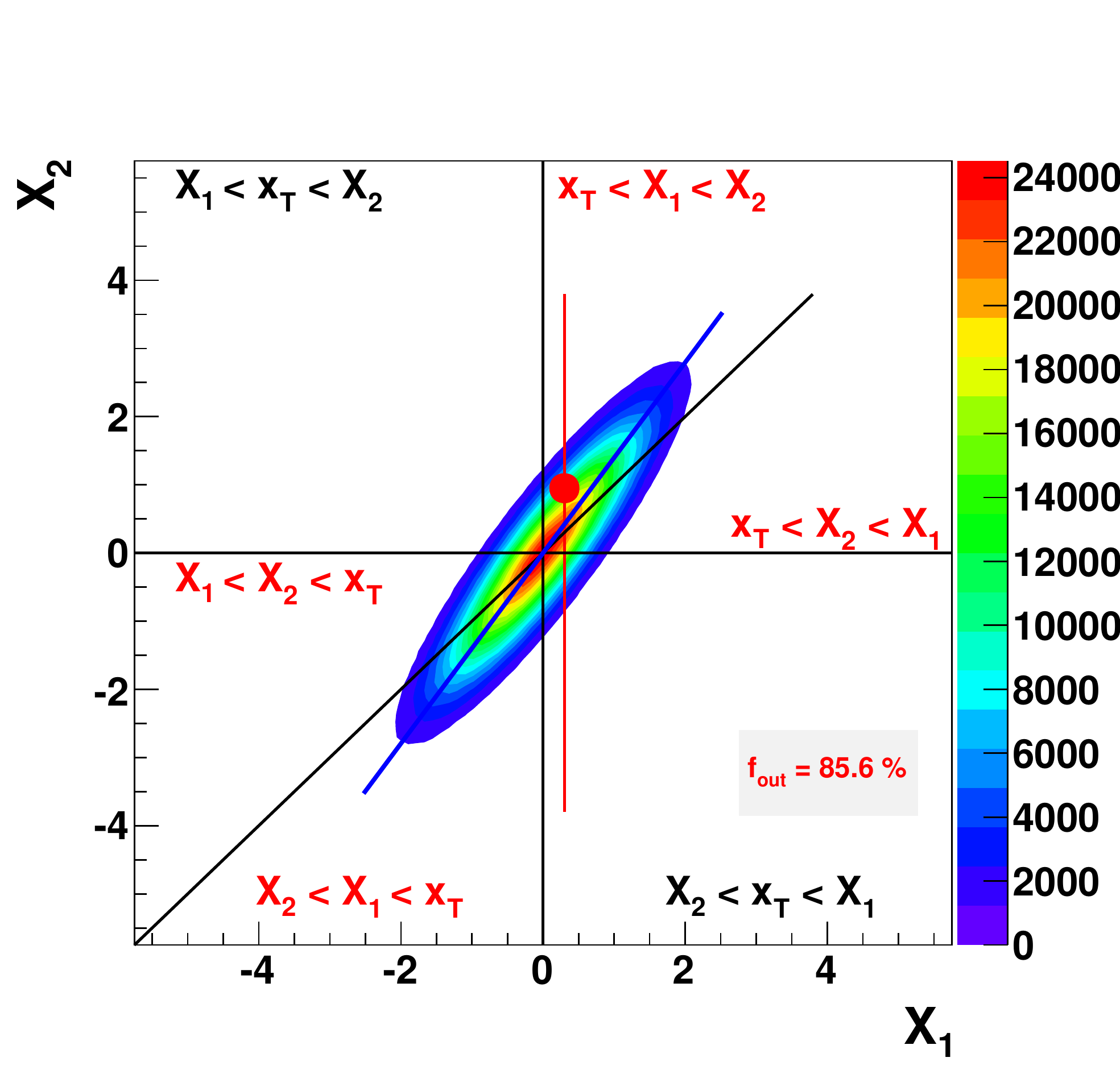}}
\subfigure[$\rhof=-0.9$]{\label{fig:Peellmr}\includegraphics[width=0.329\textwidth]{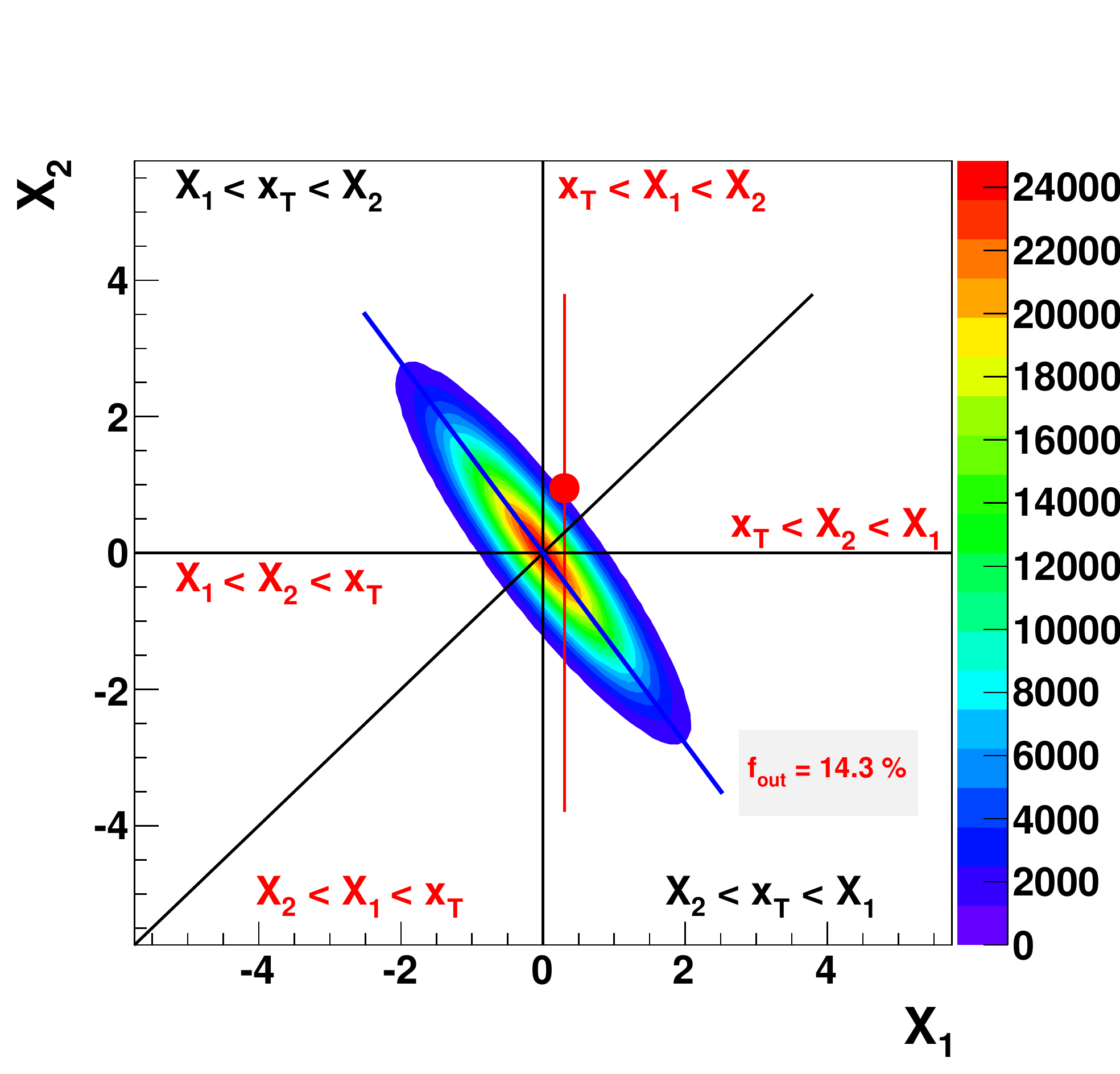}}
\caption{The two-dimensional pdf \Pdf{\Xo}{\Xt}\ for three values of the
  correlation $\rhof$ obtained using five million pairs of estimates. The black
  line corresponds to $\Xo=\Xt$, the red line to $\Xo=\xo$, and finally the dot
  to a particular pair of estimates chosen to be $\xo=0.30$ and $\xt=0.95$.
  The variable \fout\ denotes the fraction of events for which \xT\ does not lie
  within the interval spanned by the pair of estimates. 
  Shown are (a) $\rhof=0$, (b) $\rhof=0.9$, and (c) $\rhof=-0.9$. In b~(c) the
  half axes shown in blue are changed and rotated (counter) clockwise from the
  positive \Xt\ axis.}
\label{fig:Peellofr}
\end{figure*}
%%%%%%%%%%%%%%%%%%%%%%%%%%%%%%%%%%%%%%%%%%%%%%%%%%%%%%%%%%%%%%%%%%%%%%%%%%%%%%

 Simulating the two-dimensional pdf \Pdf{\Xo}{\Xt}\ using five million pairs of
 estimates, the consequences of the conditional probability for the example of
 individually unbiased estimators obeying $\langle\Xo\rangle = \langle\Xt\rangle
 = \xT = 0$ are discussed.
 For uncertainties of $\so=0.85$ and $\st=1.15$, i.e.~for $z=1.35$, the results
 are shown in Figure~\ref{fig:Peellofr} for three different values of the
 correlation, $\rhof=0, 0.9, -0.9$.
 For the uncorrelated case, Figure~\ref{fig:Peellzr}, the half axes of the
 ellipses coincide with the coordinate axes. For any value of \xo, e.g.~along
 the vertical red line shown, the conditional pdf is centred around $\Xt=\xT$.
 A hypothetical outcome, namely the pair of estimates \xo\ and \xt, is indicated
 by the red dot. Depending on the value of \rhof\ this is a more or less likely
 outcome, as can be seen from the different colours of the pdf at the location
 of the point in Figure~\ref{fig:Peellofr}. 
 Numerically, for the three scenarios $\rhof=0, 0.9, -0.9$, the value of the pdf
 at the chosen point with respect to the maximum of the pdf,
 i.e.~$\Pdf{\xo}{\xt} / \Pdf{0}{0}$, amounts to $0.67, 0.48, 0.03$.
 Since for the chosen value of \xT\ this point lies in the upper right
 (i.e.~first) quadrant, both estimates are larger than \xT. Since the point is
 above the diagonal line, \xt\ has been chosen to be larger than \xo, such that
 the order is $\xT<\xo<\xt$. This means the true value is outside the interval
 given by the two estimates.
 Analysing the entire two-dimensional pdf one finds that, even for the
 uncorrelated case, for which the pdf is equally shared by the four quadrants in
 the \Xo--\Xt plane, in half of all possible outcomes (namely in quadrants one
 and three), the true value does not fall within the interval spanned by the
 estimates, despite the fact that both estimators are unbiased and not
 correlated\footnote{Although discussed here for the two-dimensional case, it
   similarly applies to a pair of estimates from the same one-dimensional
   Gaussian estimator.}.

 The situation of largely positively correlated uncertainties with $\rhof=0.9$,
 a situation frequently referred to as Peelle's Pertinent
 Puzzle~\cite{PEL-8701,CHI-9101}, is shown in Figure~\ref{fig:Peellpr}.
 This time, due to the positive correlation, the ellipses is deformed and
 rotated clockwise from the positive \Xt\ axis with increasing rotation angle
 $\theta$ for increasing \rhof\ according to the following
 formula~\cite{BAR-8901}:
%
%%%%%%%%%%%%%%%%%%%%%%%%%%%%%%%%%%%%%%%%%%%%%%%%%%%%%%%%%%%%%%%%%%%%%%%%%%%%%%%
\begin{eqnarray}
 \tan{2\theta} &=& \frac{2\rhof z}{1-z^2}\nonumber.
\end{eqnarray}
%%%%%%%%%%%%%%%%%%%%%%%%%%%%%%%%%%%%%%%%%%%%%%%%%%%%%%%%%%%%%%%%%%%%%%%%%%%%%%%

 The shifted mean of the conditional pdf of \Xt\ given \xo\ is apparent from the
 intersection of the ellipses with the vertical red line. In this case, since
 the ellipses is mostly contained in the first and third quadrant, only in about
 $14\%$ of all cases the true value falls within the interval spanned by the two
 estimates.
 Only for negatively correlated estimates, Figure~\ref{fig:Peellmr}, for which
 the pdf mostly populates the second and fourth quadrant, the likely situation
 is that \xT\ lies within the interval spanned by the estimate, which in this
 case occurs for about $86\%$ of all cases.

 In practise, the typical situation occurring for the combination of two
 estimates of the same observable is that the estimates are positively
 correlated. This is especially likely for the situation of systematically
 dominated total uncertainties, and where both estimates suffer from the
 imperfect knowledge on the same sources of uncertainty.
 In this case the most likely place for the true value to lie is outside the
 interval spanned by the two estimates, a fact that should be kept in mind.

 The information on \xT\ that can be gained by adding the information from
 \xt\ to the one from \xo\ is discussed next.
%
%------------------------------------------------------------------------------
%
\section{The special case of two correlated estimates}
\label{sec:twoest}
%
%%%%%%%%%%%%%%%%%%%%%%%%%%%%%%%%%%%%%%%%%%%%%%%%%%%%%%%%%%%%%%%%%%%%%%%%%%%%%%%
\begin{figure*}[tbp!]
\centering
\subfigure[\bet as a function of \rhof]{
  \label{fig:beta}\includegraphics[width=0.491\textwidth]{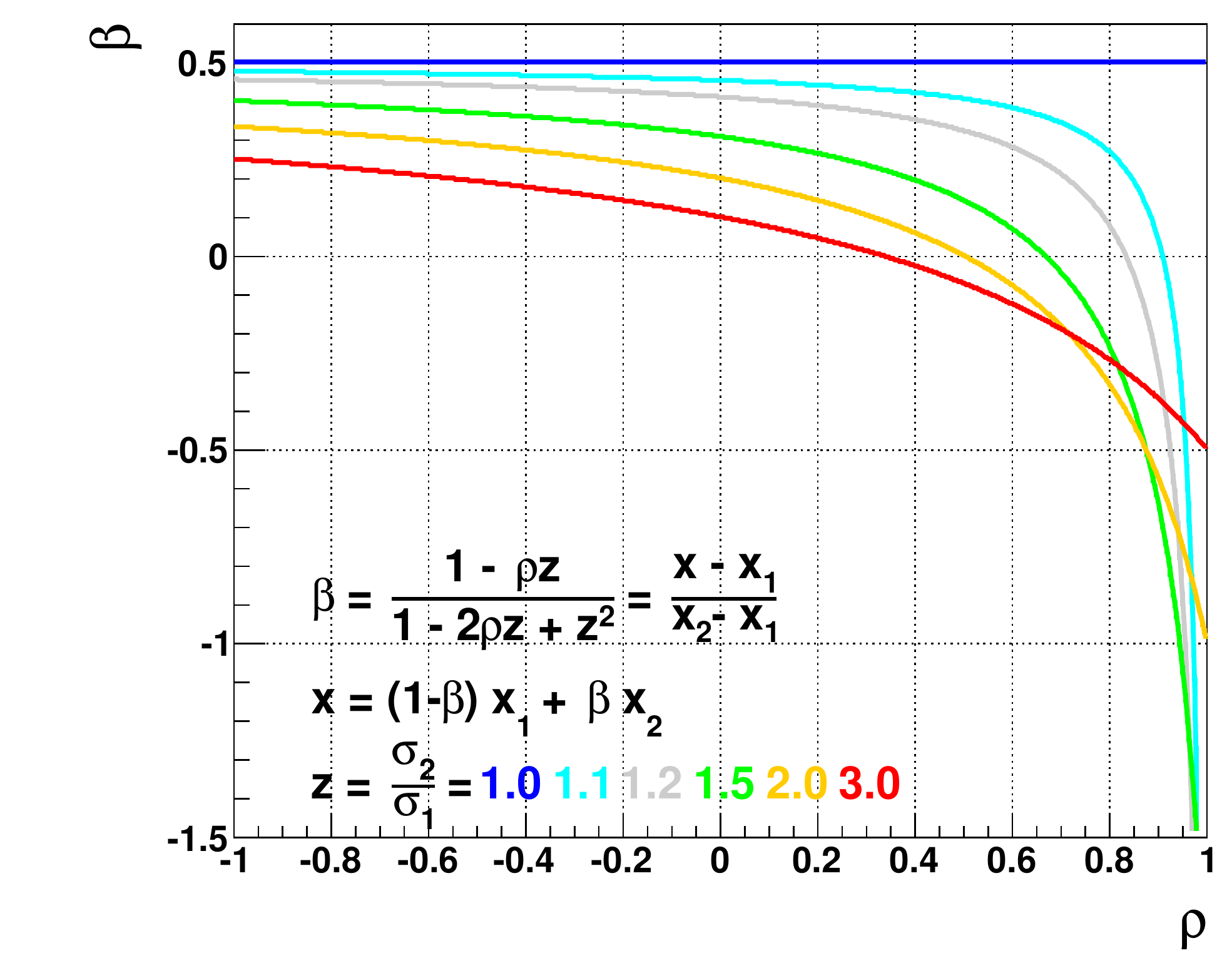}}
\subfigure[\sx/\so as a function of \rhof]{
  \label{fig:sigx}\includegraphics[width=0.491\textwidth]{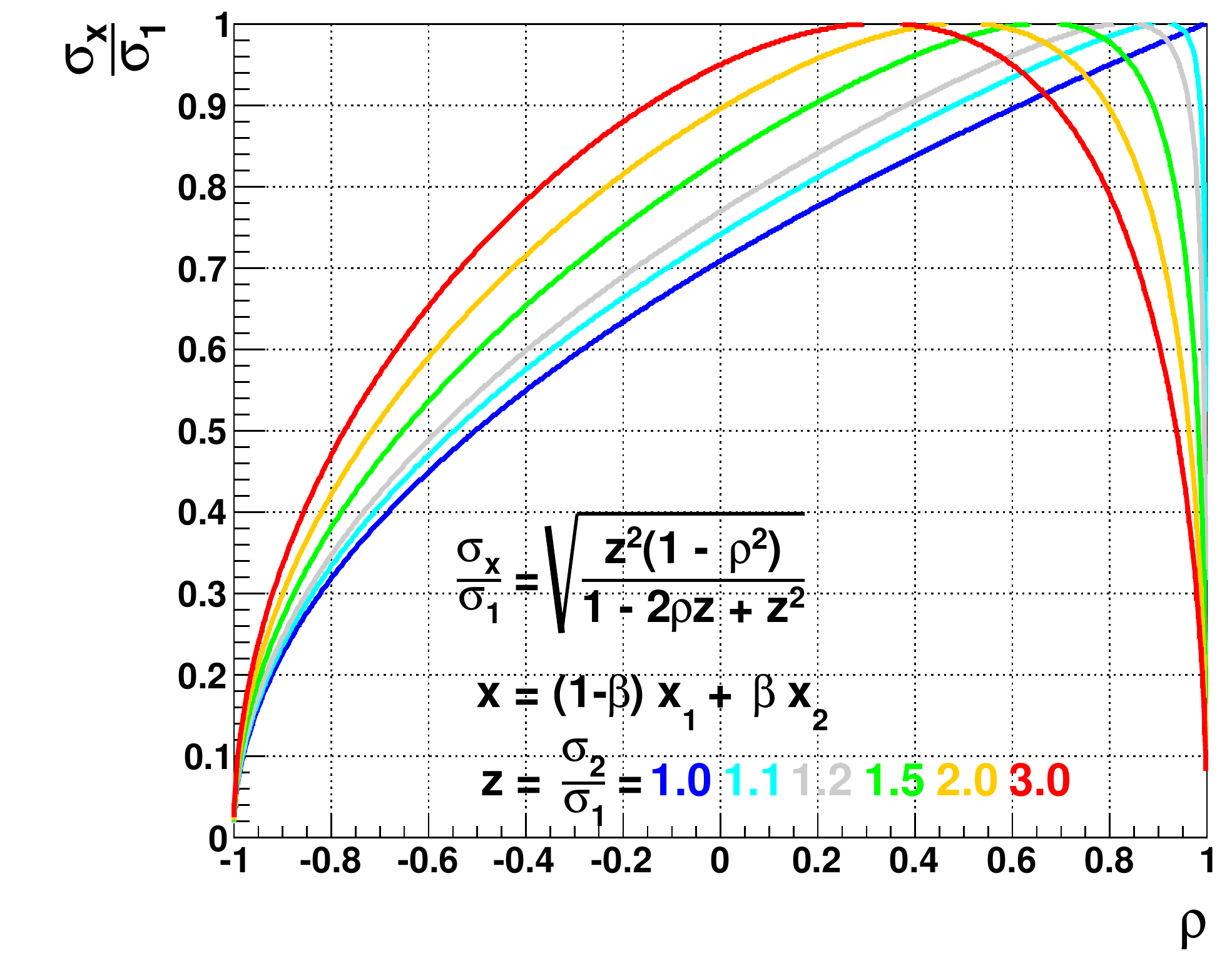}}
\subfigure[$\partial\bet/\partial\rhof$ as a function of \rhof]{
  \label{fig:dbdr}\includegraphics[width=0.491\textwidth]{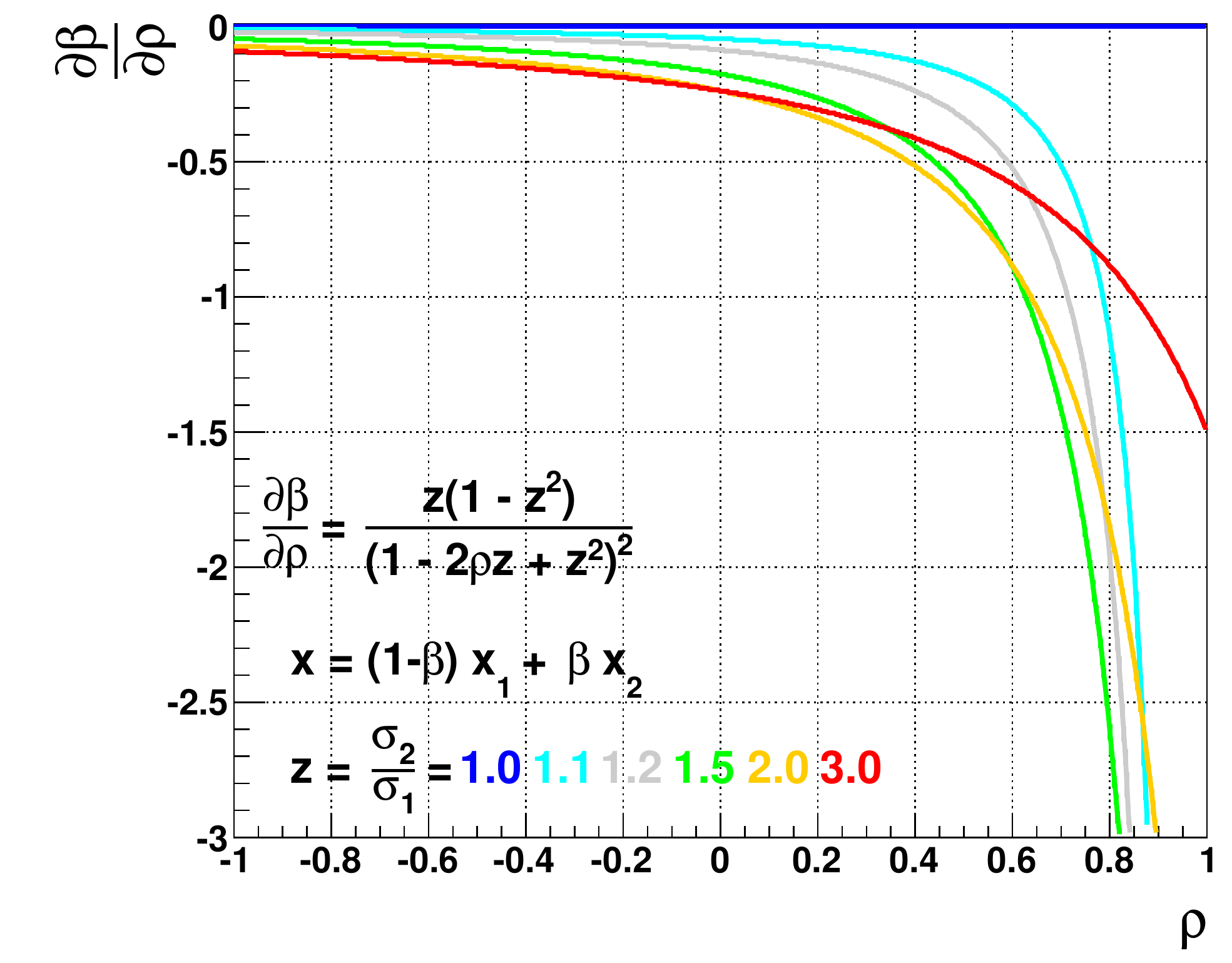}}
\subfigure[$1/\so\ \partial\sx/\partial\rhof$ as a function of \rhof]{
  \label{fig:dsdr}\includegraphics[width=0.491\textwidth]{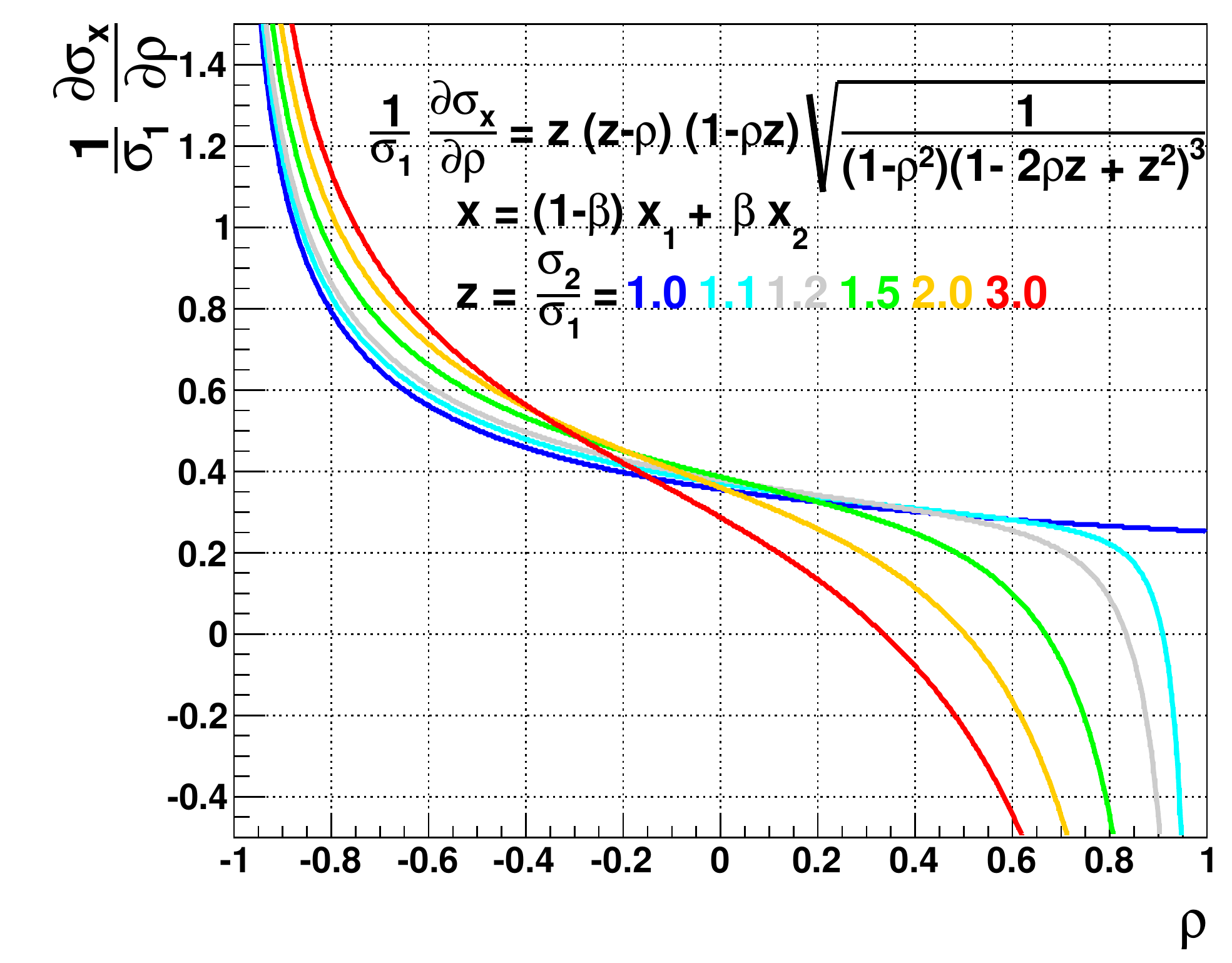}}
\caption{The results for Eqs.~\ref{eq:beta},~\ref{eq:sigx}--\ref{eq:dsdr} as
  functions of \rhof\ for a number of $z$ values. Shown are (a) \bet\ and (b)
  \sx/\so\ and their derivatives with respect to \rhof, (c)
  $\partial\bet/\partial\rhof$ and (d) $1/\so\ \partial\sx/\partial\rhof$.}
\label{fig:bsvsr}
\end{figure*}
%%%%%%%%%%%%%%%%%%%%%%%%%%%%%%%%%%%%%%%%%%%%%%%%%%%%%%%%%%%%%%%%%%%%%%%%%%%%%
%
 Again, \xo\ and \xt\ with variances \soq\ and \stq\ obeying $z=\st/\so \geq 1$
 are two Gaussian estimates from two unbiased estimators of the true value
 \xT\ of the observable, and \rhof\ denotes their total correlation with
 $-1\leq\rhof\leq 1$.
 In this situation the \BLUE\ of \xT\ is:
%
%%%%%%%%%%%%%%%%%%%%%%%%%%%%%%%%%%%%%%%%%%%%%%%%%%%%%%%%%%%%%%%%%%%%%%%%%%%%%%%
\begin{eqnarray}
 x &=& (1-\bet)\,\xo + \bet\,\xt\,,
\label{eq:linc}
\end{eqnarray}
%%%%%%%%%%%%%%%%%%%%%%%%%%%%%%%%%%%%%%%%%%%%%%%%%%%%%%%%%%%%%%%%%%%%%%%%%%%%%%%
%
 where \bet\ is the weight of the less precise estimate, and, by construction,
 the sum of weights is unity. The variable $x$ is the combined result and
 \sxq\ denotes its variance, i.e.~the uncertainty assigned to the combined value
 is \sx.

 In the following the derivation of the formulas for \bet\ and \sx/\so\ within
 the \BLUE\ formalism is repeated, see Ref.~\cite{LYO-1988}.
 The covariance matrix for the general solution of the linear combinations in
 the \BLUE\ formalism is given by Eq.~5 of Ref.~\cite{Val-0301}. For the studied
 case of two estimates of one observable it reduces to;
%
%%%%%%%%%%%%%%%%%%%%%%%%%%%%%%%%%%%%%%%%%%%%%%%%%%%%%%%%%%%%%%%%%%%%%%%%%%%%%%%
\begin{eqnarray}
\sxq &=&
        \left(\begin{array}{c}  1-\bet \\ \bet \\\end{array} \right)^T
        \cdot
        \left( \begin{array}{cc}
                    \soq & \rhof\,\so\,\st \\
          \rhof\,\so\,\st & \stq \\
          \end{array} \right)
        \cdot
        \left(\begin{array}{c}  1-\bet \\ \bet \\\end{array} \right) \,,
\label{eq:blue1}
\end{eqnarray}
%%%%%%%%%%%%%%%%%%%%%%%%%%%%%%%%%%%%%%%%%%%%%%%%%%%%%%%%%%%%%%%%%%%%%%%%%%%%%%%
%
 dividing by \soq\ and inserting $z$ yields:
%
%%%%%%%%%%%%%%%%%%%%%%%%%%%%%%%%%%%%%%%%%%%%%%%%%%%%%%%%%%%%%%%%%%%%%%%%%%%%%%%
\begin{eqnarray}
\frac{\sxq}{\soq} &=&\left(\begin{array}{c}  1-\bet \\ \bet \\\end{array} \right)^T
        \cdot
        \left( \begin{array}{cc}
                    1 & \rhof z \\
                \rhof z & z^2 \\
          \end{array} \right)
        \cdot
        \left(\begin{array}{c}  1-\bet \\ \bet \\\end{array} \right) \,,
\label{eq:blue2}
\end{eqnarray}
%%%%%%%%%%%%%%%%%%%%%%%%%%%%%%%%%%%%%%%%%%%%%%%%%%%%%%%%%%%%%%%%%%%%%%%%%%%%%%%
%
 multiplication results in:
%
%%%%%%%%%%%%%%%%%%%%%%%%%%%%%%%%%%%%%%%%%%%%%%%%%%%%%%%%%%%%%%%%%%%%%%%%%%%%%%%
\begin{eqnarray}
\frac{\sxq}{\soq} &=& (1-\bet)^2 + 2\rhof z \bet (1-\bet) + \bet^2z^2\nonumber\\
                  &=& 1 - 2\bet (1-\rhof z) + \bet^2 (1- 2\rhof z +z^2) \,,
\label{eq:blue3}
\end{eqnarray}
%%%%%%%%%%%%%%%%%%%%%%%%%%%%%%%%%%%%%%%%%%%%%%%%%%%%%%%%%%%%%%%%%%%%%%%%%%%%%%%
%
 taking the derivative with respect to \bet\ equal to zero (i.e.~the
 \chiq\ minimisation) gives:
%
%%%%%%%%%%%%%%%%%%%%%%%%%%%%%%%%%%%%%%%%%%%%%%%%%%%%%%%%%%%%%%%%%%%%%%%%%%%%%%%
\begin{eqnarray}
\frac{\partial}{\partial\,\bet}(
\frac{\sxq}{\soq}) &=& -2(1-\rhof z) + 2\bet (1- 2\rhof z + z^2) = 0.
\label{eq:blue4}
\end{eqnarray}
%%%%%%%%%%%%%%%%%%%%%%%%%%%%%%%%%%%%%%%%%%%%%%%%%%%%%%%%%%%%%%%%%%%%%%%%%%%%%%%
%
 Finally, after solving for \bet\ one obtains:
% 
%%%%%%%%%%%%%%%%%%%%%%%%%%%%%%%%%%%%%%%%%%%%%%%%%%%%%%%%%%%%%%%%%%%%%%%%%%%%%%%
\begin{eqnarray}
  \bet &=& \frac{1 - \rho z}{1 - 2\rho z + z^{2}}
        =  \frac{1 - \rho z}{(1 - \rho z)^2 + z^{2}(1-\rhof^2)} \label{eq:beta} \,
\end{eqnarray}
%%%%%%%%%%%%%%%%%%%%%%%%%%%%%%%%%%%%%%%%%%%%%%%%%%%%%%%%%%%%%%%%%%%%%%%%%%%%%%%
%
 which is valid for $-1 \leq \rhof \leq 1$ and $z \geq 1$, but for $\rhof = z = 1$.

 The last term in Eq.~\ref{eq:beta} shows that the denominator of \bet\ is
 always positive such that the sign of \bet\ is determined by the sign of the
 numerator.
 The resulting \bet\ as a function of \rhof, and for various $z$ values is shown
 in Figure~\ref{fig:beta}.
 Identifying Eq.~\ref{eq:linc} and Eq.~\ref{eq:beta} yields:
%
%%%%%%%%%%%%%%%%%%%%%%%%%%%%%%%%%%%%%%%%%%%%%%%%%%%%%%%%%%%%%%%%%%%%%%%%%%%%%%%
\begin{eqnarray}
 \frac{1}{2} \geq \bet &=&\frac{x - \xo}{\xt- \xo} = 
                       \frac{1 - \rho z}{1 - 2\rho z + z^{2}} 
                     \geq \frac{1}{1 - z}\,,
 \label{eq:differ}
\end{eqnarray}
%%%%%%%%%%%%%%%%%%%%%%%%%%%%%%%%%%%%%%%%%%%%%%%%%%%%%%%%%%%%%%%%%%%%%%%%%%%%%%%
%
 where the left limit has been derived at $\rhof\neq 1$ $z = 1$, and the right
 limit at $\rhof = 1$.
%
%%%%%%%%%%%%%%%%%%%%%%%%%%%%%%%%%%%%%%%%%%%%%%%%%%%%%%%%%%%%%%%%%%%%%%%%%%%%%%%
\begin{figure*}[tbp!]
\centering
\subfigure[\bet as a function of $z$]{
  \label{fig:bvsz}\includegraphics[width=0.491\textwidth]{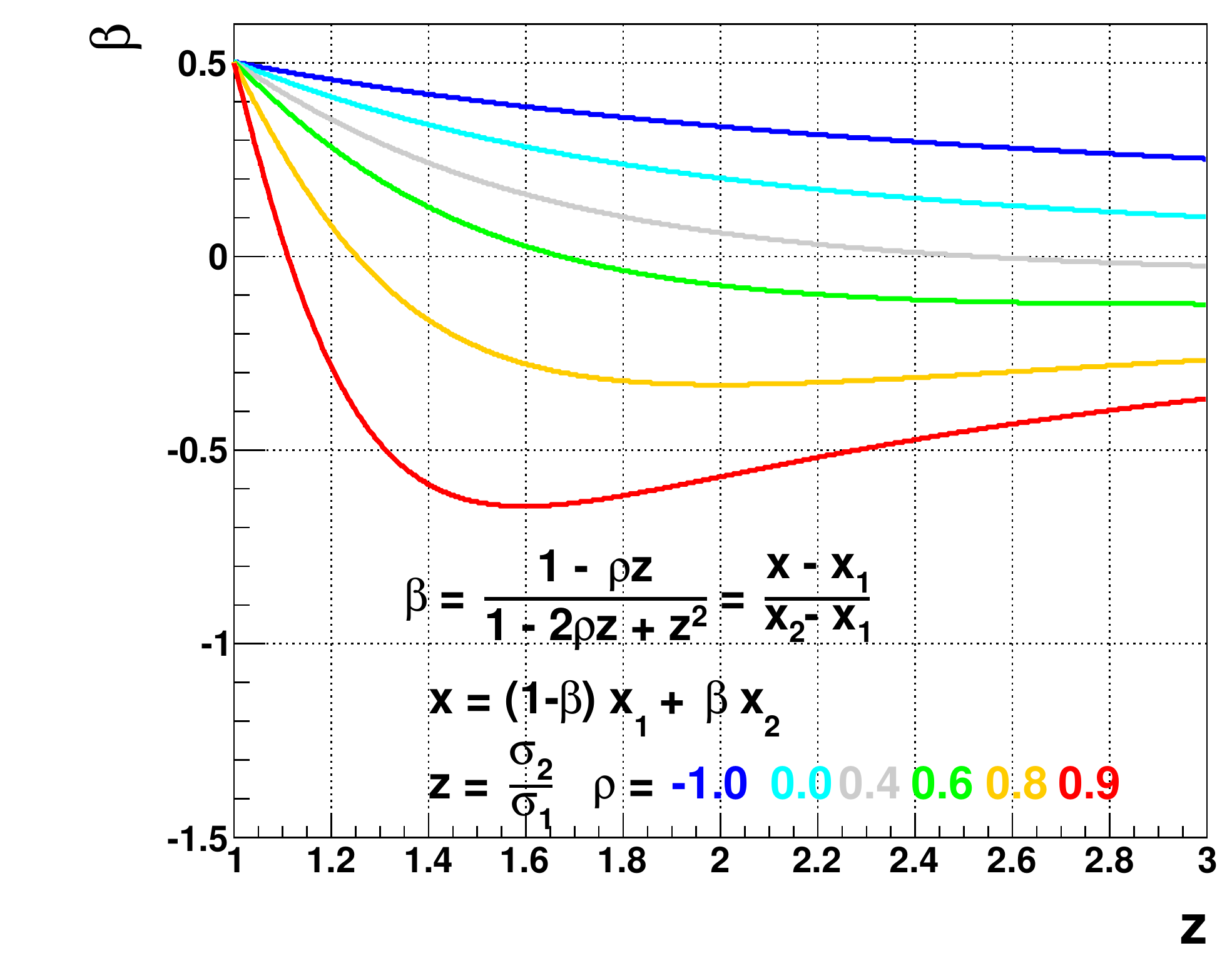}}
\subfigure[\sx/\so as a function of $z$]{
  \label{fig:svsz}\includegraphics[width=0.491\textwidth]{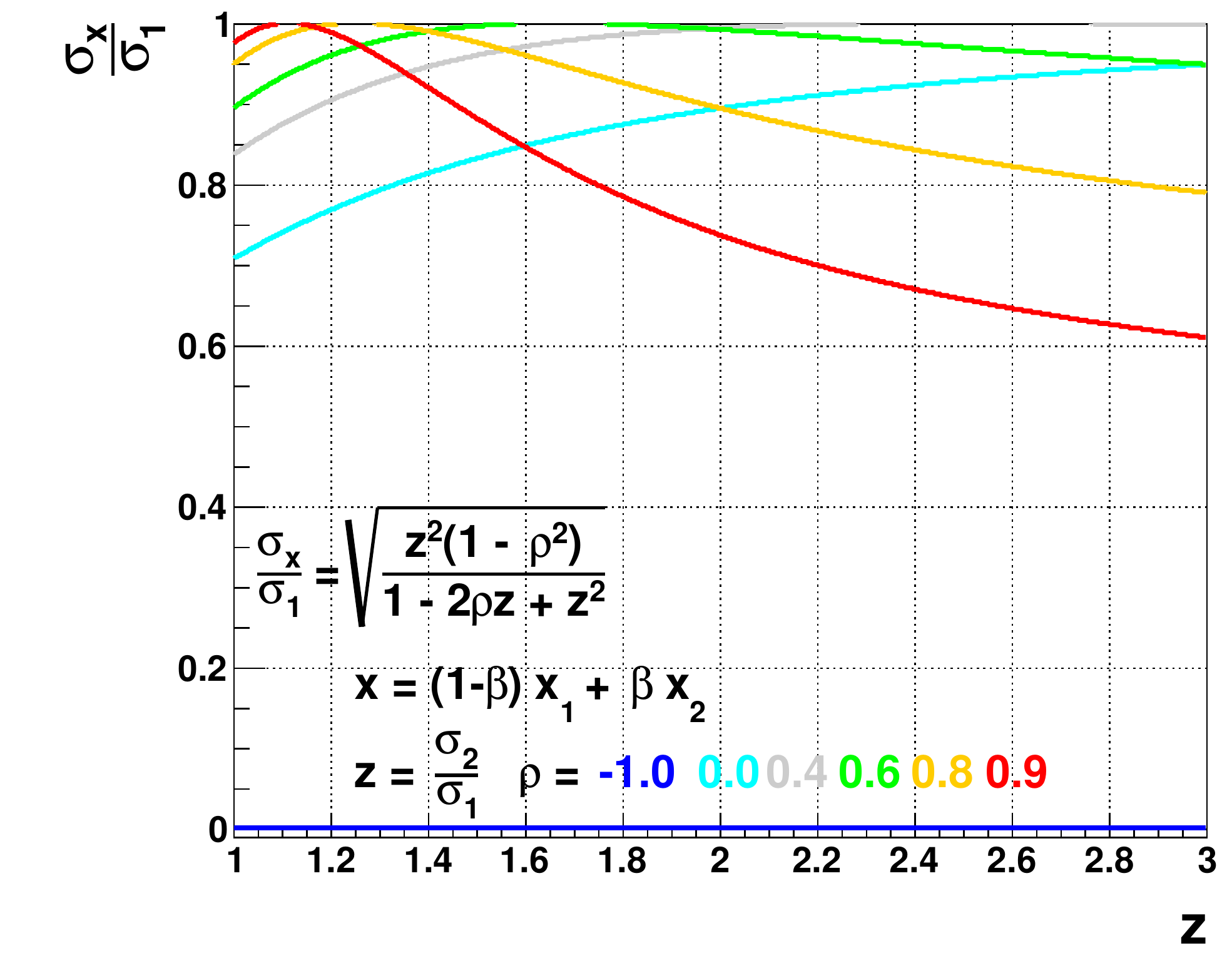}}
\subfigure[$\partial\bet/dz$ as a function of $z$]{
  \label{fig:dbdz}\includegraphics[width=0.491\textwidth]{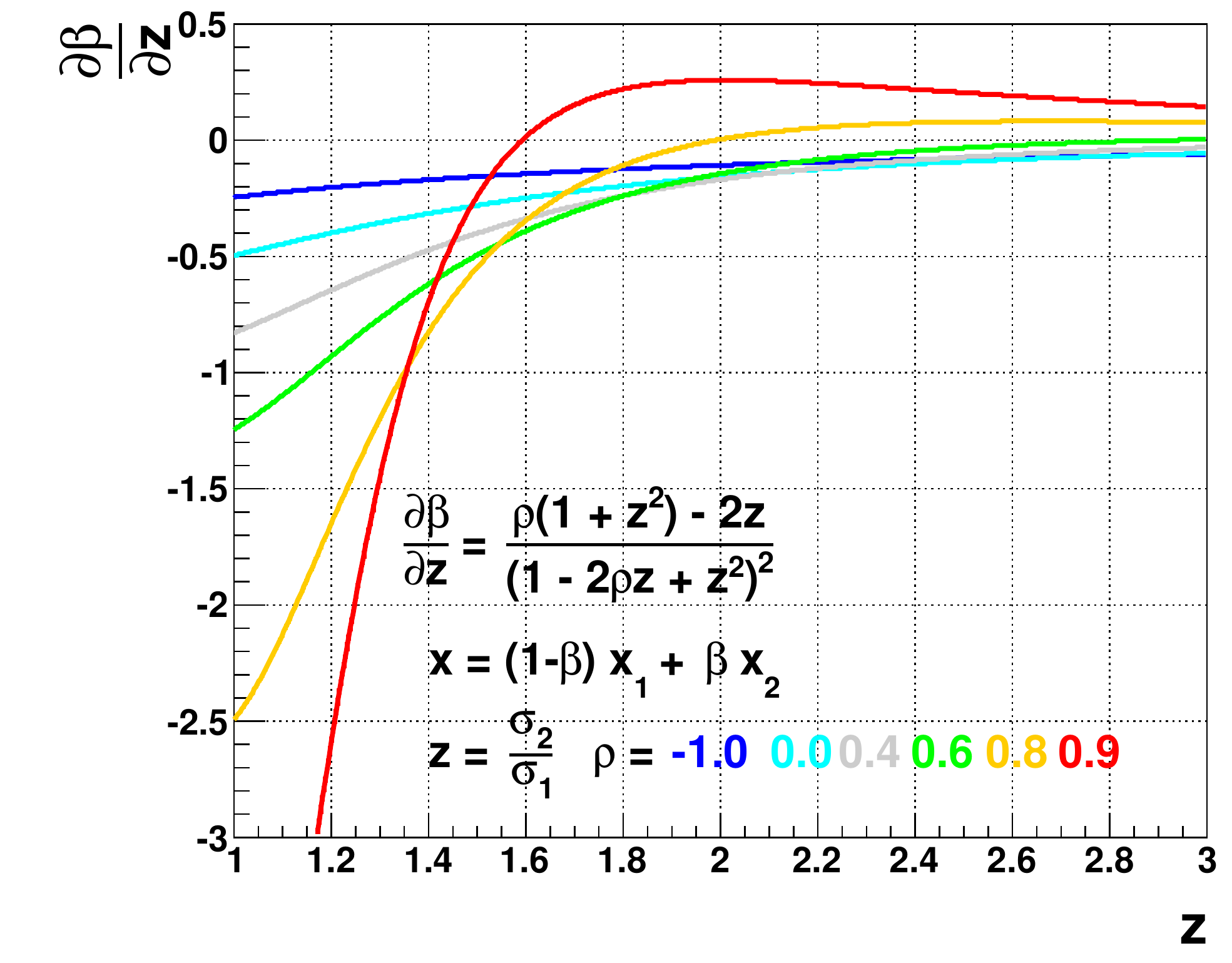}}
\subfigure[$1/\so\ \partial\sx/dz$ as a function of $z$]{
  \label{fig:dsdz}\includegraphics[width=0.491\textwidth]{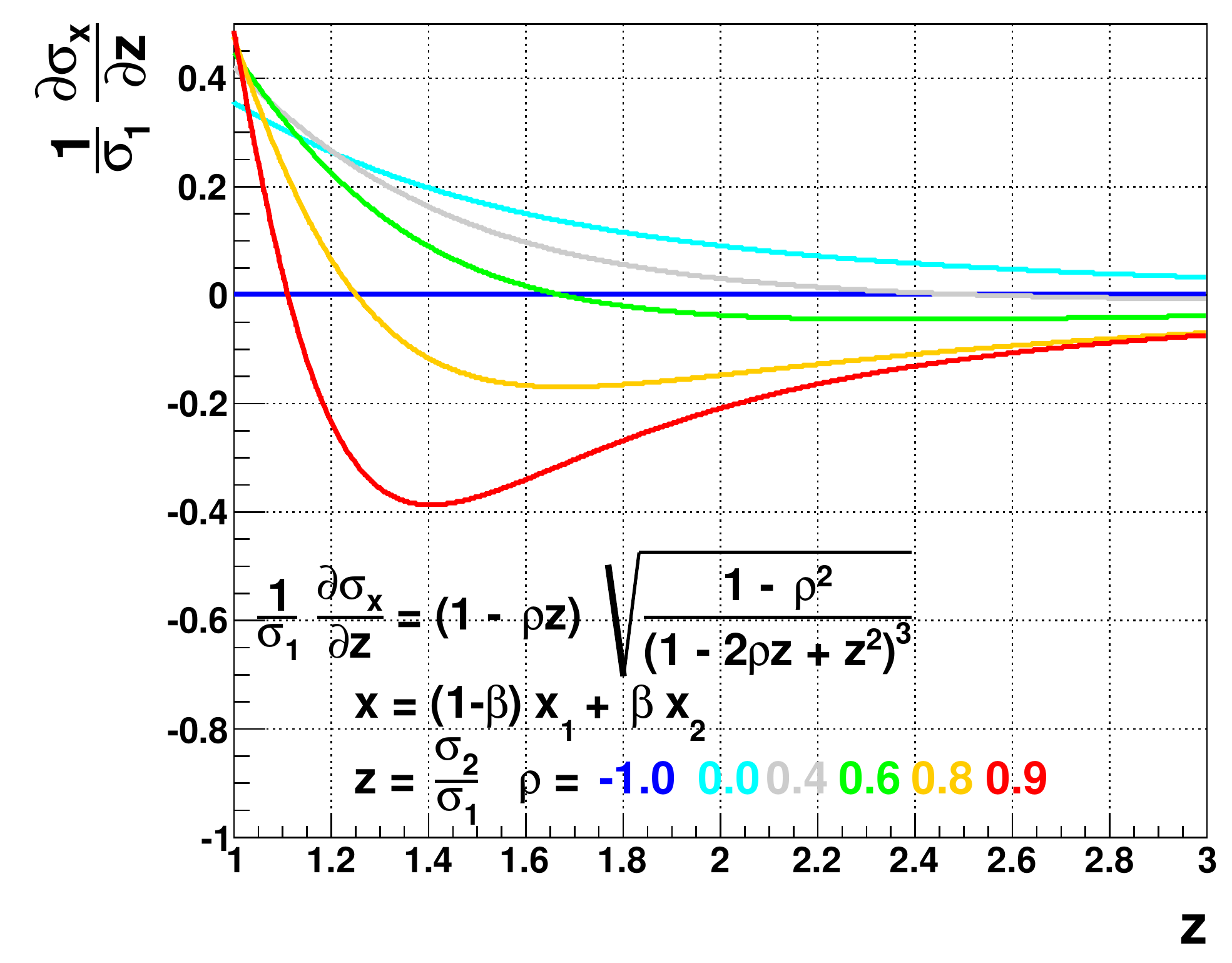}}
\caption{The results for
  Eqs.~\ref{eq:beta},~\ref{eq:sigx},~\ref{eq:dbdz}--\ref{eq:dsdz} as functions
  of $z$ for a number of \rhof\ values.  Shown are (a) \bet\ and (b)
  \sx/\so\ and their derivatives with respect to $z$, (c) $\partial\bet/\partial
  z$ and (d) $1/\so\ \partial\sx/\partial z$.}
\label{fig:bsvsz}
\end{figure*}
%%%%%%%%%%%%%%%%%%%%%%%%%%%%%%%%%%%%%%%%%%%%%%%%%%%%%%%%%%%%%%%%%%%%%%%%%%%%%

 A few features are important to understand the results of the combination.
 As expected, the value of \bet\ has to be smaller or equal than 0.5, because
 otherwise \xt\ would be the more precise estimate. Since the denominator in
 Eq.~\ref{eq:beta} is positive for all allowed values of \rhof\ and $z$, the
 function for \bet\ turns negative for $\rhof>1/z$ as shown in
 Figure~\ref{fig:beta}.
 This is exactly the point at which for a given \xo\ the conditional probability
 for \Xt\ to be even further away from \xT\ than \xo\ is, exceeds 50$\%$, see
 Section~\ref{sec:corest}.

 The first equal sign in Eq.~\ref{eq:differ} means that the value of \bet\ can
 be interpreted as the difference of the combined value from the more precise
 estimate in units of the difference of the two estimates.
 If \bet\ is positive, the signs of the numerator and denominator are identical
 and $x$ lies within the interval spanned by \xo\ and \xt. Given $\bet\leq 0.5$
 it never lies further away from the more precise estimate than half the
 difference of the two. Again, this is expected since the more precise estimate
 should dominate the combination.
 In contrast, if \bet\ is negative, the signs of the numerator and denominator
 are different. This means the value of $x$ lies on the opposite side of
 \xo\ than \xt\ does, or in other words, the combined value lies outside the
 interval spanned by the two estimates.
 Given the discussion about the conditional pdf in Section~\ref{sec:corest}, a
 very desirable feature.

 Inserting the result for \bet\ into Eq.~\ref{eq:blue3} yields:
%
%%%%%%%%%%%%%%%%%%%%%%%%%%%%%%%%%%%%%%%%%%%%%%%%%%%%%%%%%%%%%%%%%%%%%%%%%%%%%%%
\begin{eqnarray}
\frac{\sxq}{\soq} &=& 1 - 2\frac{(1-\rhof z)^2}{1- 2\rhof z + z^2}
                        +  \frac{(1-\rhof z)^2}{1- 2\rhof z + z^2}\nonumber\\
                  &=& \frac{(1- 2\rhof z + z^2)-(1-\rhof z)^2}{1-2\rhof z+z^2}\,,
\label{eq:blue5}
\end{eqnarray}
%%%%%%%%%%%%%%%%%%%%%%%%%%%%%%%%%%%%%%%%%%%%%%%%%%%%%%%%%%%%%%%%%%%%%%%%%%%%%%%
%
 which after evaluating the numerator and taking the square root gives:
%
%%%%%%%%%%%%%%%%%%%%%%%%%%%%%%%%%%%%%%%%%%%%%%%%%%%%%%%%%%%%%%%%%%%%%%%%%%%%%%%
\begin{eqnarray}
 \frac{\sx}{\so} &=& \sqrt{\frac{z^{2}(1 - \rho^{2})}{1 - 2\rhof z + z^{2}}}\,.
\label{eq:sigx}
\end{eqnarray}
%%%%%%%%%%%%%%%%%%%%%%%%%%%%%%%%%%%%%%%%%%%%%%%%%%%%%%%%%%%%%%%%%%%%%%%%%%%%%%%
%
 The resulting \sx/\so, as a function of \rhof, and for various $z$ values is
 shown in Figure~\ref{fig:sigx}.
 This variable quantifies the uncertainty of the combined value in units of the
 uncertainty of the more precise estimate, i.e.~$1-\sx/\so$ is the relative
 improvement achieved by also using \xt, i.e.~including the information
 contained in the less precise estimator.
 Consequently, \sx/\so\ can be used to decide whether it is worth combining.

 Since in the numerator of Eq.~\ref{eq:blue5} the first term is identical to the
 denominator (which is always positive, see Eq.~\ref{eq:beta}), and the second
 term is positive for all values of \rhof\ and $z$, the value of \sx/\so\ is
 always smaller or equal to unity, as shown in Figure~\ref{fig:sigx}.
 Again this is expected, since including the information from the estimate
 \xt\ should improve the knowledge on $x$, which means its precision \sx.
 Not surprisingly, the value of \sx/\so\ is exactly one for $\rhof = 1/z$,
 i.e.~for $\bet=0$. In this situation, the value of \xt\ is irrelevant in the
 linear combination of Eq.~\ref{eq:linc}, and consequently $x=\xo$ and
 $\sx=\so$.
 Finally, \sx/\so\ is exactly zero if $\rhof=\pm 1$ in accordance with the
 variance of $\Xt$ for the conditional PDF given \xo\ and \rhof, shown in
 Section~\ref{sec:corest}.
 This means that for the fully correlated or fully anti-correlated case of two
 estimators, given \xo, the result is known for sure, and the outcome of the
 second estimate has to be $\xt=\xT+\rho z(\xo-\xT)$.
 For combinations of experimental results, for which for all pairs of estimates
 there are also uncorrelated components of the uncertainty, this situation never
 happens.

 The typical situation is that both \rhof\ and $z$ are only known with some
 precision. In this situation it is essential to analyse the sensitivity of the
 central value of the combination to this imperfect knowledge that is encoded in
 the respective derivatives.
 The derivatives of \bet\ and \sx/\so\ with respect to the parameters \rhof\ and
 $z$ have been derived in this paper and are given in
 Eqs.~\ref{eq:dbdr}--\ref{eq:dsdz}.
%
%%%%%%%%%%%%%%%%%%%%%%%%%%%%%%%%%%%%%%%%%%%%%%%%%%%%%%%%%%%%%%%%%%%%%%%%%%%%%%%
\begin{eqnarray}
 \frac{\partial\,\bet}{\partial\,\rhof} &=& \frac{z(1 - z^{2})}
                              {(1 - 2\rhof z + z^{2})^{2}}
 \label{eq:dbdr}\\
 \frac{\partial\,\frac{\sx}{\so}}
      {\partial\,\rhof}   &=& z (z-\rhof) (1-\rhof z) 
                              \sqrt{\frac{1}{(1-\rhof^{2})(1-2\rhof z+z^{2})^{3}}}
 \label{eq:dsdr}\\
 \frac{\partial\,\bet}{\partial\,z} 
                          &=& \frac{\rhof(1 + z^{2}) - 2z}{(1 - 2\rhof z + z^{2})^{2}}
 \label{eq:dbdz}\\
 \frac{\partial\,\frac{\sx}{\so}}
      {\partial\,z}       &=& (1-\rhof z)
                              \sqrt{\frac{1-\rhof^{2}}{(1-2\rhof z + z^{2})^{3}}}
 \label{eq:dsdz}
\end{eqnarray}
%%%%%%%%%%%%%%%%%%%%%%%%%%%%%%%%%%%%%%%%%%%%%%%%%%%%%%%%%%%%%%%%%%%%%%%%%%%%%%%
%
 The resulting variations of the combined value, Eq.~\ref{eq:linc}, are given in
 Eqs.~\ref{eq:dxdr}--\ref{eq:dxdz}.
%
%%%%%%%%%%%%%%%%%%%%%%%%%%%%%%%%%%%%%%%%%%%%%%%%%%%%%%%%%%%%%%%%%%%%%%%%%%%%%%%
\begin{eqnarray}
\frac{\partial\,x}{\partial\,\rho} &=& (\xt-\xo)\, 
                                       \frac{\partial\,\bet}{\partial\,\rho}\label{eq:dxdr}\\
   \frac{\partial\,x}{\partial\,z} &=& (\xt-\xo)\, 
                                       \frac{\partial\,\bet}{\partial\,z}\label{eq:dxdz}
\label{eq:xderivat}
\end{eqnarray}
%%%%%%%%%%%%%%%%%%%%%%%%%%%%%%%%%%%%%%%%%%%%%%%%%%%%%%%%%%%%%%%%%%%%%%%%%%%%%%%
%
 The derivatives of \bet\ and $\sx/\so$ with respect to \rhof\ as functions of
 \rhof, and for various $z$ values, Eq.~\ref{eq:dbdr} and Eq.~\ref{eq:dsdr}, are
 shown in Figures~\ref{fig:dbdr} and~\ref{fig:dsdr}.
 The equations for \bet\ and $\sx/\so$, this time as a function of $z$ and for
 various \rhof\ values, are shown in Figures~\ref{fig:bvsz} and~\ref{fig:svsz}.
 Finally, the derivatives of \bet\ and $\sx/\so$ with respect to $z$ as
 functions of $z$, and for various \rhof\ values, Eq.~\ref{eq:dbdz} and
 Eq.~\ref{eq:dsdz}, are shown in Figures~\ref{fig:dbdz} and~\ref{fig:dsdz}.
 These derivatives can be used to visualise the sensitivity of the combined
 result to the imperfect knowledge on both the correlation \rhof\ and the
 uncertainty ratio $z$ of the individual estimators, and help to decide on
 whether to refrain from combining.
 This decision should only be based on the parameters of the combination but not
 on the outcome for a particular pair of estimates \xo\ and \xt. This is because
 these parameters are features of the underlying two-dimensional pdf of the
 estimators, whereas the two specific values are just a pair of estimates,
 i.e.~a single possible likely or unlikely outcome of results.
 A suggestion for how to proceed is given in Section~\ref{sec:decide}.
%
%------------------------------------------------------------------------------
%
\section{Estimator properties}
\label{sec:estpro}
 In general, in experimental analyses an estimator is constructed by studying
 Monte Carlo simulated events that are taken as data substitutes.
 Using those events it is verified that the estimator is unbiased. 
 By applying the method to data, the measured value of the estimator, i.e.~the
 estimate, e.g.~\xo, is obtained together with its statistical uncertainty.
 Subsequently, individual systematic uncertainties are obtained for the
 estimator and assigned to the estimate. 
 For example, in top quark mass measurements like Ref.~\cite{ATL-2012-036}, this
 is achieved, e.g.~by changing the reconstructed objects like leptons and jets
 within their uncertainties, by altering the underlying Monte Carlo model for
 the signal, and by varying the background evaluations from data or simulations.
 In these procedures, the systematic variations per source $k$ of uncertainty
 are chosen to be performed in an uncorrelated way from any other source
 $k^\prime$, and the actual values of the uncertainties are considered one
 standard deviation Gaussian uncertainties.
 Consequently, the total systematic uncertainty is calculated as the square root
 of the quadratic sum of the contributions from the individual sources.
 Finally, the result is quoted as:
%
%%%%%%%%%%%%%%%%%%%%%%%%%%%%%%%%%%%%%%%%%%%%%%%%%%%%%%%%%%%%%%%%%%%%%%%%%%%%%%%
\begin{eqnarray}
 \xo  &=& \XZ{\mathrm{value}}{\mathrm{stat}}{\mathrm{\sqrt{\sum_k{syst^2}}}}.
 \label{eq:resu}
\end{eqnarray}
%%%%%%%%%%%%%%%%%%%%%%%%%%%%%%%%%%%%%%%%%%%%%%%%%%%%%%%%%%%%%%%%%%%%%%%%%%%%%%%
%
 To enable their combination, the breakdown of systematic uncertainties is
 provided.
 Consequently, the features of the estimates are:
%
%%%%%%%%%%%%%%%%%%%%%%%%%%%%%%%%%%%%%%%%%%%%%%%%%%%%%%%%%%%%%%%%%%%%%%%%%%%%%%%
\begin{Enumerate}
 \item[1)] they are unbiased,
 \item[2)] their uncertainties are assumed to be Gaussian,
 \item[3)] the uncertainty sources are constructed to be uncorrelated.
\end{Enumerate}
%%%%%%%%%%%%%%%%%%%%%%%%%%%%%%%%%%%%%%%%%%%%%%%%%%%%%%%%%%%%%%%%%%%%%%%%%%%%%%%
%
 The property 3) relates to the correlation of two sources ($k, k^\prime$) of
 uncertainties and should not be confused with the correlation \rhoijk\ of two
 estimates ($i, j$) for the same source $k$ of uncertainty.
 If there are physics reasons to believe that two sources ($k, k^\prime$) are
 indeed correlated, it is advisable to reconsider the separation of the
 uncertainty sources, because otherwise, using the quadratic sum of
 Eq.~\ref{eq:resu} is questionable.

 When performing the combination of a pair $ij$ of estimates, for each source
 $k$ of uncertainty a correlation \rhoijk\ has to be assigned for that pair.
 The statistical uncertainties are either uncorrelated, or, for the case of two
 estimates obtained from overlapping or even the same data events, their
 correlation can be obtained within the analysis by means of pseudo-experiments,
 as described e.g.~in Ref.~\cite{ATL-2012-036}.
 For the systematic uncertainties, the value of the assigned correlation always
 is a physics motivated choice that can only be made with some uncertainty.
 The easiest case occurs if the uncertainties of the estimators have been
 determined in exactly the same way, e.g.~within one experiment while using the
 identical procedure for all estimates. In this case, the assumption of
 $\rhoijk=1$ is justified, and any observed difference in the size of
 uncertainty $\sik\ne\sjk$ is likely caused by the different sensitivities of
 the estimators to that particular source of uncertainty.
 The uncertainty of this correlation assumption can be assessed by varying the
 value of \rhoijk\ within bounds to be chosen. Given the estimator property 3),
 for each source $k$ this should be performed independently from all other
 sources.

 A more complicate situation arises however, when combining estimates obtained
 by different experiments, which even may have partly been derived without
 knowledge on the procedure applied for the respective other result.
 Given the difference in strategy, there may be a smaller correlation.
 In addition, even for $\rhoijk=1$ differences in the size of the uncertainties
 can originate from a different size of variation performed for the two
 estimators. As an example, one experiment may perform larger variations of
 Monte Carlo parameters than another, an example of which can be found in
 Ref.~\cite{ATLAS-CONF-2013-102}.
 In this situation given the different dependences of \bet\ and \sx/\so\ on
 \rhof\ and $z$, the difference can not be accounted for by changes in \rhoijk,
 but the most appropriate choice is to vary \sik\ and/or \sjk.

 Given the above, an individual assessment of the correlation assumptions per
 source $k$, as is performed e.g.~in Ref.~\cite{ATLAS-CONF-2013-098}, is
 strongly preferred. In contrast, any automated procedure of simultaneous
 variations very likely can not properly account for the specific situations of
 all sources $k$. This is discussed in more detail in Section~\ref{sec:maxvar}.
 In any case, all systematic variations on the assumptions should be performed
 obeying the features of the estimators listed above.

 Frequently, the question arises whether a pair ($i, j = 1, 2$) of estimates is
 compatible. This can be decided upon using a \chiq\ that is defined as the
 squared ratio of the difference of the estimates, $\Delta$, and its
 uncertainty, $\sigma_\Delta$:
%
%%%%%%%%%%%%%%%%%%%%%%%%%%%%%%%%%%%%%%%%%%%%%%%%%%%%%%%%%%%%%%%%%%%%%%%%%%%%%%%
\begin{eqnarray}
 \chiq(\xo, \xt) &=& \left(\frac{\Delta}{\sigma_\Delta}\right)^2 
                  =   \frac{\left(\xo-\xt\right)^2}{\soq+\stq-2\rho\so\st}\,,
 \label{eq:chiq}
\end{eqnarray}
%%%%%%%%%%%%%%%%%%%%%%%%%%%%%%%%%%%%%%%%%%%%%%%%%%%%%%%%%%%%%%%%%%%%%%%%%%%%%%%
%
 which is the significance of the difference of the estimates of being
 incompatible with zero.
 Alternatively, one may exploit the related \chiq\ probability for one degree of
 freedom, \Prob{\chiq}{1}, defined as the integral:
%
%%%%%%%%%%%%%%%%%%%%%%%%%%%%%%%%%%%%%%%%%%%%%%%%%%%%%%%%%%%%%%%%%%%%%%%%%%%%%%%
\begin{eqnarray}
 \Prob{\chiq}{1} &=& \int_{\chiq}^{\infty} \Pdf{\chiq}{1}\, d\chiqp\,,
 \label{eq:chiqprob}
\end{eqnarray}
%%%%%%%%%%%%%%%%%%%%%%%%%%%%%%%%%%%%%%%%%%%%%%%%%%%%%%%%%%%%%%%%%%%%%%%%%%%%%%%
%
 which is the probability for an even larger \chiq\ to occur for any other
 pair~\cite{BAR-8901}.

 Ideally, only compatible estimates should be combined, otherwise the combined
 result is not trustworthy.
 Unfortunately, given the statistical nature of the problem, the question of
 compatibility of a given pair of estimates can not be answered unambiguously,
 i.e.~for a single pair of estimates it is impossible to decide whether this is
 an unlikely case given the underlying pdf, or an incompatible case.
 In turn this also means that no conclusions on properties of the estimator
 distribution or even the combination method can be drawn solely based on
 specific pairs of estimates and the result of their combination.

 For the situation of a larger number of estimates to be combined, it is
 advisable to inspect the distribution of the \chiq\ values of the pairwise
 compatibility tests calculated from Eq.~\ref{eq:chiq} that should resemble a
 \chiq\ distribution for one degree of freedom.
 For pairs resulting in large \chiq\ values, the analysis procedures applied in
 obtaining the uncertainties should be investigated in detail for detecting
 possible incompatibilities.
 The outcome for a specific example is discussed in the next section.

 Finally, the global \chiq\ of the combination, i.e.~the quantity minimised, for
 $i, j=1,\ldots, m$ estimates of a single observable \xT, yielding a combined
 value $x$ and with an inverse covariance matrix $V^{-1}$ is defined as:
%
%%%%%%%%%%%%%%%%%%%%%%%%%%%%%%%%%%%%%%%%%%%%%%%%%%%%%%%%%%%%%%%%%%%%%%%%%%%%%%%
\begin{eqnarray}
 \chiq &=& \sum_{i=1}^{m}\sum_{j=1}^{m} \left(\xii-x\right) V_{ij}^{-1} \left(\xj-x\right)\,.
 \label{eq:chiqtot}
\end{eqnarray}
%%%%%%%%%%%%%%%%%%%%%%%%%%%%%%%%%%%%%%%%%%%%%%%%%%%%%%%%%%%%%%%%%%%%%%%%%%%%%%%
%
 This \chiq\ is a single number per combination that should be small for a
 compatible set of estimates for the observable under investigation.
%
%------------------------------------------------------------------------------
%
\section{Relative uncertainties}
\label{sec:relunc}
 The formulas described above are only valid for Gaussian estimators with
 absolute uncertainties \sik\ for all sources.
 Here, the term absolute uncertainty means that the value of the uncertainty is
 identical for all possible values of the estimator pdf, i.e.~it is independent
 of the actual value of the estimate.
 This means it is the same for the actual estimate, any combined value, and the
 true value, such that $\si=\si(\xii)=\si(x)=\si(\xT)$.
 Therefore, irrespectively of whether it was calculated for the estimate, it
 also applies to the combined value.
 In contrast, a relative uncertainty\footnote{Sometimes in the literature the
   terms additive (absolute) and multiplicative (relative) uncertainties are
   used instead.} (e.g.~of some percent) depends on the actual value of \xT.
 Consequently, for relative uncertainties, the uncertainty assigned to the
 estimate, $\si=\si(\xii)$, is formally incorrect, since it should correspond to
 the uncertainty of the estimator pdf, i.e.~$\si=\si(\xT)$, which has a
 different value.

 Within the \BLUE\ method this can be accounted for approximately by performing
 the combination in an iterative way, see Ref.~\cite{LYO-9001,CHI-9101}.
 In this procedure, starting from the initially assigned value, after each
 iteration the uncertainty is replaced by the expected uncertainty of the true
 value \xT, approximated by the one of the combined value $x$.
 For most applications, for a given source $k$ of systematic uncertainty, a
 linear dependence of the uncertainty \sik\ on $x$ is
 assumed\footnote{Typically, e.g.~for counting experiments, the estimate is
   proportional to the observed number of events $N$, whereas the statistical
   uncertainty scales with $\sqrt{N}$, i.e.~it is not linear in the estimate.},
 however, there also exist more complicate cases like the one discussed in
 Ref.~\cite{LYO-9001}.
%
%------------------------------------------------------------------------------
%
\begin{table*}[tbp!]
\caption{Comparison of the combinations for Peelle's Pertinent Puzzle for the
  \BLUE\ method with absolute and relative uncertainties using various scenarios
  for the estimates and their correlation.
 The five scenarios analysed are: \sA the original values for the estimates
 $i=1, 2$, uncertainties, $k=0, 1$ and correlations \rhootk\ with $\rhootz=0$
 and $\rhooto=1$, $\sB=\sA$ but with all uncertainties scaled by a factor two,
 $\sC~(\sD)=\sA$ but with a changed value for the second estimate and with the
 original (rescaled) uncertainties, and $\sE=\sA$ but with a decreased value of
 the assumed correlation for the systematic uncertainty, i.e.~for $k=1$.
 The estimates are listed together with their uncertainties. In addition given
 are the parameters and results of the combination.}
\label{tab:BluePeel}
\begin{center}
\begin{tabular}{rrrrrrrrrrrrrrr}
\hline\noalign{\smallskip}
\multicolumn{15}{c}{Estimates}\\\noalign{\smallskip}\hline\noalign{\smallskip}
           & Value & Stat & Syst & Full & &&&&&&&\\\noalign{\smallskip}\hline\noalign{\smallskip}
           &  \xii & \sinv  & \sio  & $\si$ &\multicolumn{10}{l}{Comment} \\\noalign{\smallskip}\hline\noalign{\smallskip}
       \sA &  1.00 & 0.10 & 0.20 & 0.22 & &&&&&& \\
           &  1.50 & 0.15 & 0.30 & 0.34 & &&&&&& \\
       \sB &  1.00 & 0.20 & 0.40 & 0.45 &\multicolumn{10}{l}{All uncertainties multiplied by two} \\
           &  1.50 & 0.30 & 0.60 & 0.67 & &&&&&& \\
       \sC &  1.00 & 0.10 & 0.20 & 0.22 &\multicolumn{10}{l}{Changed value for \xt\ with unchanged uncertainties} \\
           &  1.25 & 0.15 & 0.30 & 0.34 & &&&&&& \\
       \sD &  1.00 & 0.10 & 0.20 & 0.22 &\multicolumn{10}{l}{Changed value for \xt\ with rescaled uncertainties} \\
           &  1.25 & 0.13 & 0.25 & 0.28 & &&&&&& \\
       \sE &  1.00 & 0.10 & 0.20 & 0.22 &\multicolumn{10}{l}{Changed correlation for the systematic uncertainty 
                                                             $\rho_{121} = 0.05\quad$} \\
           &  1.50 & 0.15 & 0.30 & 0.34 & &&&&&& \\
\noalign{\smallskip}\hline\noalign{\smallskip}
\multicolumn{15}{c}{Combined results}\\\noalign{\smallskip}\hline\noalign{\smallskip}
           &  $x, y$  & \ssta  & \ssys & \sx & \chiq\ & \rhof\ & $z$  & $\bet$  &
  $\frac{\sx}{\so}$ & $\frac{d\bet}{d\rhof}$ & $\frac{1}{\so}\frac{d\sx}{d\rhof}$ 
                    & $\frac{d\bet}{dz}$     & $\frac{1}{\so}\frac{d\sx}{dz}$ \\
\noalign{\smallskip}\hline\noalign{\smallskip}
\multicolumn{15}{c}{\BLUE}\\\noalign{\smallskip}\hline\noalign{\smallskip}
       \sA &  0.88 & 0.13 & 0.18 & 0.22 & 5.9 & 0.80 & 1.50 & -0.24 &  0.98 & -2.60 & -0.45 & -0.55 & -0.15 \\
       \sB &  0.88 & 0.26 & 0.35 & 0.44 & 1.5 & 0.80 & 1.50 & -0.24 &  0.98 & -2.60 & -0.45 & -0.55 & -0.15 \\
       \sC &  0.94 & 0.13 & 0.18 & 0.22 & 1.5 & 0.80 & 1.50 & -0.24 &  0.98 & -2.60 & -0.45 & -0.55 & -0.15 \\
       \sD &  1.00 & 0.10 & 0.20 & 0.22 & 2.2 & 0.80 & 1.25 &  0.00 &  1.00 & -2.22 &  0.00 & -1.42 &  0.00 \\
       \sE &  1.15 & 0.08 & 0.17 & 0.19 & 1.6 & 0.04 & 1.50 &  0.30 &  0.85 & -0.19 &  0.37 & -0.29 &  0.17 \\
\noalign{\smallskip}\hline\noalign{\smallskip}
\multicolumn{15}{c}{Relative uncertainties}\\\noalign{\smallskip}\hline\noalign{\smallskip}
       \sA &  1.25 & 0.09 & 0.25 & 0.27 & 8.0 & 0.80 & 1.00 &  0.50 &  0.95 &  0.00 &  0.26 & -2.50 &  0.47 \\
       \sB &  1.25 & 0.18 & 0.50 & 0.53 & 2.0 & 0.80 & 1.00 &  0.50 &  0.95 &  0.00 &  0.26 & -2.50 &  0.47 \\
       \sC &  1.02 & 0.10 & 0.21 & 0.23 & 2.3 & 0.80 & 1.20 &  0.08 &  1.00 & -1.95 &  0.09 & -1.66 &  0.06 \\
       \sD &  1.12 & 0.08 & 0.23 & 0.24 & 2.5 & 0.80 & 1.00 &  0.50 &  0.95 &  0.00 &  0.26 & -2.50 &  0.47 \\
       \sE &  1.25 & 0.09 & 0.18 & 0.20 & 1.7 & 0.04 & 1.00 &  0.50 &  0.72 &  0.00 &  0.35 & -0.52 &  0.36 \\
\noalign{\smallskip}\hline
\end{tabular}
\end{center}
\end{table*}
%
%------------------------------------------------------------------------------

 It is worth noticing that during the iterations the originally assigned
 uncertainties of the estimates are altered, albeit at unchanged correlation
 assumptions.
 For example, when using the same linear dependence for all estimates $i$ and a
 given source of uncertainty $k$, this means that after the first iteration the
 uncertainty from this source is identical for all estimates, and finally, at
 convergence its value amounts to a given fraction of the combined result.
 Assuming this behaviour for all uncertainties of a pair of estimates leads to
 $z=1$. This results in $\bet=0.5$, see Eq.~\ref{eq:beta}, for all possible
 values of \rhof, and the combination reduces to averaging the estimates,
 i.e.~$x=(\xo+\xt)/2$, irrespectively of their correlation.
 Solely the uncertainty \sx\ depends on the value of the correlation,
 i.e.~Eq.~\ref{eq:sigx} reduces to $\sx/\so\ = \sqrt{(1+\rhof)/2}$. 
 An example of this situation is Peelle's Pertinent Puzzle.

 Numerically, the difference of using absolute or relative uncertainties rarely
 is of importance, especially so when combining consistent precision
 measurements.
 This is because a difference of $n\%$ between the estimates and the combined
 value only results in a relative change of $n\%$ in \sik. Given that \sik\ in
 itself is small compared to \xii, this likely ends up in very small differences
 in $x$ and \sx, in any case well below the size of the respective uncertainty.

 At first sight a counter example is the original formulation of Peelle's
 Pertinent Puzzle~\cite{PEL-8701,CHI-9101}\footnote{The puzzle was introduced in
   an internal memorandum~\cite{PEL-8701}. The originally used numerical values
   can be found in Ref.~\cite{CHI-9101}.}, for which the estimates are given in
 Table~\ref{tab:BluePeel} scenario \sA.
 This puzzle, however restricted to situations like scenario \sA\ and
 investigating different models for the uncertainties, has been discussed in the
 literature, see e.g.~\cite{CHI-9101,BLO-0301,HAN-0501}.

 For scenario \sA, the statistical uncertainties are uncorrelated and the
 systematic uncertainties are fully correlated, which results in $\rhof =
 0.8$. Given that a percentage uncertainty of $10\%\, (20\%)$ is quoted for the
 statistical (systematic) uncertainty, the ratio of the total uncertainty equals
 the ratio of the estimates, i.e.~$z = 1.5$.
 The \chiq\ of the two estimates, calculated from Eq.~\ref{eq:chiq}, is large,
 i.e.~$\chiq(\xo, \xt)=5.9$ and $\Prob{\chiq}{1}=1.5\%$, which means whatever
 method is used, a combination of this pair of estimates is questionable.

 Given the procedures applied to obtain the systematic uncertainty it should be
 possible to decide whether this source is an absolute or relative uncertainty.
 Here, the combination is performed for both assumptions, i.e.~using either
 absolute and relative uncertainties for all sources of uncertainty, see also
 Ref.~\cite{CHI-9101}. The results are listed in Table~\ref{tab:BluePeel},
 scenario \sA.
 In the case of relative uncertainties, given the combined value, the final
 statistical (systematic) uncertainties assigned to the estimates are $0.13$
 ($0.25$), i.e.~they are equal for both estimates and different from the values
 quoted in the upper part of the table. The resulting corresponding
 uncertainties of the combined result are $0.09$ and $0.25$, respectively.
 Due to the changes in uncertainties, for the \BLUE\ method with relative
 uncertainties the \chiq\ of the two estimates, calculated from the finally
 assigned uncertainties, is even larger, i.e.~$\chiq(\xo, \xt)=8.0$ and
 $\Prob{\chiq}{1}=0.5\%$. As explained above, by construction, the combined
 result is the mean of the two estimates.
%
%%%%%%%%%%%%%%%%%%%%%%%%%%%%%%%%%%%%%%%%%%%%%%%%%%%%%%%%%%%%%%%%%%%%%%%%%%%%%%%
\begin{figure*}[tbp!]
\centering
\subfigure[ $\Pdf{\Xo}{\Xt}$]{\label{fig:Peellhyp:a}\includegraphics[width=0.491\textwidth]{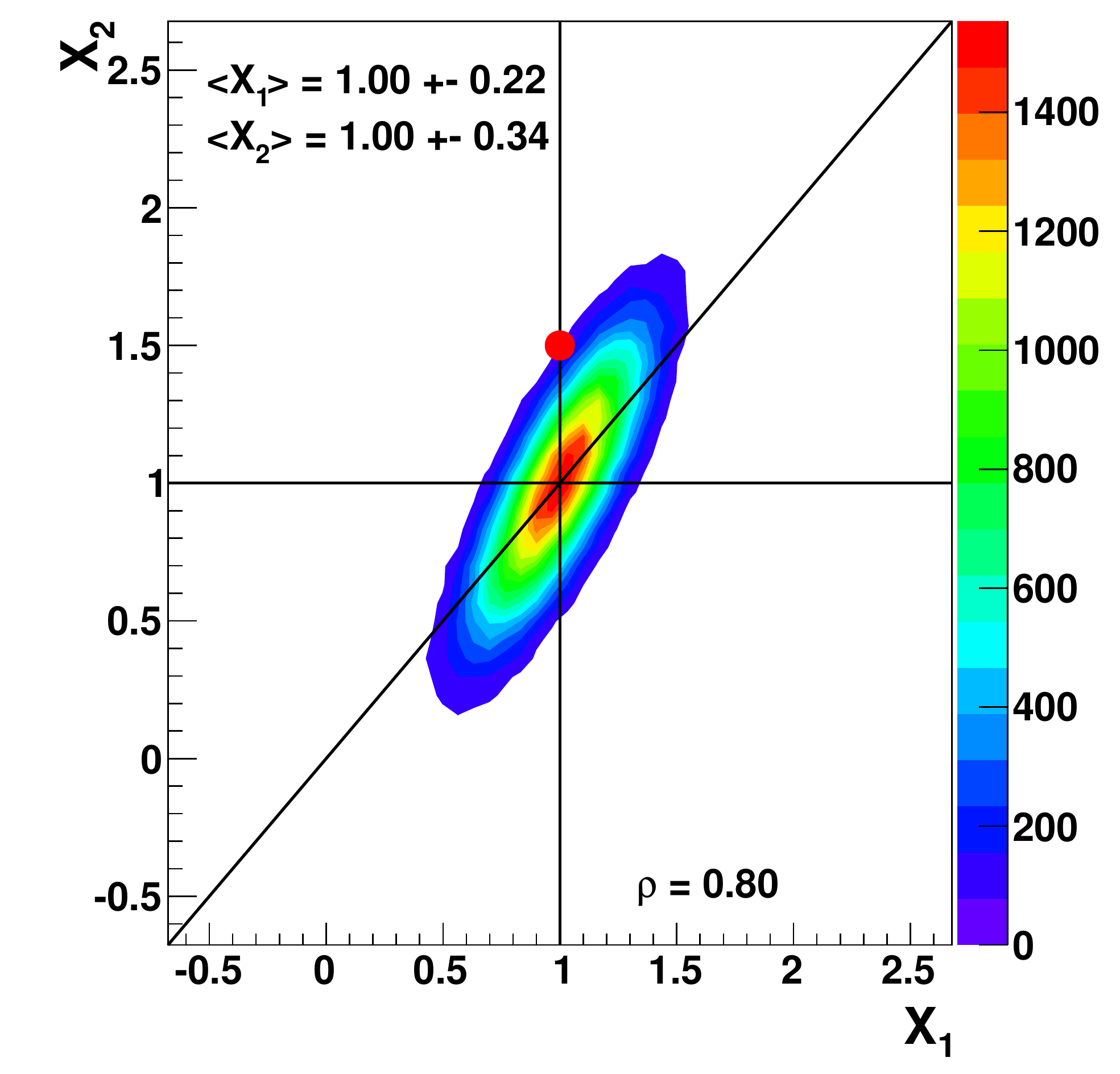}}
\subfigure[$\chiq(\Xo, \Xt)$]{\label{fig:Peellhyp:b}\includegraphics[width=0.491\textwidth]{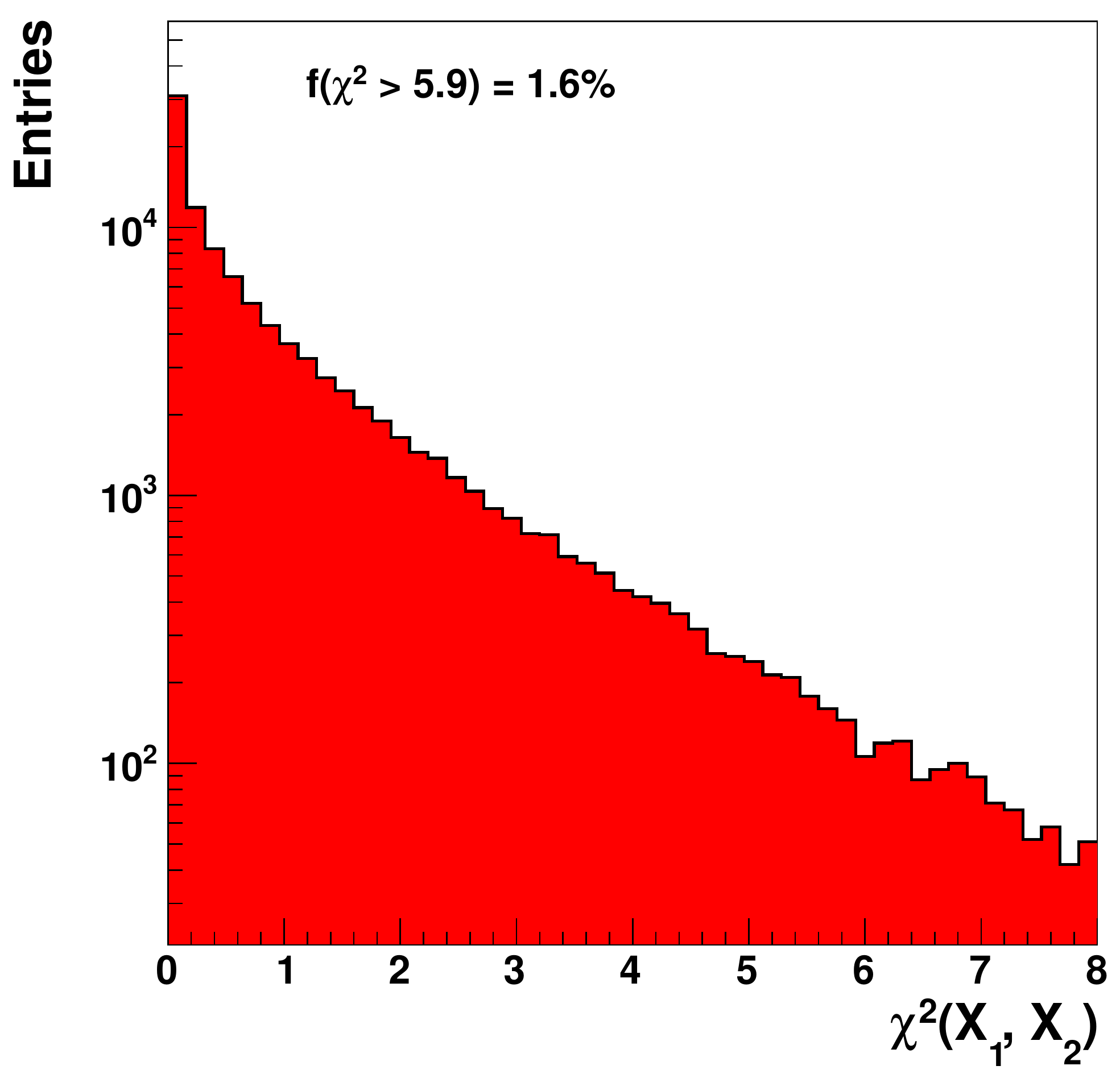}}
\subfigure[     $\Pdf{X}{Y}$]{\label{fig:Peellhyp:c}\includegraphics[width=0.491\textwidth]{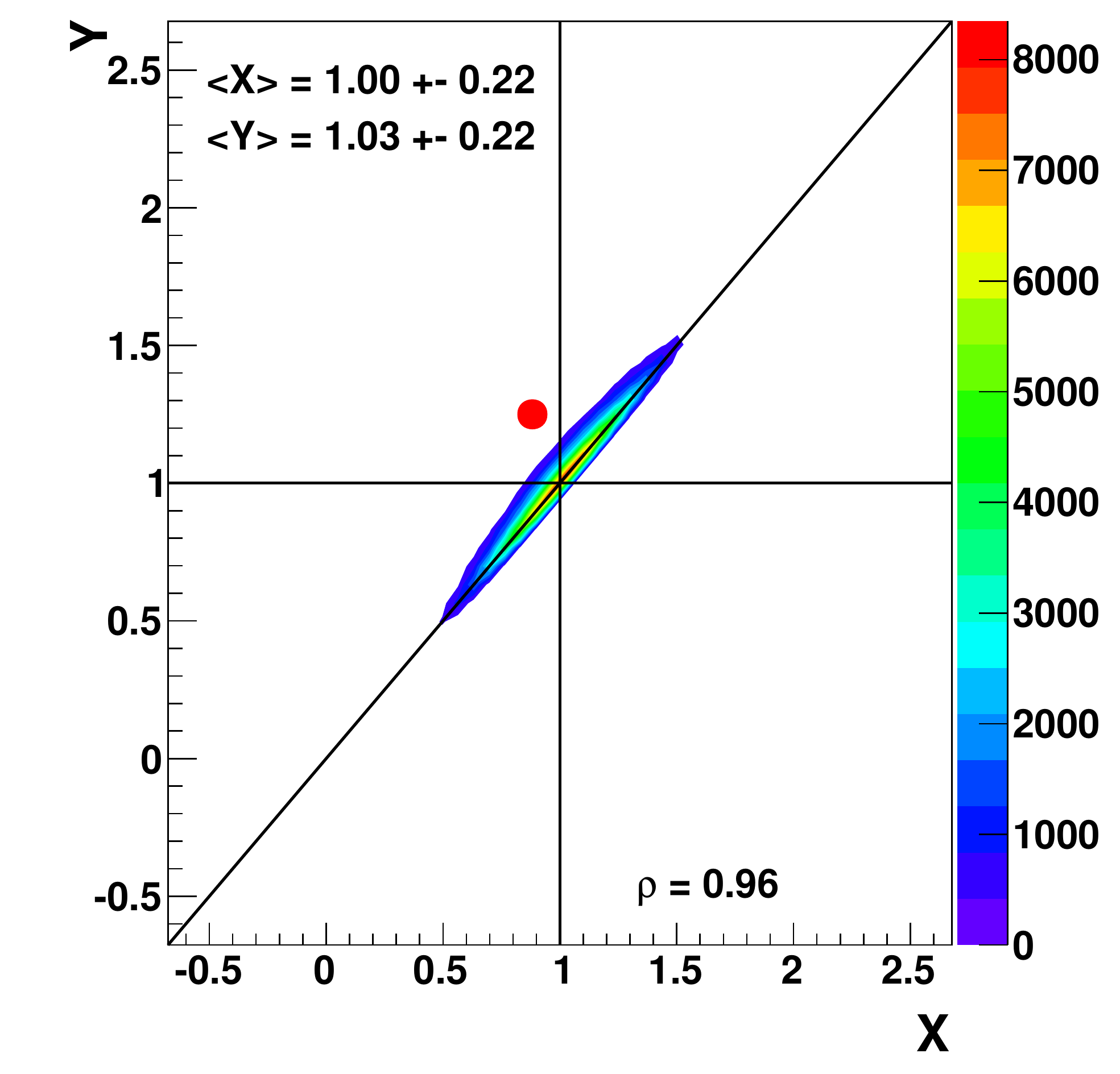}}
\subfigure[    $\chiq(X, Y)$]{\label{fig:Peellhyp:d}\includegraphics[width=0.491\textwidth]{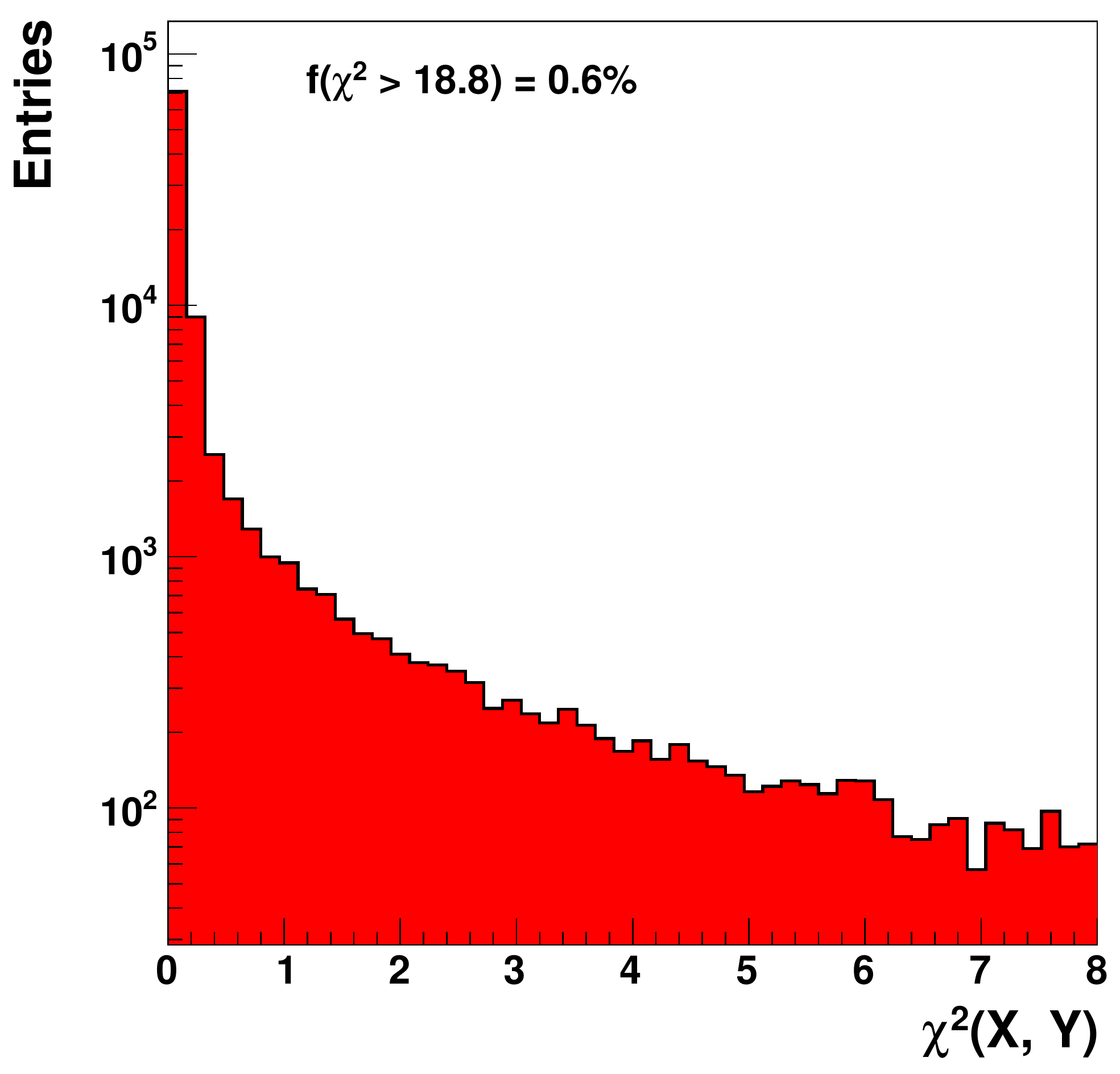}}
\caption{Results of Peelle's Pertinent Puzzle for scenario \sA for one hundred
  thousand pairs of estimates. The simulation is based on a hypothetical
  two-dimensional pdf assuming $\xT=1$, using the uncertainties and correlation
  of the estimates from this scenario, and simulating absolute uncertainties.
  Shown are (a) the two-dimensional distribution of the pairs of estimates, (b)
  the $\chiq(\Xo, \Xt)$ of the pairs of estimates, (c) the two-dimensional
  distribution of the pairs of combined results when using either absolute
  uncertainties (X), or relative uncertainties (Y), and (d) the $\chiq(X, Y)$ of
  the pairs of results.
  Both \chiq\ distributions are truncated at $\chiq=8$.
  The red points correspond to the estimates (a) and combined results (c) for
  this scenario, see Table~\ref{tab:BluePeel}.
  In addition listed for the estimates are in (a) their mean values and
  uncertainties together with their correlation, and in (b) the fraction of
  pairs for which the \chiq\ value exceeds the one observed for this scenario.
  The analogous quantities for the combined results are given in (c) and (d),
  respectively.
}
\label{fig:Peellhyp}
\end{figure*}
%%%%%%%%%%%%%%%%%%%%%%%%%%%%%%%%%%%%%%%%%%%%%%%%%%%%%%%%%%%%%%%%%%%%%%%%%%%%%

 To assess the significance of the difference of the two combined results
 obtained with the two combination methods, utilising the \chiq\ of
 Eq.~\ref{eq:chiq}, the correlation of the two results has to be calculated.
 In general, given the iterative procedure of the \BLUE\ method with relative
 uncertainties this can not be calculated analytically from the inputs to the
 combination, but has to be obtained numerically by performing numerous
 combinations.
 To do so an underlying estimator distribution \Pdf{\Xo}{\Xt}\ has to be
 constructed, see Eq.\ref{eq:cond1}.
 All models of \Pdf{\Xo}{\Xt}\ investigated here are based on the uncertainties
 and the correlations \rhootk\ of the estimates for the two sources $k$ of
 uncertainty as given in Table~\ref{tab:BluePeel}.
 In addition, a true value has to be assumed, together with an uncertainty
 model, based on either absolute or relative uncertainties.
 To ensure that the conclusions are neither biased towards the uncertainty model
 chosen, nor to a specific value of \xT, six estimator distributions
 \Pdf{\Xo}{\Xt}\ are investigated.
 They assume either absolute or relative uncertainties for three assumptions on
 the true value, namely $\xT = 0.75, 1, 1.25$, thereby spanning the entire
 range of results obtained for all scenarios listed in Table~\ref{tab:BluePeel}
 and both uncertainty models.

 Technically, the pdfs are based on Eq.\ref{eq:cond1} at a given value of
 \xT. The values for the uncertainties are taken from the upper part of
 Table~\ref{tab:BluePeel}.
 When simulating the absolute uncertainty model those uncertainties are taken at
 face value, whereas for the relative uncertainty model the fractions are
 retained, i.e.~the uncertainties from Table~\ref{tab:BluePeel} are scaled to
 the corresponding value of \xT.
 Finally, the correlation of the estimators is obtained from the covariance and
 the total uncertainties assigned.
 For a given pair of estimates generated, before performing the combination,
 uncertainties have to be assigned to the estimates.
 When simulating the absolute uncertainty model, the uncertainties from the pdf
 are kept.
 When instead simulating the relative uncertainty model, the uncertainties are
 rescaled to the estimates to be combined.

 As an example, for scenario \sA, for $\xT = 1$ and assuming the model of
 absolute uncertainties the results are visualised in Figure~\ref{fig:Peellhyp}.
 Figure~\ref{fig:Peellhyp:a} shows the predicted two-dimensional distribution
 for one hundred thousand pairs of estimates given the model. The red point in
 Figure~\ref{fig:Peellhyp:a} indicates the pair of estimates from the original
 Puzzle, which, if it is assumed to stem from this pdf, is an unlikely outcome.
 In addition listed in Figure~\ref{fig:Peellhyp:a} are the mean values and
 uncertainties of the estimator distributions together with their correlation.
 By construction they coincide with the values in Table~\ref{tab:BluePeel},
 proving the consistency of the simulation.
 The corresponding \chiq\ distribution of the pairs of estimates shown in
 Figure~\ref{fig:Peellhyp:b} exhibits the steep fall off expected for pairs of
 estimates consistent with stemming from this two-dimensional pdf.
 Here, this is achieved by construction. For a set of compatible experimental
 estimates to be combined, ideally a similar distribution for the pairwise
 \chiq\ values obtained from Eq.~\ref{eq:chiq} should be observed.
 In comparison, the corresponding \chiq\ value for the original pair of
 estimates is rather large, which makes it an unlikely case given this pdf,
 i.e.~only in about 1.6$\%$ of the cases a larger \chiq\ will be observed.
 This observation holds for both uncertainty models, and also does not depend on
 the chosen value of \xT\ since this only moves the ellipses along the diagonal.

 For a given combination, the combined results are denoted by $x$ ($y$) when
 assuming absolute (relative) uncertainties in the combination procedure (which
 are assigned irrespectively of the assumed uncertainty model of the pdf).
 Their two-dimensional distribution \Pdf{X}{Y}\ is shown in
 Figure~\ref{fig:Peellhyp:c}.
 It is found that for each estimator distribution chosen, the respective
 combination method is unbiased, whereas the other method shows a bias.
 For the example shown, assuming the uncertainty model with absolute
 uncertainties results in $\langle X \rangle = \xT$ for the \BLUE\ combination,
 whereas, in this case the \BLUE\ method with relative uncertainties has a bias,
 i.e.  $\langle Y \rangle = \xT + 0.03$. This is caused by the fact that, given
 the underlying absolute uncertainty model of the pdf, the wrong uncertainty
 model is assumed when performing the combination.
 However, this bias is insignificant, given the size of the statistical
 uncertainty. This conclusion applies to all scenarios and all six models for
 the estimator distribution.
 In all cases the combined results from the two methods are highly correlated,
 and the mean values differ by less than the statistical uncertainty of the
 combination method that shows the bias.

 The red point in Figure~\ref{fig:Peellhyp:c} denotes the pair of combined
 results from the original Puzzle, which lies far away from the ellipses.
 The correlation of the combined results from the two methods is deduced from
 all pairs of estimates, combining them with both prescription, and calculating
 the correlation of the two-dimensional distribution \Pdf{X}{Y}. For this pdf
 the correlation amounts to 0.96.
 Figure~\ref{fig:Peellhyp:d} shows the \chiq\ distribution for all pairs of
 results, again a steeply falling distribution. Using the correlation obtained
 from the simulation, the resulting value for the original pair is $\chiq(x, y)
 = 18.8$ which sits in the tail of this distribution, i.e.~only in about 0.6$\%$
 of the cases a larger \chiq\ will be observed.

 Applying all six models to scenario \sA, the correlation of \Pdf{X}{Y}\ varies
 from $0.92-0.98$, and the corresponding $\chiq(x, y)$ values for the original
 puzzle range from $11-27$.
 Given this, for all models the pair of results is not very likely or
 incompatible. However, as demonstrated for example by the results in
 Figure~\ref{fig:Peellhyp}, this is a mere consequence of the very unlikely or
 incompatible input and not of the differences of the method.
 This can be more clearly seen by analysing the additional scenarios $\sB-\sE$
 given in Table~\ref{tab:BluePeel}. They are designed to artificially improve
 the compatibility of the input, while using different aspects of the estimates.
 The parameters of the combinations depend on \rhof\ and $z$ such that they only
 change, if one of those changes.
 The value of \rhof\ is defined by the scenario, and, due to the simultaneous
 scaling of both uncertainty sources, in this case is not altered by any of the
 methods, see Table~\ref{tab:BluePeel}.
 In contrast, for given initial values of the uncertainties and correlations per
 source, the value of $z$ of the \BLUE\ method with absolute uncertainties is
 altered by the \BLUE\ method with relative uncertainties.
 This is caused by the dependence of the estimator uncertainties on the combined
 value, as  can be seen  e.g.~by comparing the  $z$ values for both  methods for
 scenario \sC.
 Given this, the uncertainty of the combined result of the \BLUE\ method with
 relative uncertainties depends on the values of the estimates, i.e.~on the
 likeliness of this particular experimental outcome, given an underlying pdf.

 In these additional scenarios, the estimates are altered by either changing:
 \sB\ the size of the uncertainties, \sC, \sD\ the value of the less precise
 estimate, and \sE\ the correlation of the systematic uncertainties. The target
 value of the estimate compatibility for the \BLUE\ method with absolute
 uncertainties was a $\chiq(\xo, \xt)$ of about 1.5.

 For scenario \sB\ the uncertainties are doubled. For none of the methods does
 this change the relative importance of the estimates, however it improves their
 compatibility.
 For scenarios \sC, \sD\ the value of the less precise estimate \xt\ is reduced
 to make it more compatible with \xo.
 The difference of the two scenarios is that, motivated by the absolute
 uncertainty model, in \sC\ the changed value for \xt\ is considered another
 possible outcome, namely a value consistent with the conditional pdf for
 \Xt. Consequently, the originally assigned uncertainties are kept.
 In contrast, for scenario \sD, this time motivated by the relative uncertainty
 model, the uncertainties are scaled to amount to the same fractional
 uncertainties as were originally assumed in \sA.
 Again the compatibility of the estimates and of the combined results is
 improved.
 For scenario \sD, by construction, all parameters of the combined result
 obtained using relative uncertainties are identical to the ones in scenario
 \sA. The combined value and its uncertainty are different, because the mean is
 changed due to the changed estimate \xt.
 For scenario \sE\ the correlation is reduced, yielding a similar level of
 agreement of the estimates. Here, again by construction, the combined result
 obtained using relative uncertainties is identical to the one in scenario \sA,
 but for its uncertainty which is reduced due to the smaller correlation of the
 estimates.

 The parameters of the combination in Table~\ref{tab:BluePeel} show that for the
 \BLUE\ method the sensitivity of scenarios $\sA-\sC$ on \rhof\ and $z$ are
 identical, such that the related conclusions drawn will not depend on the
 scenario.
 In addition, the derivatives reveal the fact that for the \BLUE\ combination
 with relative uncertainties, for all scenarios that retain the initial relative
 uncertainties, the weights of the estimates are independent of \rhof\ but have
 a large sensitivity to $z$. For the \BLUE\ combination the situation is rather
 different. Here, the weights have a much larger dependence on \rhof\ than on
 $z$.

 For the additional scenarios, the resulting compatibilities of the combined
 results are estimated as described above for scenario \sA. As an example of the
 six estimator distributions investigated, the $\chiq(x, y)$ values for $\xT =
 1$ and using the absolute uncertainty model are: $18.8, 1.5, 1.4, 3.4, 1.3$ for
 scenarios \sA, \sB, \sC, \sD, \sE, which means that the differences of the
 methods strongly diminish when using a more compatible input.
 This observation does only weakly depend on the underlying estimator
 distribution, i.e.~although the \chiq\ values differ, the pattern of the
 \chiq\ values for the different scenarios is very similar for all six cases.
 For all estimator distributions, firstly, the by far largest \chiq\ value is
 observed for scenario \sA. Secondly, for the remaining scenarios with the same
 correlation of the estimators, i.e.~for scenarios $\sB-\sD$, scenario \sC\ in
 all but one case has the smallest \chiq\ value. Thirdly, within a given
 uncertainty model for the estimator distribution, scenario \sE, has either the
 smallest or the second smallest \chiq\ value of all scenarios.

 Finally, applying all models to scenarios $\sB-\sE$, the correlation of
 \Pdf{X}{Y}\ varies from $0.78-0.98$ ($0.94-0.97$), for the estimator
 distributions with absolute (relative) uncertainties.
 The corresponding $\chiq(x, y)$ values for the remaining scenarios in
 Table~\ref{tab:BluePeel} range from $0.7-4.8$ and ($1.2-4.3$), respectively,
 i.e.~they are much smaller than what was observed for scenario \sA.
 Consequently, the apparently large difference observed for the two combined
 results for scenario \sA\ is not caused by the differences in the methods, but
 by the unlikeliness of the specific pair of estimates for all scenarios,
 i.e.~the incompatibility of the input to the \BLUE\ combination.

 As an alternative solution, for each scenario in Table~\ref{tab:BluePeel}, the
 most likely \xT\ given the estimates \xii, their uncertainties \si\ and
 correlation \rhof\ is obtained from a maximum likelihood fit using
 Eq.\ref{eq:cond1} for $\Xii=\xii$ as the likelihood function.
 Two likelihood functions are constructed. The result for \xT\ of those should
 be compared to the combined values $x$ from the \BLUE\ method with absolute and
 relative uncertainties, respectively.
 The first likelihood uses constant values for the \si. In contrast, for the
 second likelihood, in view of the relative uncertainty model, the uncertainties
 are chosen to depend on \xT\ according to the given fractional uncertainties
 for the scenarios in Table~\ref{tab:BluePeel}, such that
 $\si=\si(\xT)=\xT\,\si/\xii$ varies with \xT.
 By construction, the results for \xT\ from the first likelihood are identical
 to the combined values $x$ of the \BLUE\ method with absolute uncertainties,
 since the likelihood is a Gaussian, i.e.~it corresponds to the situation for
 which the \BLUE\ formulas were derived.
 The second likelihood has non Gaussian tails, and consequently, the results for
 \xT\ differ from the combined values $y$ of the \BLUE\ method with relative
 uncertainties, which is only an approximation.
 The results of the second likelihood are $\xT=1.53, 1.25, 1.03, 1.15, 1.24$ for
 scenarios \sA, \sB, \sC, \sD, \sE.
 The corresponding symmetrised uncertainties are $0.34, 0.44, 0.22, 0.23, 0.19$.
 Apart from the unlikely scenario \sA, the values for \xT\ nicely agree with the
 combined values from the \BLUE\ combination with relative uncertainties, see
 Table~\ref{tab:BluePeel}.  This demonstrates the quality of the approximation
 for consistent pairs of estimates.
 The uncertainties obtained from the likelihood and the \BLUE\ combination with
 relative uncertainties differ more strongly for scenarios \sA, \sB\ that have
 the largest non Gaussian contributions, whereas for the remaining scenarios
 they are almost identical.
 This ends the discussion of Peelle's Pertinent Puzzle.

 The definition of whether a given source of uncertainty is an absolute and
 relative uncertainty has to be made in view of the actual procedure followed to
 determine this uncertainty.
 Nevertheless, as purely numerical examples, and without any physics motivation,
 for a number of examples of publicly available combinations, to evaluate the
 numerical importance for real applications, the results for both assumptions
 are given below.
 All values quoted follow the convention of Eq.~\ref{eq:resu}.
 The two examples for which originally relative uncertainties are assigned are
 the combination of lifetimes of $B$ mesons~\cite{LYO-9001}, and of the
 cross-section for single top quark production at the
 LHC~\cite{ATLAS-CONF-2013-098}.
 In these cases for comparison absolute uncertainties are assumed for all
 sources.
 The two examples for which originally absolute uncertainties are assigned are
 the latest combinations of the measurements of the top quark mass \mt,
 performed at the Tevatron~\cite{TEV-1301} and the
 LHC~\cite{ATLAS-CONF-2013-102}. 
 In these cases for comparison relative uncertainties are assumed for all
 sources of systematic uncertainties.

 The corresponding results are for the $B$-lifetime:
% 
%%%%%%%%%%%%%%%%%%%%%%%%%%%%%%%%%%%%%%%%%%%%%%%%%%%%%%%%%%%%%%%%%%%%%%%%%%%%%%%
\begin{eqnarray}
 \tau~[ps]  &=& \XZ{1.13}{0.09}{0.11} \quad \mbox{(relative)} \nonumber\\
 \tau~[ps]  &=& \XZ{1.13}{0.09}{0.09} \quad \mbox{(absolute)}\,,
 \label{eq:Blife}
\end{eqnarray}
%%%%%%%%%%%%%%%%%%%%%%%%%%%%%%%%%%%%%%%%%%%%%%%%%%%%%%%%%%%%%%%%%%%%%%%%%%%%%%%
%
 for the single top quark production cross-section:
%
%%%%%%%%%%%%%%%%%%%%%%%%%%%%%%%%%%%%%%%%%%%%%%%%%%%%%%%%%%%%%%%%%%%%%%%%%%%%%%%
\begin{eqnarray}
 \sigma~[pb]  &=& \XZ{85.3}{4.1}{11.5} \quad \mbox{(relative)} \nonumber\\
 \sigma~[pb]  &=& \XZ{83.7}{4.6}{11.2} \quad \mbox{(absolute)}\,,
 \label{eq:sigt}
\end{eqnarray}
%%%%%%%%%%%%%%%%%%%%%%%%%%%%%%%%%%%%%%%%%%%%%%%%%%%%%%%%%%%%%%%%%%%%%%%%%%%%%%%
%
 for \mt\ measured at the Tevatron:
%
%%%%%%%%%%%%%%%%%%%%%%%%%%%%%%%%%%%%%%%%%%%%%%%%%%%%%%%%%%%%%%%%%%%%%%%%%%%%%%%
\begin{eqnarray}
 \mt~[\gev]  &=& \XZ{173.21}{0.51}{0.71} \quad \mbox{(absolute)}  \nonumber\\
 \mt~[\gev]  &=& \XZ{173.26}{0.51}{0.71} \quad \mbox{(relative)}\,,
 \label{eq:mtoptev}
\end{eqnarray}
%%%%%%%%%%%%%%%%%%%%%%%%%%%%%%%%%%%%%%%%%%%%%%%%%%%%%%%%%%%%%%%%%%%%%%%%%%%%%%%
%
 and finally, for \mt\ measured at the LHC:
%
%%%%%%%%%%%%%%%%%%%%%%%%%%%%%%%%%%%%%%%%%%%%%%%%%%%%%%%%%%%%%%%%%%%%%%%%%%%%%%%
\begin{eqnarray}
 \mt~[\gev]  &=& \XZ{173.29}{0.23}{0.92} \quad \mbox{(absolute)}  \nonumber\\
 \mt~[\gev]  &=& \XZ{173.30}{0.23}{0.92} \quad \mbox{(relative)}\,.
 \label{eq:mtoplhc}
\end{eqnarray}
%%%%%%%%%%%%%%%%%%%%%%%%%%%%%%%%%%%%%%%%%%%%%%%%%%%%%%%%%%%%%%%%%%%%%%%%%%%%%%%
%
 In all cases the difference of the pair of results is small compared to their
 statistical uncertainties. The results on \mt\ are almost indistinguishable,
 and at the quoted precision the uncertainties are identical.

 This ends the discussion about relative uncertainties. In the remainder of the
 paper only absolute uncertainties are considered.
%
%------------------------------------------------------------------------------
%
\section{The concept of reduced correlations}
\label{sec:redcor}
 Reduced correlations postulate that for each pair of estimates, e.g.~the pair
 (1, 2), that are positively correlated for a given source of uncertainty $k$,
 i.e.~$\rhootk>0$, the smaller of the individual uncertainties,
 e.g.~$\sok<\stk$, is fully correlated, and the remainder is uncorrelated.
 This replaces the covariance $\rhootk \sok \stk$ by the square of the smaller
 of the individual uncertainties, e.g.~\sokq for this source, see
 e.g.~Ref.~\cite{OPALPR404}. This is equivalent to assuming the correlation to
 amount to the ratio of the smaller to the larger uncertainty,
 $\rhootk=\sok/\stk=1/\zotk$.
 
 The impact of this concept can be seen by analysing the contribution of the
 source $k$ to the covariance matrix separated into the postulated uncorrelated
 (u) and correlated (c) parts that reads:
%
%%%%%%%%%%%%%%%%%%%%%%%%%%%%%%%%%%%%%%%%%%%%%%%%%%%%%%%%%%%%%%%%%%%%%%%%%%%%%%%
\begin{eqnarray}
V_{k} &=& \left(\begin{array}{cc} 
                              \sokq & \rhootk \sok \stk\\
                 \rhootk \sok \stk  &  \stkq \\
                \end{array} \right) \nonumber\\
       &=&
\left( \begin{array}{cc} 0 & 0 \\
                         0 & \stkq-\sokq \\\end{array} \right)_\mathrm{u} +
\left( \begin{array}{cc} \sokq & \sokq\\
                         \sokq & \sokq \\
\end{array}\right)_\mathrm{c} 
\label{eq:redcmat}
\end{eqnarray}
%%%%%%%%%%%%%%%%%%%%%%%%%%%%%%%%%%%%%%%%%%%%%%%%%%%%%%%%%%%%%%%%%%%%%%%%%%%%%%%
%
 By construction, this effectively replaces one source of uncertainty by two and
 assigns zero (full) correlation to the first (second),
 i.e.~$\sokq=1\cdot\sok\sok$.
 Typically, it is suggested to apply this concept to sources for which the
 initially assigned correlation of the estimates was $\rhootk=1$, or at least
 $\rhootk>1/\zotk$. This is because in this situation the correlation is always
 reduced with respect to the initial value, hence the name.

 If this source is the only uncertainty, this will lead to $\bet=0$.
 For the case in which $\rhootk\ge 0$ for all $k$, with an arbitrary number of
 sources, and applying the concept to all sources with $\rhootk > 0$ (i.e.~an
 unfavourite situation in which the correlation is partly even increased), the
 covariance with reduced correlations reads:
%
%%%%%%%%%%%%%%%%%%%%%%%%%%%%%%%%%%%%%%%%%%%%%%%%%%%%%%%%%%%%%%%%%%%%%%%%%%%%%%%
\begin{eqnarray}
\rhored \so \st &=&\! \sum_{\sokq<\stkq}  \!\! \sokq + 
                   \! \! \sum_{\stkq<\sokq}  \!\! \stkq \le 
                      \! \sum_{\rhootk> 0} \!\! \sokq \le \soq 
\label{eq:redccov}
\end{eqnarray}
%%%%%%%%%%%%%%%%%%%%%%%%%%%%%%%%%%%%%%%%%\%%%%%%%%%%%%%%%%%%%%%%%%%%%%%%%%%%%%%%
%
 where \rhored\ is the total reduced correlation of the pair of estimates.
 The first (second) term sums the variances of the sources for which initially
 the estimates were positively correlated and for which \xo\ (\xt) has the
 smaller uncertainty.
 If the second estimate does not have a smaller uncertainty for any of these
 sources, for the first inequality the equal sign is realised, otherwise
 replacing some \stkq\ by \sokq\ in the second sum will increase the covariance.
 Finally, if there are also no sources of uncertainty for which initially the
 estimates were taken as uncorrelated, for the second inequality the equal sign
 is valid.
 In any case, comparing the first and last terms the result is:
%
%%%%%%%%%%%%%%%%%%%%%%%%%%%%%%%%%%%%%%%%%%%%%%%%%%%%%%%%%%%%%%%%%%%%%%%%%%%%%%%
\begin{eqnarray}
\rhored &\le& \frac{\so}{\st} = \frac{1}{z}\,, 
\label{eq:redcval}
\end{eqnarray}
%%%%%%%%%%%%%%%%%%%%%%%%%%%%%%%%%%%%%%%%%\%%%%%%%%%%%%%%%%%%%%%%%%%%%%%%%%%%%%%%
%
 which means $\bet\ge 0$ is ensured by the method.
 This is also true if initially the total correlation is smaller, i.e.~there are
 in addition sources for which the estimates are negatively correlated, or if
 the method is only applied to sources with $\rhootk> 1/\zotk$.
 As a consequence, by construction $x$ is always within \xo\ and \xt. 
 However, as has been shown above, due to the conditional probability, the true
 value \xT\ is outside this interval in the majority of all cases.

 Apart from this deficiency, also from physics arguments this procedure is
 questionable as can be seen from an example.
 Lets assume there are two estimates of the same experiment, which suffer from
 the same source of uncertainty (lets say an energy scale uncertainty), but
 apply different phase space requirements, e.g.~on the jet transverse momentum
 \pt.
 Typically, the uncertainty on these scales decrease with increasing \pt, such
 that the estimate with the stronger requirement will have the smaller
 uncertainty.
 The method now effectively assigns a correlation to the uncertainty from this
 source, which is zero (one) for $\pt<\ptmin$ ($\pt>\ptmin$), where \ptmin\ is
 the larger of the two minimum transverse momenta required for the two
 estimates, see Eq.~\ref{eq:redcmat}.
 As a result, firstly, the limit of the correlation for $\pt=\ptmin$ from above
 and below is different. Secondly, the uncertainty slightly below \ptmin\ is by
 construction independent of the one slightly above. Given that the value of
 \ptmin\ is arbitrary, and that the facts that lead to the uncertainty of the
 energy scale do not disappear across the threshold, an unphysical situation.
 This is an example, where for $\rhootk=1$ a difference in \zotk\ is attempted
 to be cured by an ad hoc change in \rhootk. However, the dependence of
 \bet\ and \sx/\so\ on \rhof\ and $z$ are different.

 It is worth noticing that this method is not a {\it conservative} approach, in
 the sense that applying it would always lead to an increased variance of the
 combined result \sxq.
 If, for a given $z$, e.g.~the value of \rhof\ is only slightly larger than
 $1/z$, the resulting \rhored\ may be much smaller than $1/z$, such that
 \sx/\so\ is actually reduced from its initial value, see Eq.~\ref{eq:sigx} and
 Figure~\ref{fig:sigx}.
 Consequently, the uncertainty assigned to the combined value by using reduced
 correlations may be either larger or smaller, depending on the initial value of
 \rhof\ and the size of the reduction.
 For a specific example the impact is evaluated below.
%
%------------------------------------------------------------------------------
%
\section{Methods to maximise the variance}
\label{sec:maxvar}
 On top of the reduced correlations discussed in the previous section, an even
 more rigorous way to avoid estimates with negative \BLUE\ weights is the choice
 to simply exclude those estimates from the combination. 
 A recipe of how to proceed if this is desired is given in Ref.~\cite{VAL-1301}.
 However, this ad hoc choice does not respect the consequence of the conditional
 probability and consequently is disfavoured.

 In addition to the above, a number of methods have been suggested to arrive at
 a {\it conservative} combined estimate, i.e.~to maximise the variance of the
 combined result \sxq.
 All attempts work by reducing the correlation in an artificial, but controlled
 way.
 Given Figure~\ref{fig:sigx} they will only be active for $\rho>1/z$ which means
 $\bet<0$.
 The three methods suggested in Ref.~\cite{VAL-1301} multiply the initially
 assigned correlations per source $k$ for any pair $ij$ of estimates by factors
 \fijk.
 These factors are either chosen:
%
%%%%%%%%%%%%%%%%%%%%%%%%%%%%%%%%%%%%%%%%%%%%%%%%%%%%%%%%%%%%%%%%%%%%%%%%%%%%%%%
\begin{Enumerate}
\item[i)] globally, $\fijk = f$ for all $i,j,k$,
\item[ii)] per uncertainty source, $\fijk = \fk$ for all $i,j$,
\item[iii)] per pair of estimates, $\fijk = \fij$ for all $k$.
\end{Enumerate}
%%%%%%%%%%%%%%%%%%%%%%%%%%%%%%%%%%%%%%%%%%%%%%%%%%%%%%%%%%%%%%%%%%%%%%%%%%%%%%%
%
 All methods are not flexible enough to incorporate the different knowledge on
 the correlations that will be available for different pairs of estimates and
 different sources of uncertainty. In addition, they do not obey some of the
 properties of the estimates outlined in Section~\ref{sec:estpro}.
 More specifically, by varying the correlation for all sources simultaneously
 the method i) does not obey property 3) of the estimates, namely that all
 sources of uncertainties are assumed to be uncorrelated. It also does not take
 into account that the knowledge on the correlation may differ from source to
 source.
 As an example, for the combination of \mt\ in Ref.~\cite{ATLAS-CONF-2013-102},
 the uncertainties related to the colour reconnection and to the background
 determined from Monte Carlo are both assumed to be fully correlated between all
 estimates.
 However, there is no physics reason to believe that the two sources of
 uncertainty are correlated. Consequently, if the correlation assumption is
 changed e.g.~for the colour reconnection by using $f=0.9$, there is no reason
 to simultaneously apply the same factor to the uncertainty from the background
 determinations, which would however be enforced when using method i).
 In contrast, if there are physics arguments to vary two sources of uncertainty
 simultaneously, i.e.~there are reasons to believe that two sources ($k,
 k^\prime$) are correlated, it is preferred to reconsider the separation of the
 uncertainty sources, see Section~\ref{sec:estpro}.

 Method ii) does not take into account that the uncertainties on \rhoijk\ likely
 are better known for pairs of estimates from the same experiment, than for
 pairs of estimates from different experiments, or even obtained at different
 colliders. Given this, although, a correlated variation for some pairs (i, j)
 can be well justified, applying this to all pairs is not flexible enough.

 Method iii), although calculated per pair, in reality corresponds to specific
 \rhoijk\ values. Since the variation is done per pair, e.g.~($i, j$) or
 ($i^\prime,j$), where $i, i^\prime$ are assumed to be estimates from the same
 experiment and $j$ from another experiment, this very likely leads to very
 different assumptions on the correlation for the source $k$ across experiments.
 Again, the available knowledge on this can not be respected by this automated
 procedure.
%
%------------------------------------------------------------------------------
\begin{table*}[tbp!]
\caption{Combinations of two correlated estimates using the \BLUE\ method for
  different scenarios $\sA-\sD$, and using the different methods described in
  the text. The two estimates used are given together with their uncertainties.
  The four scenarios analysed for the estimates $i = 1, 2$, and uncertainties,
  $k = 0, 1, 2$ with correlations \rhootk\ are: \sA\ the default values of the
  uncertainties with two fully correlated systematic uncertainties, $\sB=\sA$
  but the first systematic uncertainty is assumed to be uncorrelated,
  $\sC~(\sD)=\sA$ but for the second systematic uncertainty the smaller (larger)
  of the two values is taken for both estimates. For the maximisation of the
  variance no values are given for scenarios $\sB-\sD$, since they coincide with
  the \BLUE\ results.}
%
%-------------------------------------------------------
\label{tab:BlueRes}
\begin{center}
\begin{tabular}{rrrrrrrrrrrrrrr}
\hline\noalign{\smallskip}
\multicolumn{15}{c}{Estimates}\\\noalign{\smallskip}\hline\noalign{\smallskip}
           &  Value & Stat &Syst1 &Syst2 & Syst & Full & &&&&&&\\\noalign{\smallskip}\hline\noalign{\smallskip}
           &  \xii  &\sinv & \sio & \sit &\sisys& \si  &\multicolumn{8}{l}{Comment} \\
\noalign{\smallskip}\hline\noalign{\smallskip}
      \sA  & 172.10 & 0.60 & 0.50 & 0.70 & 0.86 & 1.05 & \multicolumn{8}{l}{Correlations of $\rho_{120} =0,\, \rho_{121} = \rho_{122}=1$} \\
           & 173.10 & 0.40 & 0.70 & 1.40 & 1.57 & 1.62 & &&&&&&& \\
      \sB  & 172.10 & 0.60 & 0.50 & 0.70 & 0.86 & 1.05 & \multicolumn{8}{l}{Changed correlation for the first systematics $\rho_{121} = 0$} \\
           & 173.10 & 0.40 & 0.70 & 1.40 & 1.57 & 1.62 & &&&&&&& \\
      \sC  & 172.10 & 0.60 & 0.50 & 0.70 & 0.86 & 1.05 & \multicolumn{8}{l}{Changed second systematics $\stt = 0.7$} \\
           & 173.10 & 0.40 & 0.70 & 0.70 & 0.99 & 1.07 & &&&&&&& \\
      \sD  & 172.10 & 0.60 & 0.50 & 1.40 & 1.49 & 1.60 & \multicolumn{8}{l}{Changed second systematics $\sot = 1.4$} \\
           & 173.10 & 0.40 & 0.70 & 1.40 & 1.57 & 1.62 & &&&&&&& \\
\noalign{\smallskip}\hline\noalign{\smallskip}
\multicolumn{15}{c}{Combined results}\\\noalign{\smallskip}\hline\noalign{\smallskip}
           & $x$ &\ssta & \ssyso & \ssyst & \ssys & \sx & \rhof\ & $z$  & $\bet$  &
   $\frac{\sx}{\so}$ & $\frac{d\bet}{d\rhof}$ & $\frac{1}{\so}\frac{d\sx}{d\rhof}$ 
                     & $\frac{d\bet}{dz}$     & $\frac{1}{\so}\frac{d\sx}{dz}$ \\
\noalign{\smallskip}\hline\noalign{\smallskip}
\multicolumn{15}{c}{\BLUE}\\\noalign{\smallskip}\hline\noalign{\smallskip}
       \sA & 171.88 & 0.74 & 0.46 & 0.55 & 0.71 & 1.02 &   0.78 & 1.54 & -0.22 &  0.98 & -2.32 & -0.42 & -0.48 & -0.14 \\
       \sB & 172.17 & 0.56 & 0.47 & 0.75 & 0.88 & 1.04 &   0.58 & 1.54 &  0.07 &  1.00 & -0.84 &  0.10 & -0.45 &  0.04 \\
       \sC & 172.56 & 0.37 & 0.59 & 0.70 & 0.92 & 0.99 &   0.75 & 1.02 &  0.46 &  0.94 & -0.14 &  0.27 & -1.96 &  0.43 \\
       \sD & 171.56 & 0.37 & 0.59 & 1.40 & 1.52 & 1.57 &   0.89 & 1.01 &  0.46 &  0.98 & -0.33 &  0.26 & -4.59 &  0.45 \\
\noalign{\smallskip}\hline\noalign{\smallskip}
\multicolumn{15}{c}{Reduced correlations}\\\noalign{\smallskip}\hline\noalign{\smallskip}
       \sA & 172.26 & 0.51 & 0.51 & 0.73 & 0.89 & 1.02 &   0.44 & 1.54 &  0.16 &  0.97 & -0.51 &  0.21 & -0.39 & 0.10 \\
       \sB & 172.32 & 0.47 & 0.42 & 0.75 & 0.86 & 0.98 &   0.29 & 1.54 &  0.22 &  0.94 & -0.34 &  0.29 & -0.34 & 0.14 \\
       \sC & 172.57 & 0.37 & 0.55 & 0.70 & 0.89 & 0.96 &   0.66 & 1.02 &  0.47 &  0.92 & -0.08 &  0.28 & -1.45 & 0.43 \\
       \sD & 172.57 & 0.37 & 0.55 & 1.40 & 1.50 & 1.55 &   0.85 & 1.01 &  0.47 &  0.97 & -0.18 &  0.26 & -3.38 & 0.45 \\
\noalign{\smallskip}\hline\noalign{\smallskip}
\multicolumn{15}{c}{Maximisation of the variance}\\\noalign{\smallskip}\hline\noalign{\smallskip}
       \sA & 172.10 & 0.60 & 0.50 & 0.70 & 0.86 & 1.05 &   0.65 & 1.54 &  0.00 &  1.00 & -1.13 &  0.00 & -0.47 & 0.00 \\
\noalign{\smallskip}\hline
\end{tabular}
\end{center}
\end{table*}
%------------------------------------------------------------------------------
%
%
%------------------------------------------------------------------------------
%
\section{A hypothetical example}
\label{sec:example}
 The impact of the reduced correlations and the three ways to maximise the
 variance of the combined result are discussed on the basis of a hypothetical
 example, motivated by typical estimates occurring in top quark mass
 measurements.
 For simplicity, only two estimates and three uncertainty sources are used. The
 extension to more estimates and uncertainty sources is straight forward.
%
%%%%%%%%%%%%%%%%%%%%%%%%%%%%%%%%%%%%%%%%%%%%%%%%%%%%%%%%%%%%%%%%%%%%%%%%%%%%%%%
\begin{figure*}[tbp!]
\centering
\subfigure[]{\label{fig:inspect:a}\includegraphics[width=0.245\textwidth]{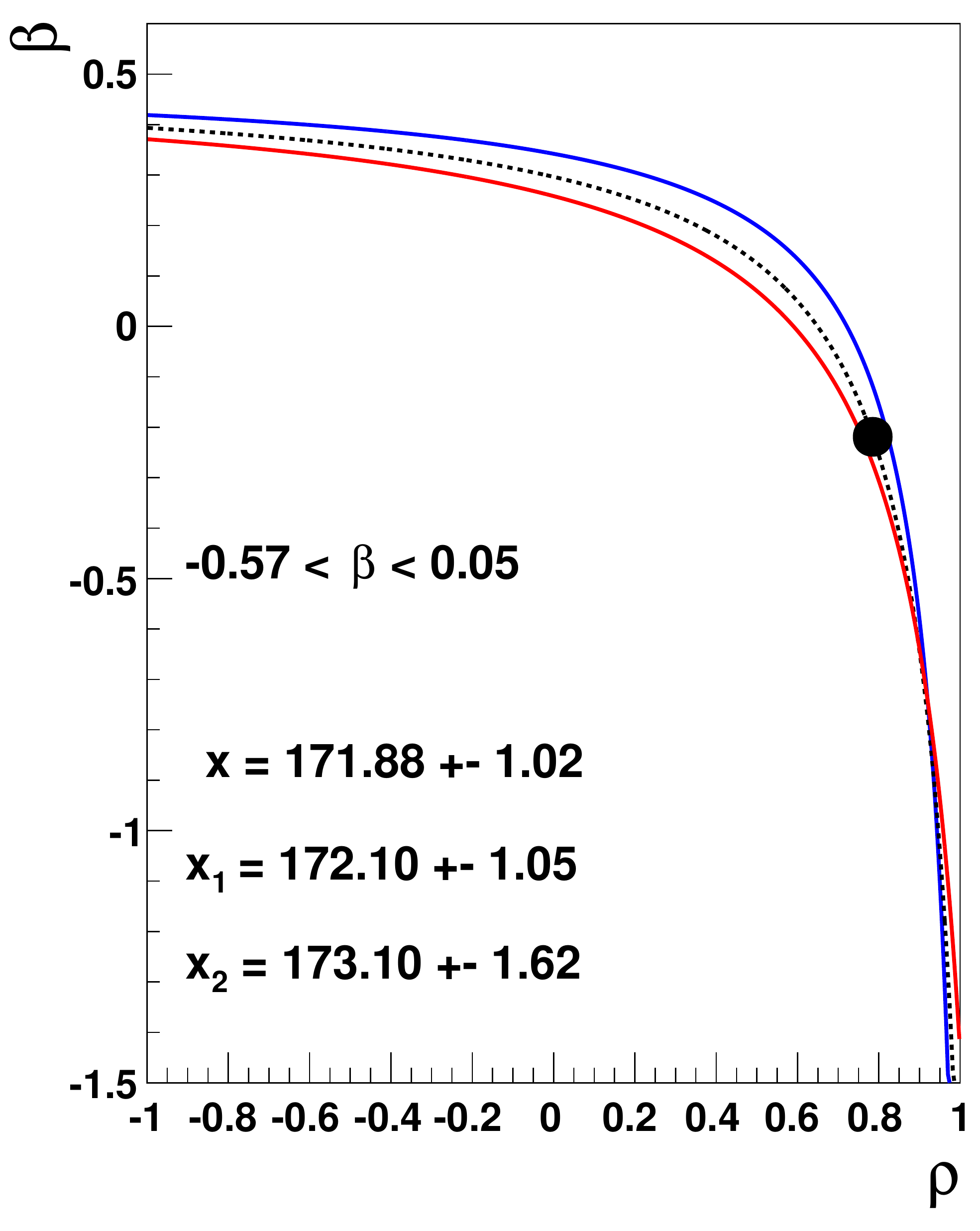}}
\subfigure[]{\label{fig:inspect:b}\includegraphics[width=0.245\textwidth]{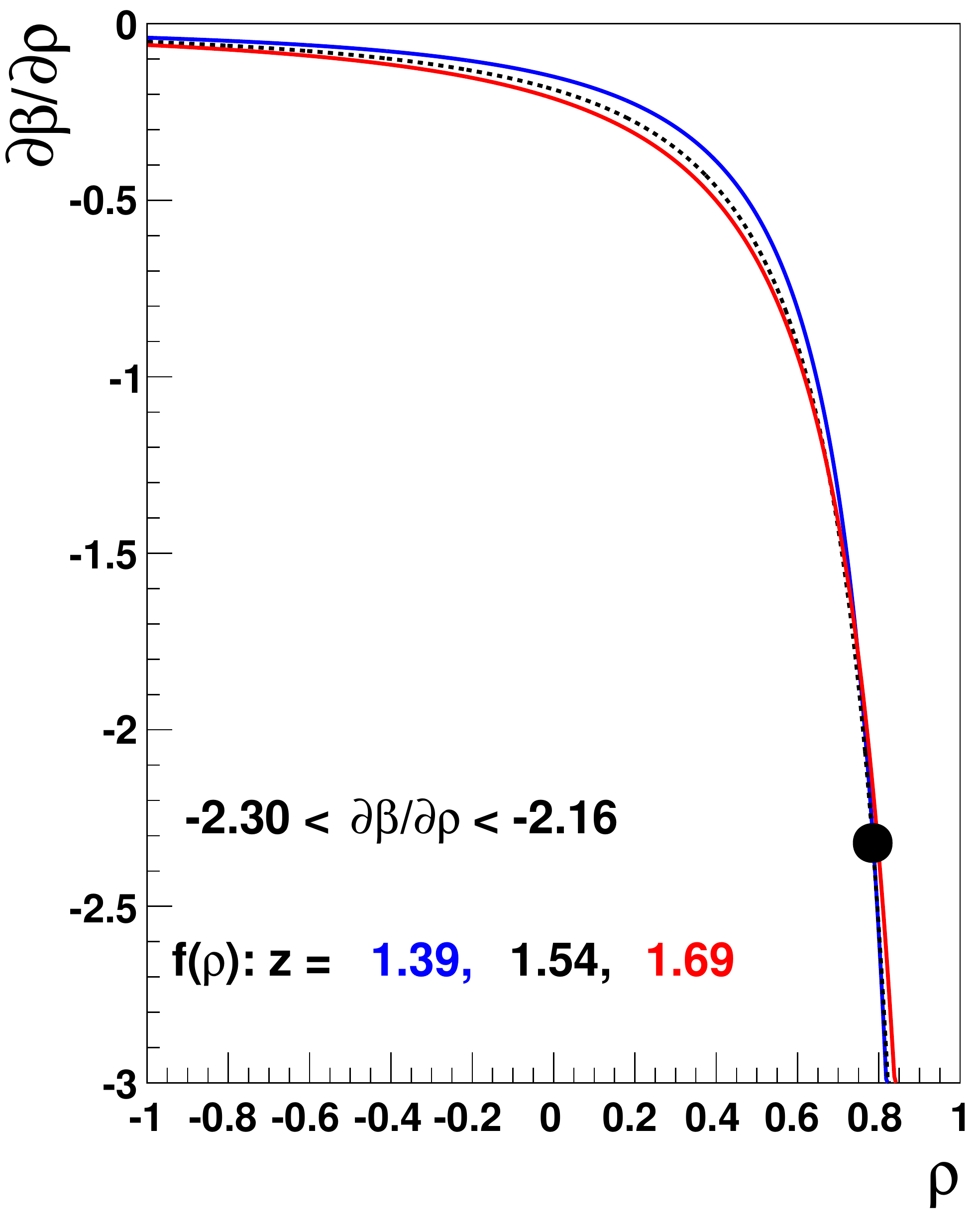}}
\subfigure[]{\label{fig:inspect:c}\includegraphics[width=0.245\textwidth]{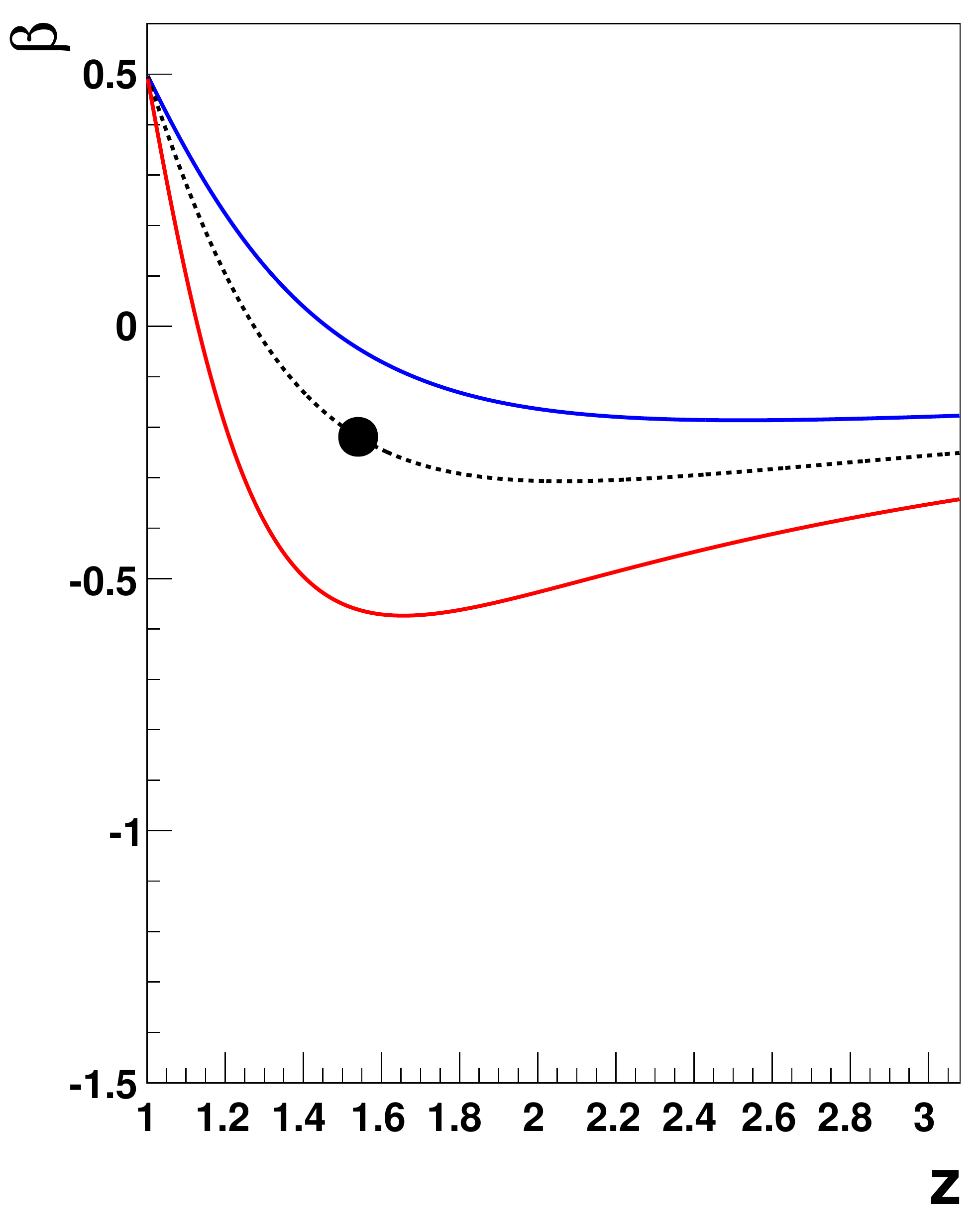}}
\subfigure[]{\label{fig:inspect:d}\includegraphics[width=0.245\textwidth]{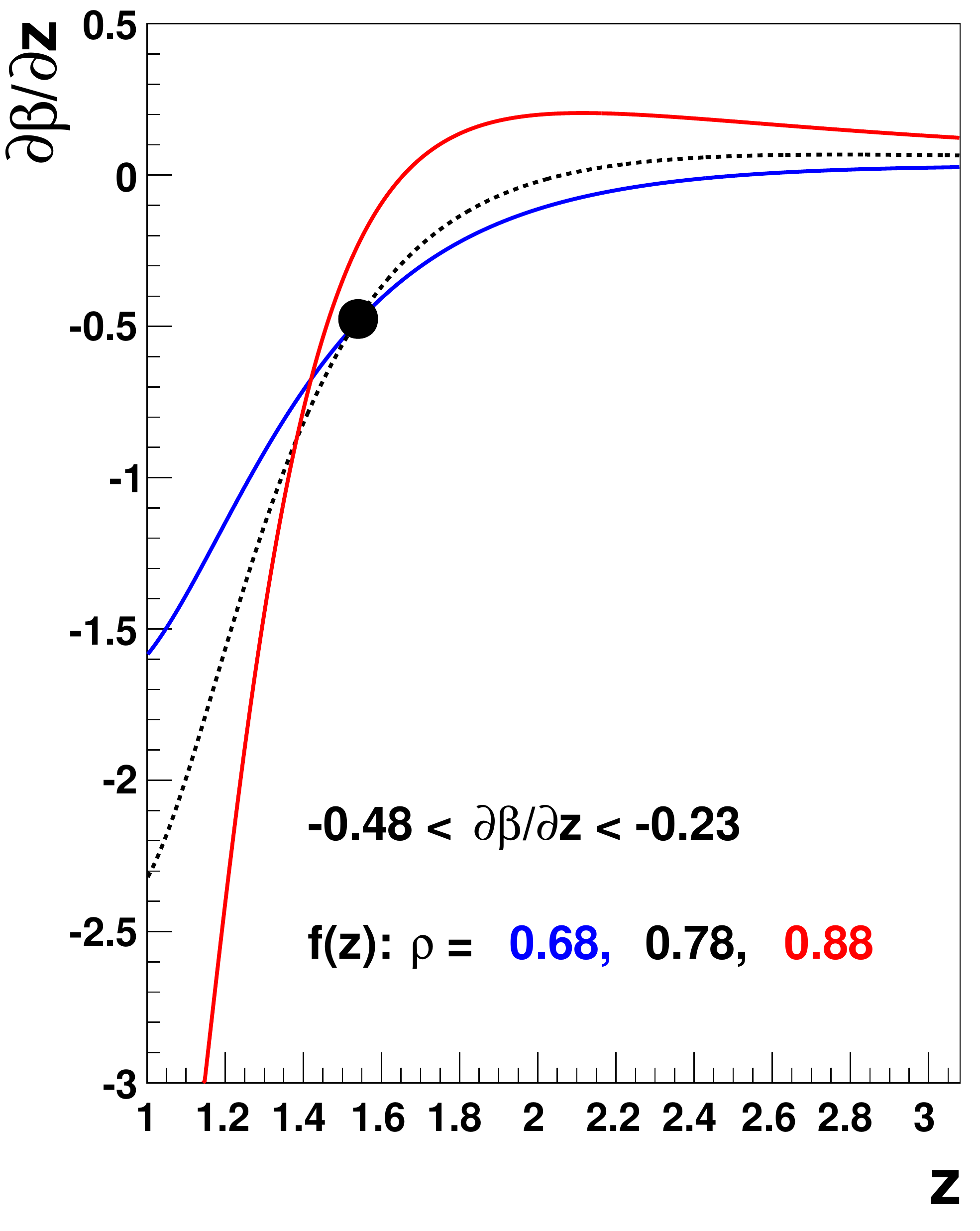}}
\subfigure[]{\label{fig:inspect:e}\includegraphics[width=0.245\textwidth]{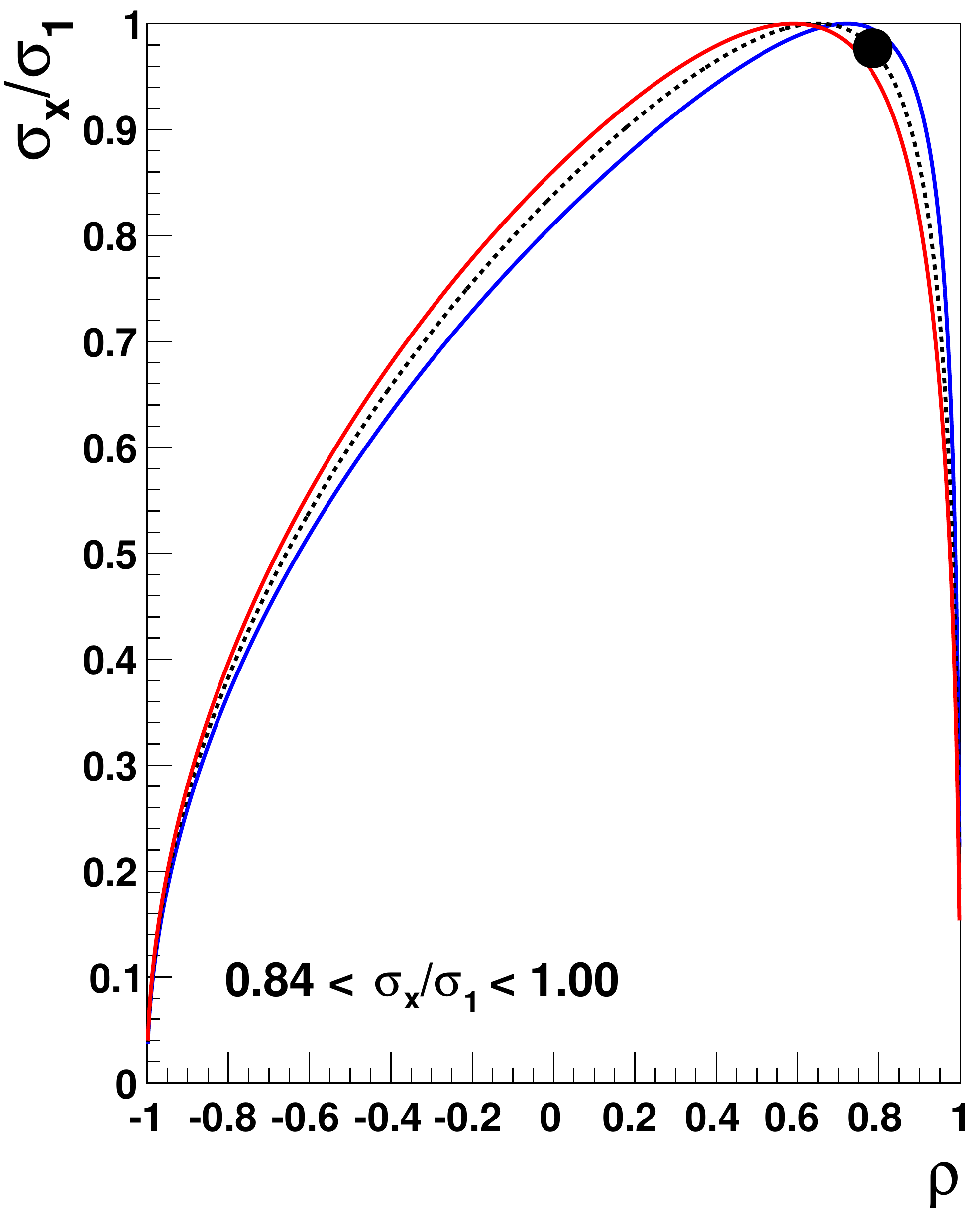}}
\subfigure[]{\label{fig:inspect:f}\includegraphics[width=0.245\textwidth]{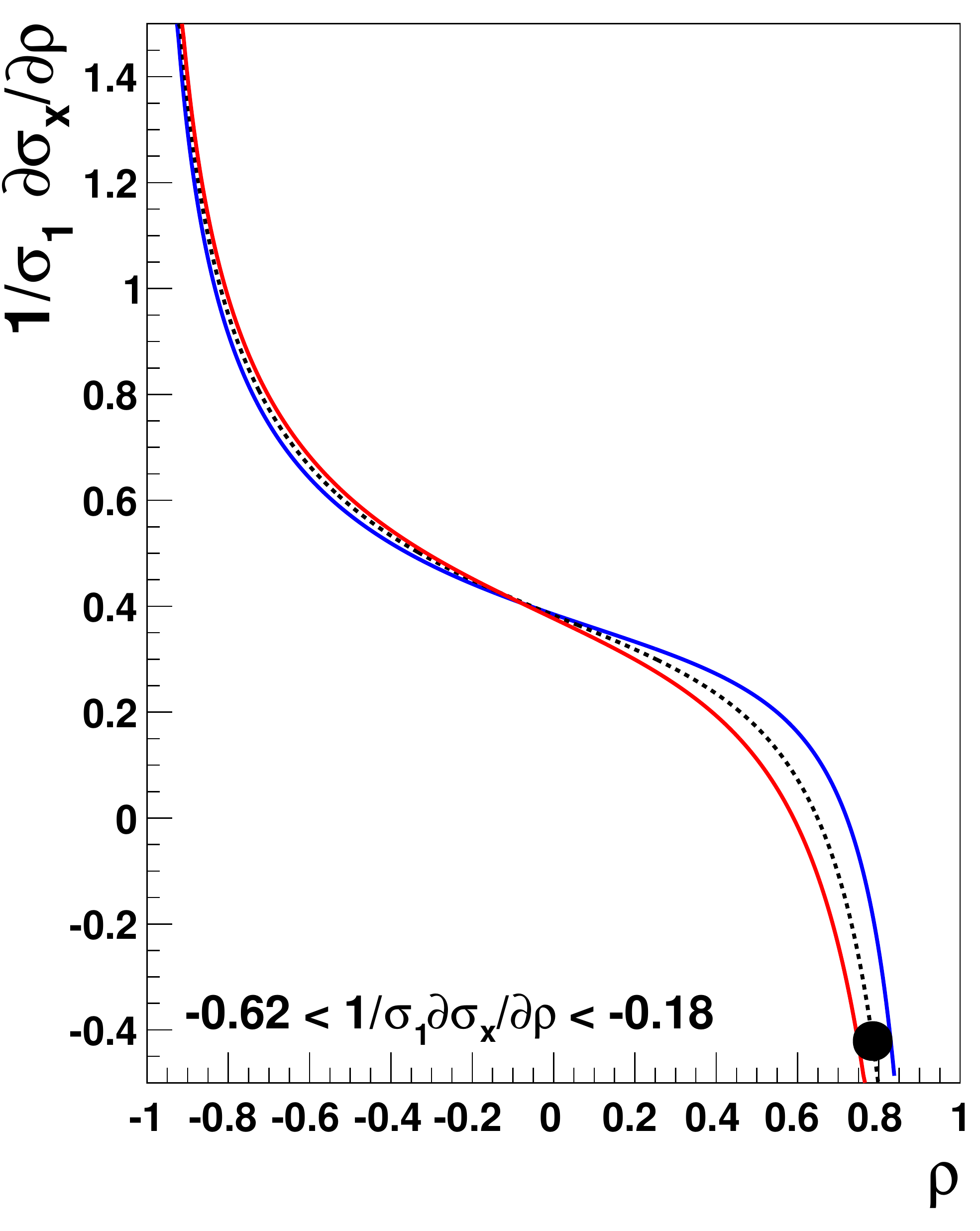}}
\subfigure[]{\label{fig:inspect:g}\includegraphics[width=0.245\textwidth]{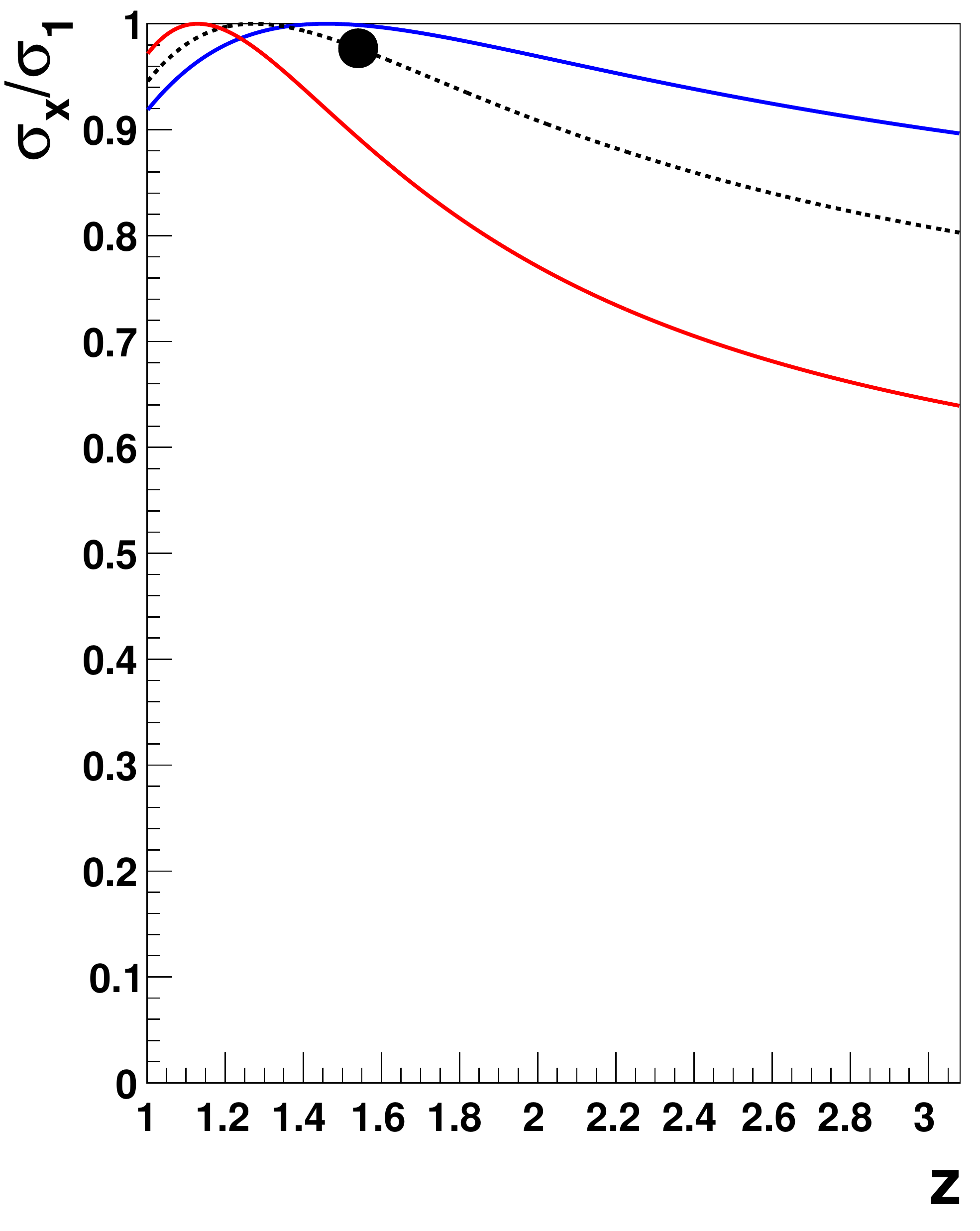}}
\subfigure[]{\label{fig:inspect:h}\includegraphics[width=0.245\textwidth]{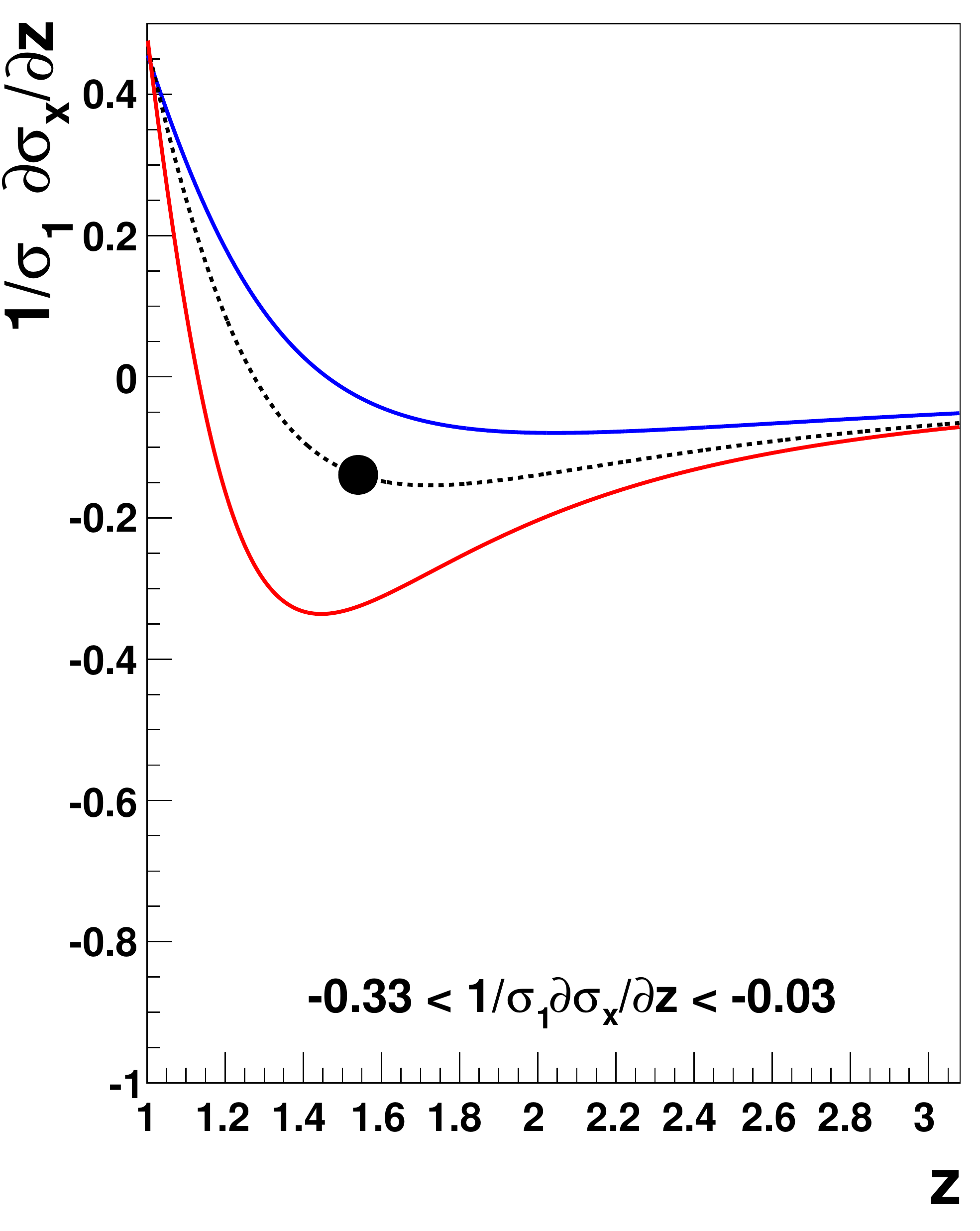}}
\caption{Results for the Blue combination using the hypothetical example from
  Table~\protect\ref{tab:BlueRes}, scenario \sA.
  The sub-figures (a)--(h) correspond to
  Figures~\protect\ref{fig:bsvsr}--\protect\ref{fig:bsvsz} for the pair of
  estimates investigated. The black points represent the actual values of the
  parameter shown at the given values of \rhof\ and $z$.
  In (a) also the estimates \xo\ and \xt, as well as the combined value $x$,
  together with their uncertainties, are listed.
  In each sub-figure three curves are shown in which, for parameters shown as a
  function of \rhof\ (or $z$), the value of $z$ (or \rhof) is varied.
  The curves corresponding to the minimum/central/maximum value of this
  variation are shown in blue/black/red, and the three values used for $z$ and
  \rhof\ are given in (b) and (d), respectively.
  For the derivatives of \bet\ and \sx/\so\ with respect to \rhof\ and $z$, for
  each sub-figure the range of observed parameter values is given. This range is
  obtained for the three curves shown, while keeping the respective value of the
  other parameter.
  As an example in (b) the range in $\partial\,\bet/\partial\,\rho$ at
  $\rhof=0.78$ is quoted observed when changing $z$ from 1.39 to 1.69.
  Finally, for \bet\ and \sx/\so\ their full range is quoted in (a) and
  (e). This range is obtained using all nine possible pairs of the \rhof\ and
  $z$ values.}
\label{fig:inspect}
\end{figure*}
%%%%%%%%%%%%%%%%%%%%%%%%%%%%%%%%%%%%%%%%%%%%%%%%%%%%%%%%%%%%%%%%%%%%%%%%%%%%%

 The two estimates are given in Table~\ref{tab:BlueRes}. They are analysed for
 four different scenarios in which the assumption on either the correlation, or
 the size of the uncertainty for one of the sources, is changed one at a time.
 Using Eq.~\ref{eq:chiq} and calculating \Prob{\chiq}{1}, the compatibility of
 the estimates is assessed for the \BLUE\ method and for all
 scenarios\footnote{The reduced correlations and the methods to maximise the
   variance reduce the value of \rhof\ while keeping $\rhof>0$. Consequently,
   the resulting \chiq\ values obtained from Eq.~\protect\ref{eq:chiq} are
   always smaller.}.
 The values obtained with this procedure are $\Prob{\chiq}{1}=0.33, 0.45, 0.18,
 0.18$, for scenarios \sA, \sB, \sC, \sD.

 Given the assigned correlations per source and $z=1.54$, scenario
 \sA\ corresponds to a situation where $\rho=0.78>1/z=0.65$ and consequently
 $\bet=-0.22<0$.
 This situation is visualised in Figure~\ref{fig:inspect}, where the eight
 sub-figures correspond to
 Figures~\protect\ref{fig:bsvsr}--\protect\ref{fig:bsvsz}, and the black points
 to the pair of estimates investigated.
 Consequently, in Figure~\ref{fig:inspect:e}, the point is to the right of the
 peak which sits at $\bet=0$.

 In Figure~\ref{fig:inspect}, the sensitivity of the combination to variations
 of \rhof\ and $z$ is visualised by the three curves per sub-figure.
 For a given functional dependence of one of the functions, e.g.~$\bet(\rhof)$,
 they show the sensitivity to the respective other parameter, here $z$, using
 the actual value (black dashed line) and two changed values (coloured full
 lines).
 For the pair of changed values, either $z$ is multiplied by $0.9$ (blue line)
 or $1.1$ (red line), Figures~\ref{fig:inspect:a}, \ref{fig:inspect:b},
 \ref{fig:inspect:e} and \ref{fig:inspect:f}, or \rhof\ is changed by $\pm 0.1$,
 Figures~\ref{fig:inspect:c}, \ref{fig:inspect:d}, \ref{fig:inspect:g} and
 \ref{fig:inspect:h}. This indicates the impact of $10\%$ uncertainties on their
 respective initial values.
 The figure shows that the dependence of \bet\ and \sx/\so\ and their
 derivatives on one of the parameters strongly depends on the value the
 respective other parameter has.
 As an example, the sensitivity to \rhof\ of the derivative of \bet\ with
 respect to $z$, visualised by the spread of the three lines in
 Figure~\ref{fig:inspect:d}, varies strongly with $z$. For the chosen example it
 is smallest close to the black point, i.e.~to the actual pair of values of
 \rhof\ and $z$. For two estimates with $z=1.1$ the sensitivity to \rhof\ would
 be much larger. In contrast, for the derivative of \sx/\so\ with respect to
 $z$, Figure~\ref{fig:inspect:h}, the chosen point in phase space lies close to
 the region with the largest spread of the curves, signalling a large
 \rhof\ dependence.
 The quoted derivatives of \sx/\so\ in Figure~\ref{fig:inspect:f} and
 \ref{fig:inspect:h}, show that a $10\%$ change in \rhof\ has a much larger
 impact on the uncertainty of the combined value than the corresponding change
 in $z$, which means that for this particular case, it is more important to
 correctly determine \rhof\ rather than $z$.
 The values of all parameters and for all scenarios investigated are listed in
 Table~\ref{tab:BlueRes}.

 Given the initial correlation assumptions, the reduced correlations act on both
 systematic uncertainties and yield $\rhof = 0.44$. As a result of this strong
 reduction of the correlation, the resulting value of \sx/\so\ is lower than for
 the initially assigned correlations, see Table~\protect\ref{tab:BlueRes}.
 Because there is a non zero uncorrelated component to the uncertainty for both
 estimates, the reduced correlations can not switch off \xt\ completely, as it
 would otherwise do, see Eq.~\ref{eq:redccov}.

 For this example, the three methods for maximising the variance, at the quoted
 precision, all give the same combined result, which is achieved for $f = 0.83$,
 $\fk = 0.34, 1$ (or $\fk = 1, 0.77$) for $k=1, 2$, and finally, $\fij = f =
 0.83$, respectively. Consequently, with these algorithms, the second estimate
 is switched off in different ways, i.e.~they all give $\bet=0$ and $x=\xo$, as
 it would be the case if estimates with negative weights would be ignored.

 For scenario \sB\ the systematic uncertainty $k=1$ is assumed to be
 uncorrelated rather than fully correlated.
 By this assumption the correlation is reduced such that the point moves to the
 left of the peak in Figure~\ref{fig:inspect:e} and the \BLUE\ combination
 results in a positive value for \bet. Given that \bet\ is very close to zero
 the estimate \xt would improve \xo\ by less than $1\%$.
 For the reduced correlations, which now only act on the source $k=2$, the
 correlation is further decreased, such that the predicted improvement in
 precision of $6\%$ is even larger than for scenario \sA.
 In contrast, since the maximisation of the variance is only attempted to the
 right of the peak in Figure~\ref{fig:inspect:e} none of the algorithms
 $i)-iii)$ is proposing any change.

 The scenarios \sC\ and \sD\ implement the situation in which for $\rhootk = 1$
 for $k=2$ the difference has been caused by the use of different
 procedures. Either estimate \xo\ has a 'too crude' procedure assigned such that
 not all features of this source are accounted for, and the quoted uncertainty
 is underestimated, scenario \sC, or for estimate \xt\ a 'too generous'
 variation was performed such that the quoted uncertainty is overestimated,
 scenario \sD.
 In these scenarios the \BLUE\ combinations give significant and different
 improvements. This means it is worth investigating whether the difference in
 uncertainty is caused by different sensitivities of the estimators used, or by
 different procedures followed and in the latter case if possible, to harmonise
 those.

 For the reduced correlations, given the assigned identical uncertainties, the
 source $k=2$ is not altered, see Section~\ref{sec:redcor}. Because in addition,
 the uncertainties for $k=1$ are much smaller than those for $k=2$, the method
 is almost switched off, i.e.~$\rhored \approx \rhof$.
 It is worth noticing that the results of the \BLUE\ method for scenarios
 \sC\ and \sD\ are much different from the result of the reduced correlations
 for scenario \sA, exemplifying the different sensitivities to \rhof\ and
 $z$. For the \BLUE\ method, and at the quoted precision, the values of \bet\ in
 \sC\ and \sD\ are identical, and much different from the one for scenario
 A. They also differ strongly from the value obtained by applying the reduced
 correlations for scenario \sA.

 Again, since the maximisation of the variance is only attempted for $\beta<0$,
 also for scenarios \sC\ and \sD\ all algorithms $i)-iii)$ are inactive.
%
%------------------------------------------------------------------------------
%
\section{How to decide on and perform a combination}
\label{sec:decide}
 The proposed procedure is described for the situation of $m$ estimates of the
 same observable and fully respects the properties of the estimates given in
 Section~\ref{sec:estpro}. The extension to more than one observable is straight
 forward.
 As an example, the procedure is applied to the input of the latest combination
 of \mt\ measurements performed at the Tevatron~\cite{TEV-1301}.
 Based on the initial input and the default assumptions on the correlations, the
 following questions are addressed:
%
%%%%%%%%%%%%%%%%%%%%%%%%%%%%%%%%%%%%%%%%%%%%%%%%%%%%%%%%%%%%%%%%%%%%%%%%%%%%%%%
\begin{Enumerate}
 \item [\,\,\, I)] Are the estimates compatible?
 \item [\,\,II)] Which estimates are worth combining?
 \item [III)] What are the consequences of varying \rhoijk?
 \item [\,IV)] What are the consequences of varying \zijk?
\end{Enumerate}
%%%%%%%%%%%%%%%%%%%%%%%%%%%%%%%%%%%%%%%%%%%%%%%%%%%%%%%%%%%%%%%%%%%%%%%%%%%%%%%
%
 Clearly, the outcome of the combination depends on the initial correlation
 assignments in Ref.~\cite{TEV-1301} that are kept to obtain the central
 combined result.

 For answering I), the compatibility is addressed by the \chiq\ defined in
 Eq.~\ref{eq:chiq}, and calculating \Prob{\chiq}{1}.
 Incompatible sets of estimates should not be combined, instead the reason for
 this should be searched for.
 From the 66 \chiq\ values of the pairwise compatibility tests for the twelve
 estimates from Ref.~\cite{TEV-1301}, 18 are above one, of which one (one) is
 above two (three), the smallest value being about $\Prob{\chiq}{1}=8\%$,
 resulting in a reasonable distribution of \chiq\ values.
 In addition, the global \chiq\ of the combination, see Eq.~\ref{eq:chiqtot},
 amounts to $\chiq=8.5$ for eleven degrees of freedom yielding a
 \chiq\ probability of 0.67.

 For answering II), starting from the most precise estimate $i$ it is proposed
 to rank the estimates $j\ne i$ by their importance. 
 Here, the importance of estimate $j$ is defined as the potential improvement in
 the most precise estimate $i$ by including the estimate $j$, irrespectively of
 the existence of any other estimate, calculated using Eq.~\ref{eq:sigx} and
 identifying $12 = ij$.
 The most precise estimate is chosen since it is special in the sense that, if
 no combination is performed, it represents the best knowledge of the
 observable, and the aim of any combination is to improve this information.
 The proposed procedure takes into account the correlation and the relative
 uncertainty of the two estimates, but is deliberately independent from the
 existence of all other estimates.
 This suggestion is motivated by the aim to only include the estimate $j$ if it
 on its own significantly improves the most precise estimate of \xT,
 irrespectively of the information contained in other estimates.
 By construction, this definition is a subjective and not unique choice, and
 other measures of importance could be taken.
%
%------------------------------------------------------------------------------
\begin{table*}[tbp!]
\caption{The list of estimates of \mt\ from Ref.~\cite{TEV-1301}. The most
  precise estimate is CDF(II) l+j. The other estimates are listed according to
  their importance, defined as the achieved improvement of the combined
  uncertainty with respect to the most precise estimate, obtained by performing
  pairwise combinations of each estimate with the most precise one.
  The correlation \rhof\ and relative uncertainties $z$ are given together with
  the two main parameters of the combination, \bet\ and \sx/\so\ and their
  derivatives with respect to \rhof\ and $z$.
 Entries quoted as $0.00$ mean that the absolute value of the actual number was
 below $0.005$.}
\label{tab:BlueResTeV}
\begin{center}
\begin{tabular}{rrrrrrrrrrrr}
\hline\noalign{\smallskip}
    Estimate &  Value & Stat &  Syst &$\rhof$ & $z$ & $\bet$ &
   $\frac{\sx}{\so}$ & $\frac{d\bet}{d\rhof}$ & $\frac{1}{\so}\frac{d\sx}{d\rhof}$ 
                     & $\frac{d\bet}{dz}$     & $\frac{1}{\so}\frac{d\sx}{dz}$ \\
\noalign{\smallskip}\hline\noalign{\smallskip}
 CDF(II) l+j & 172.85 &  0.52 & 0.99 &      &       &       &      &       &       &       &       \\
  D0(II) l+j & 174.94 &  0.83 & 1.24 & 0.30 &  1.34 &  0.30 & 0.91 & -0.27 &  0.31 & -0.46 &  0.20 \\
  CDF(I) l+j & 176.10 &  5.10 & 5.31 & 0.49 &  6.61 & -0.06 & 0.93 & -0.19 & -0.44 &  0.01 & -0.01 \\
 CDF(II) Met & 173.95 &  1.26 & 1.37 & 0.32 &  1.67 &  0.17 & 0.96 & -0.41 &  0.25 & -0.29 &  0.10 \\
 CDF(II) had & 172.47 &  1.43 & 1.41 & 0.29 &  1.80 &  0.15 & 0.96 & -0.39 &  0.24 & -0.23 &  0.08 \\
\noalign{\smallskip}\hline\noalign{\smallskip}
  D0(II) dil & 174.00 &  2.36 & 1.49 & 0.18 &  2.51 &  0.09 & 0.98 & -0.33 &  0.20 & -0.09 &  0.03 \\
  CDF(I) had & 186.00 & 10.00 & 5.72 & 0.30 & 10.34 & -0.02 & 0.98 & -0.11 & -0.22 &  0.00 &  0.00 \\
 CDF(II) dil & 170.28 &  1.95 & 3.13 & 0.48 &  3.31 & -0.07 & 0.98 & -0.43 & -0.25 & -0.01 & -0.02 \\
  CDF(I) dil & 167.40 & 10.30 & 4.90 & 0.29 & 10.24 & -0.02 & 0.98 & -0.11 & -0.22 &  0.00 &  0.00 \\
 CDF(II) Lxy & 166.90 &  9.00 & 2.82 & 0.08 &  8.46 &  0.00 & 1.00 & -0.12 &  0.04 &  0.00 &  0.00 \\
   D0(I) dil & 168.40 & 12.30 & 3.61 & 0.11 & 11.51 &  0.00 & 1.00 & -0.09 & -0.03 &  0.00 &  0.00 \\
   D0(I) l+j & 180.10 &  3.60 & 3.87 & 0.22 &  4.75 &  0.00 & 1.00 & -0.22 & -0.02 &  0.01 &  0.00 \\
\noalign{\smallskip}\hline
\end{tabular}
\end{center}
\end{table*}
%------------------------------------------------------------------------------

 After producing this list, a combination is performed by using the most precise
 estimate and adding one additional estimate at a time following that list.
 Finally, setting a threshold for the minimum relative improvement required, it
 can be decided which estimates to use, and for which it is not worth to perform
 the difficult task of finding the appropriate variations in \rhoijk\ and
 \zijk\ for assessing the stability of the combined result.
 If a selection of estimates is not attempted and all measurements are retained,
 the definition of importance is irrelevant.

 The details of the hypothetical pairwise combinations are listed in
 Table~\ref{tab:BlueResTeV}. Looking at the parameters of the combination it is
 apparent that the importance of the exact knowledge of \rhof\ and $z$ strongly
 depends on the pair of estimate under consideration.
 As an example, the derivatives of \sx/\so\ with respect to \rhof\ vary by about
 a factor of 10-20 in absolute size. In addition, they have different signs,
 such that for some estimates the uncertainty on the combined result is reduced
 when reducing the correlation, for others it is instead increased.
 Using this information it becomes apparent for which estimate the proper
 assignment of the correlation is most important. The derivatives nicely show
 the sensitivity around the chosen default assumption.
 For example for the estimates CDF(I)~l+j and CDF(II)~Met, the sensitivity of
 the combination with the most precise estimate to \rhof\ is almost twice as
 large for the former than for the latter, and it is even larger than the one of
 the most important additional estimate D0(II)~l+j. In addition, increasing
 \rhof\ for the estimate CDF(I)~l+j would decrease the uncertainty of the
 combined result, whereas for the estimate CDF(II)~Met it would instead be
 increased.
%
%%%%%%%%%%%%%%%%%%%%%%%%%%%%%%%%%%%%%%%%%%%%%%%%%%%%%%%%%%%%%%%%%%%%%%%%%%%%%%%
\begin{figure}[tbp!]
\centering
\includegraphics[width=0.50\textwidth]{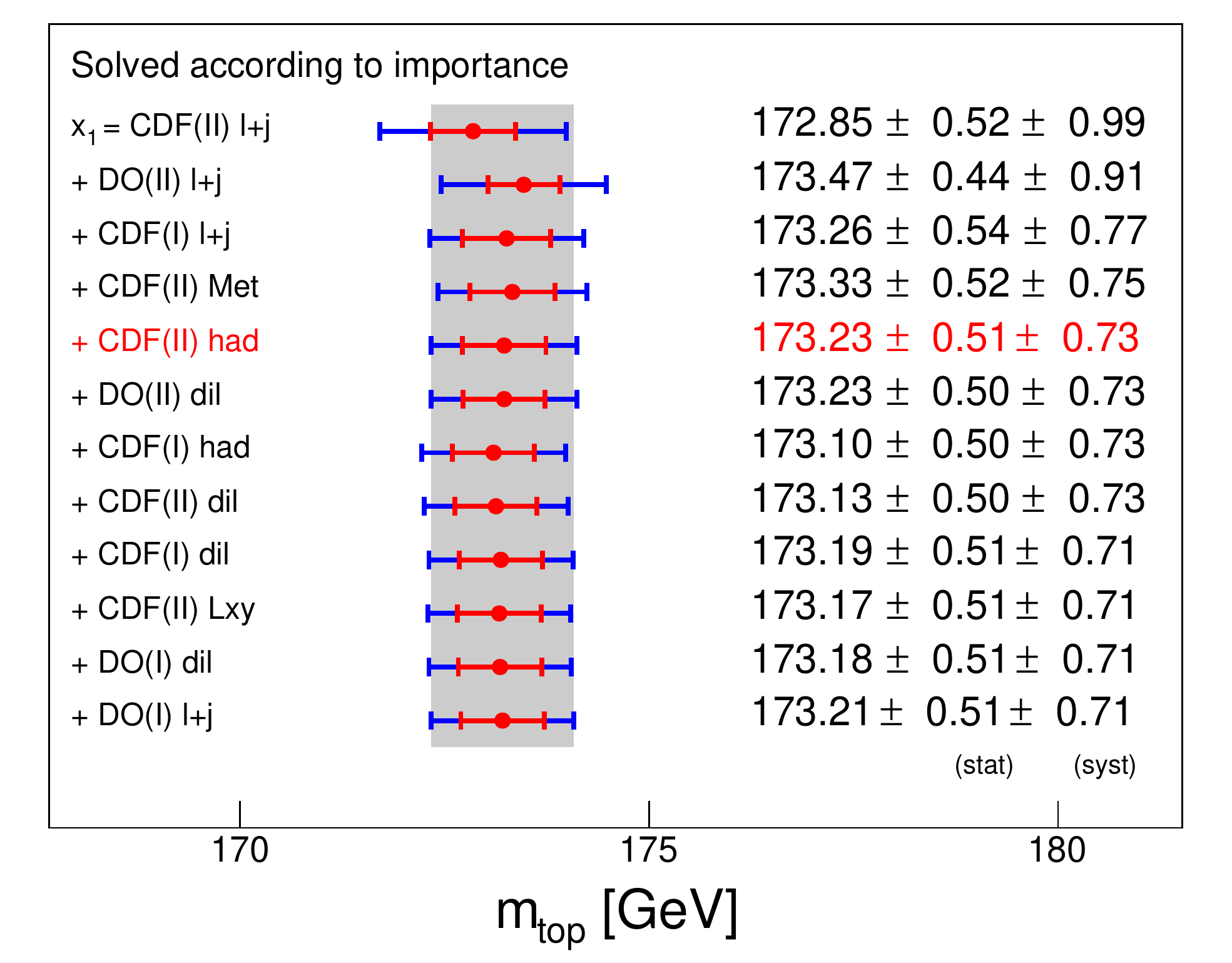}
\caption{Results of successive combinations according to importance of the
  estimates of \mt\ from~\protect\cite{TEV-1301}.
  The first line shows the result of the most precise estimate. All following
  lines report the combined result after adding the estimate listed to the
  previously accumulated list. Combinations, below the line with its \mt\ value
  given in red never improve the total uncertainty by more than $1\%$.
}
\label{fig:accimp}
\end{figure}
%%%%%%%%%%%%%%%%%%%%%%%%%%%%%%%%%%%%%%%%%%%%%%%%%%%%%%%%%%%%%%%%%%%%%%%%%%%%%

 The result of applying the proposed procedure to the input of the latest
 combination of \mt\ measured at the Tevatron~\cite{TEV-1301} is shown in
 Figure~\ref{fig:accimp}.
 The first line in Figure~\ref{fig:accimp} shows the result of the most precise
 estimate. All following lines report the results of successive combinations
 after adding the estimate listed to the previously accumulated list.
 If, as an example, an improvement in the total uncertainty of at least $1\%$
 for each individual remaining estimate to be included is desired, only the
 first five estimates should be combined.

 If the estimates were sorted according to their absolute \BLUE\ weights for the
 combination based on all estimates, which takes into account the correlations
 of all estimates (and the fact that the uncertainty is reduced on both sides of
 $\sx/\so = 1, \bet=0$), the same five estimates would have been chosen,
 i.e.~the combined result is the same.
 If instead the estimates were sorted by their inverse variance 1/\siq, which
 deliberately ignores all correlations and weights the estimates as if $\rhoij =
 0$, a slightly different list would be used. In the latter case, as can be seen
 from the values of $z$ reported in Table~\ref{tab:BlueResTeV}, the estimate
 CDF(I)~l+j would not be used, but D0(II)~dil would be used instead, despite the
 fact that looking at \sx/\so\ its impact is much smaller, demonstrating the
 large importance of the correlation.

\begin{figure*}[tbp!]
\centering
\subfigure[]{\label{fig:inspectmt:a}\includegraphics[width=0.245\textwidth]{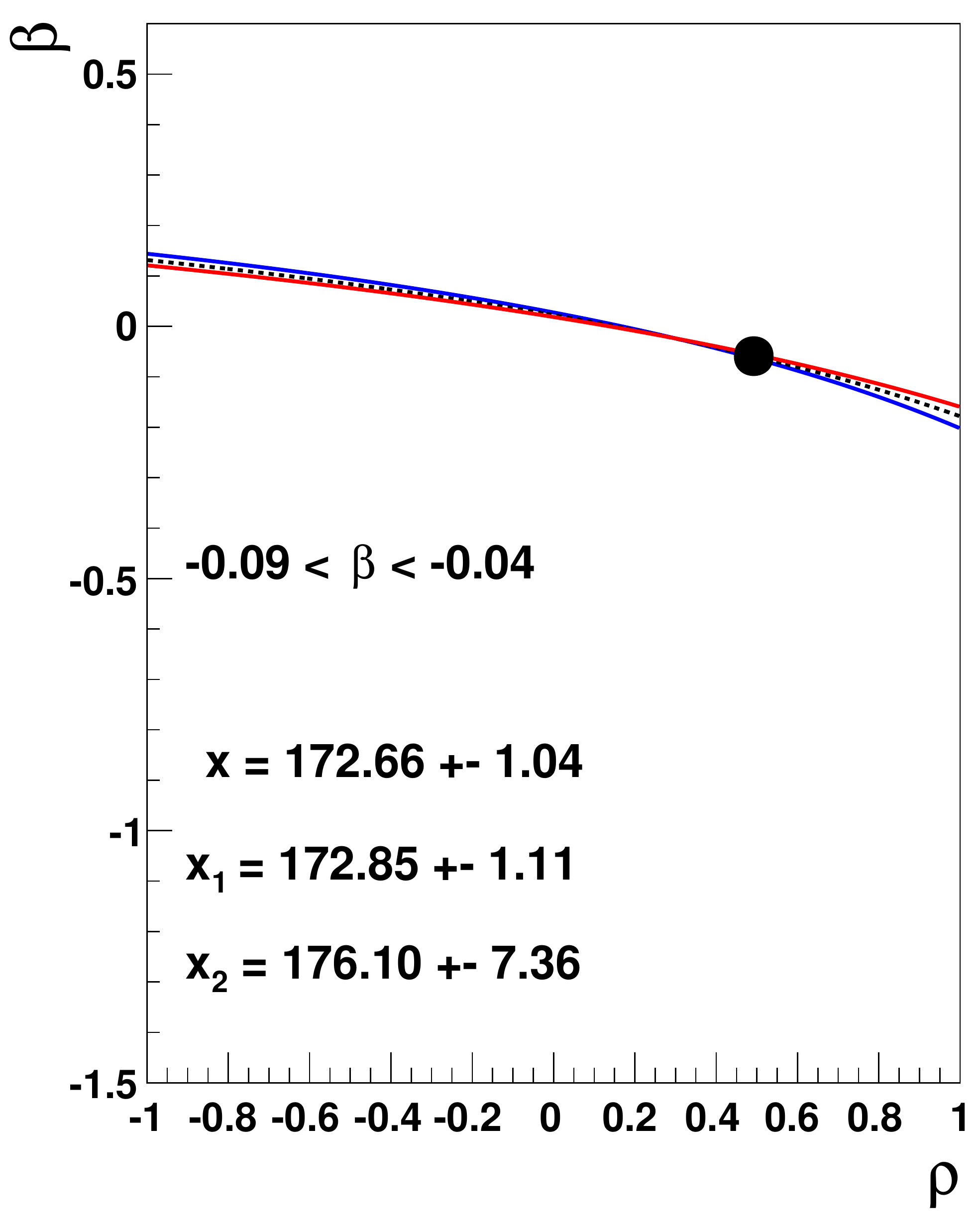}}
\subfigure[]{\label{fig:inspectmt:b}\includegraphics[width=0.245\textwidth]{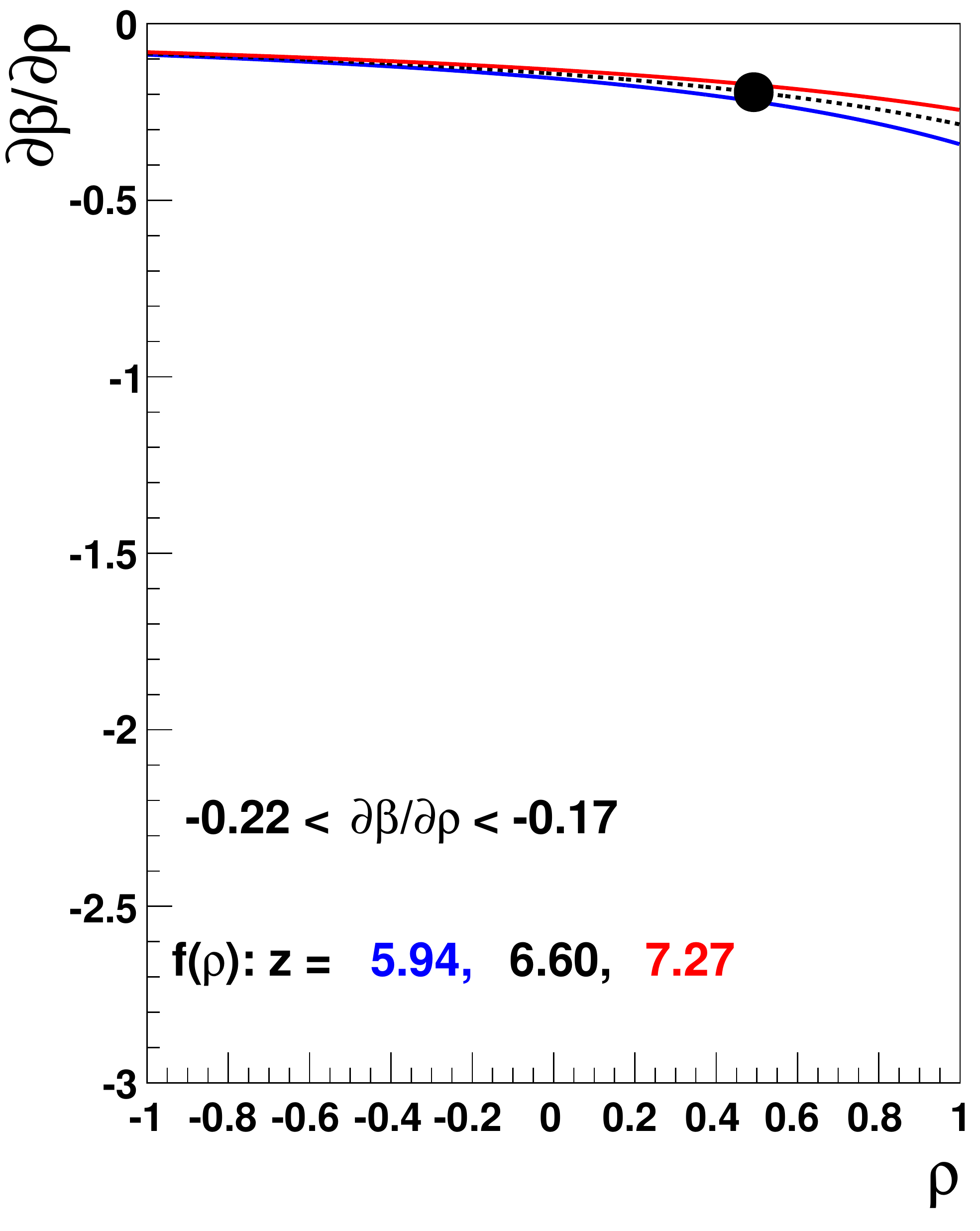}}
\subfigure[]{\label{fig:inspectmt:c}\includegraphics[width=0.245\textwidth]{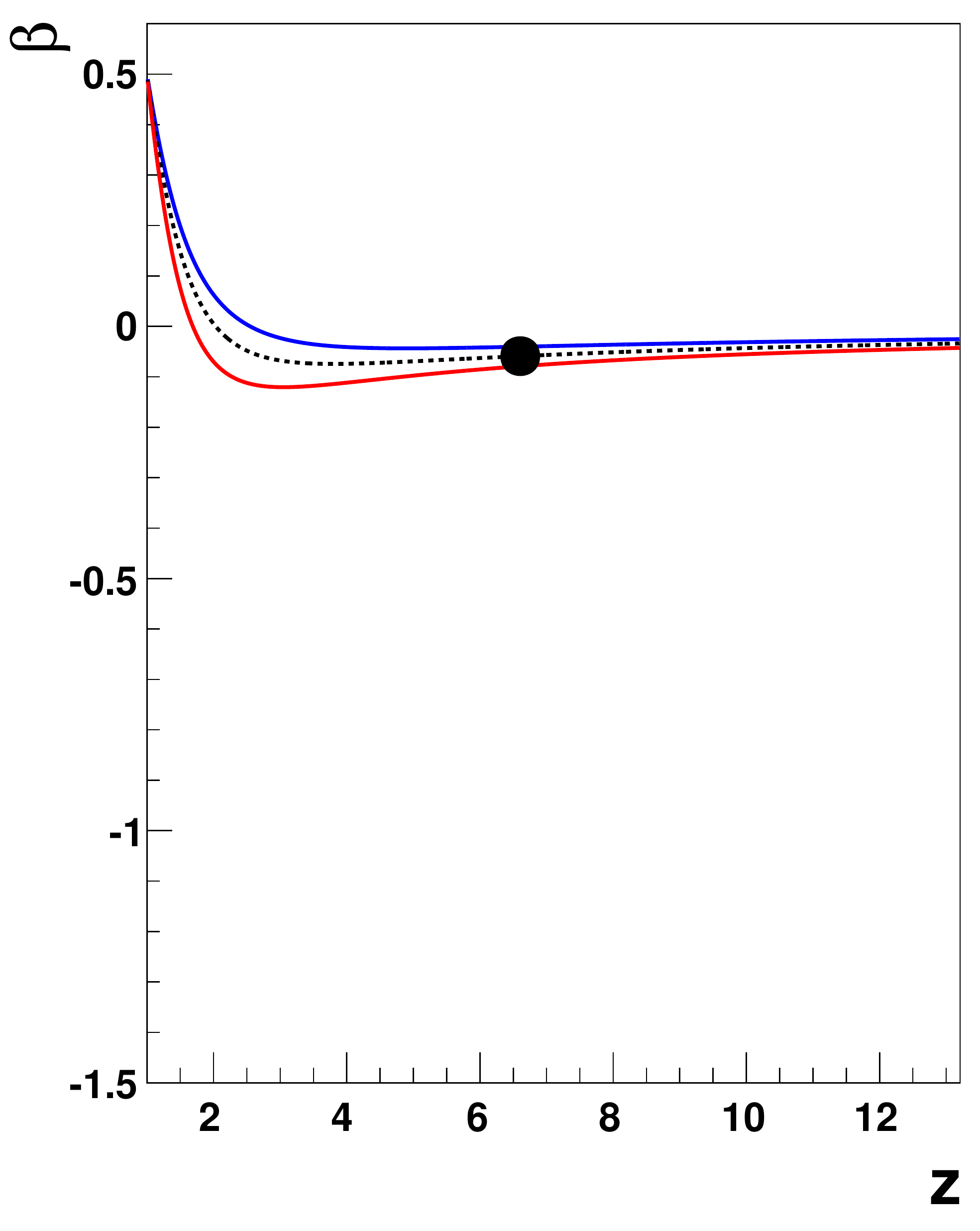}}
\subfigure[]{\label{fig:inspectmt:d}\includegraphics[width=0.245\textwidth]{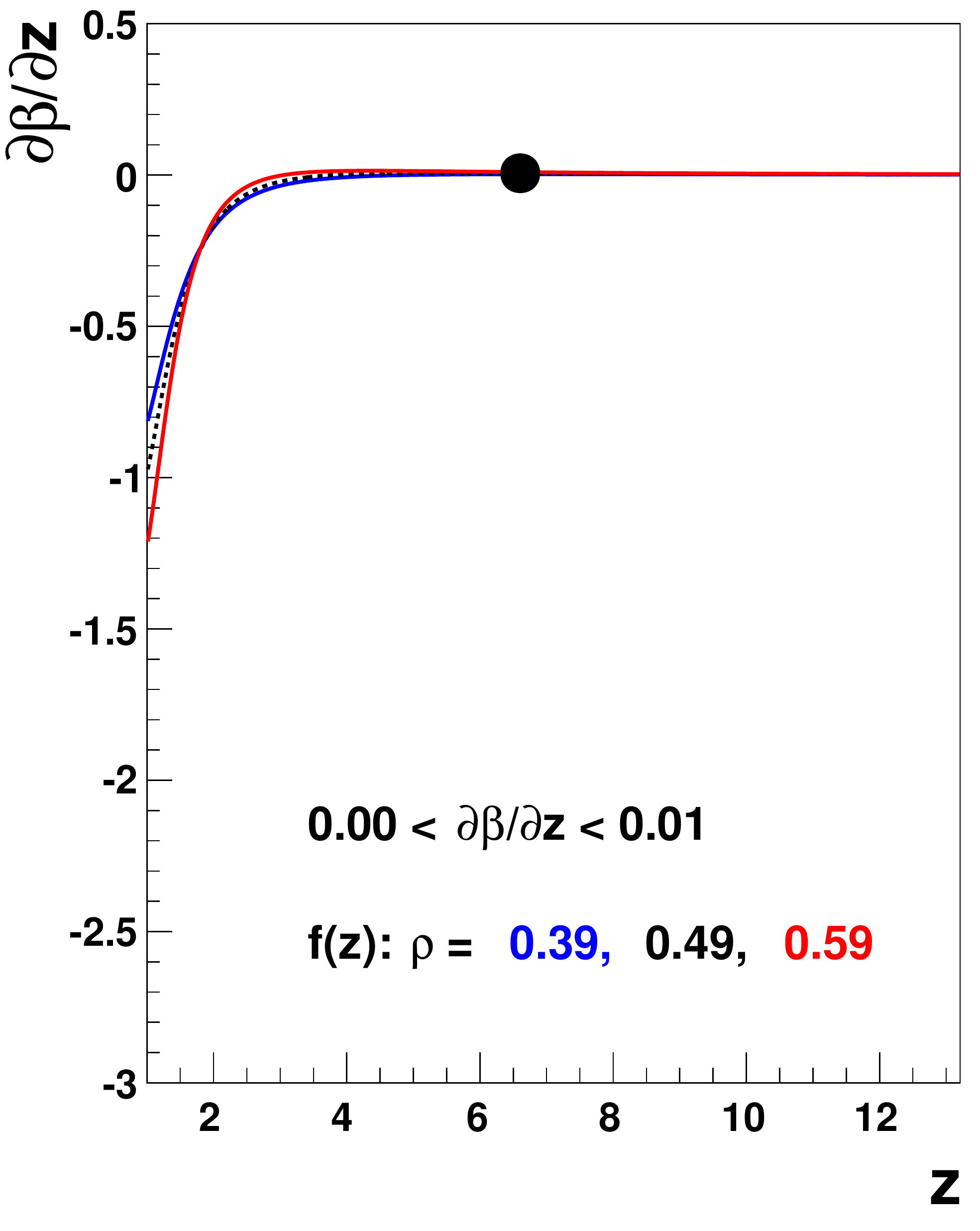}}
\subfigure[]{\label{fig:inspectmt:e}\includegraphics[width=0.245\textwidth]{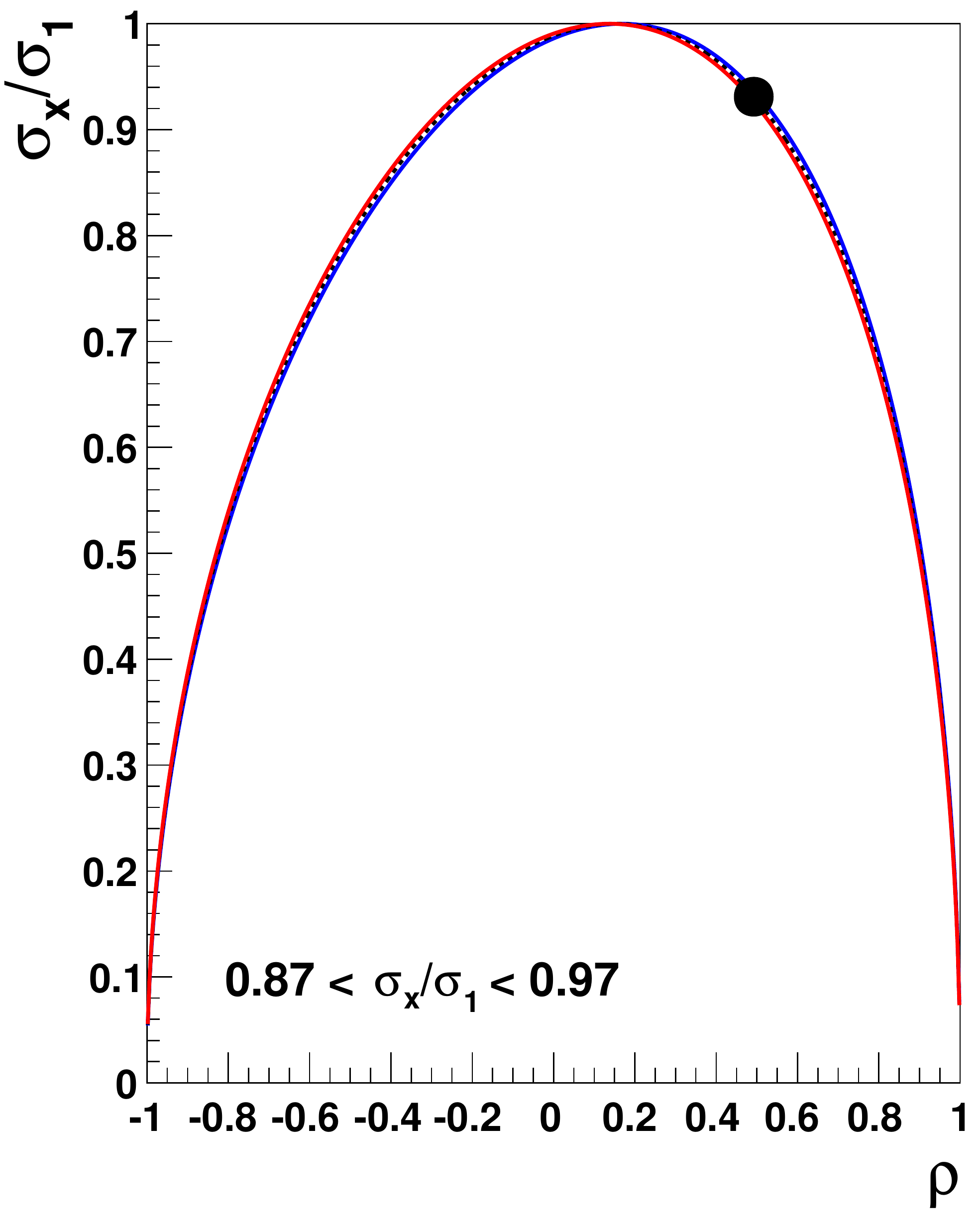}}
\subfigure[]{\label{fig:inspectmt:f}\includegraphics[width=0.245\textwidth]{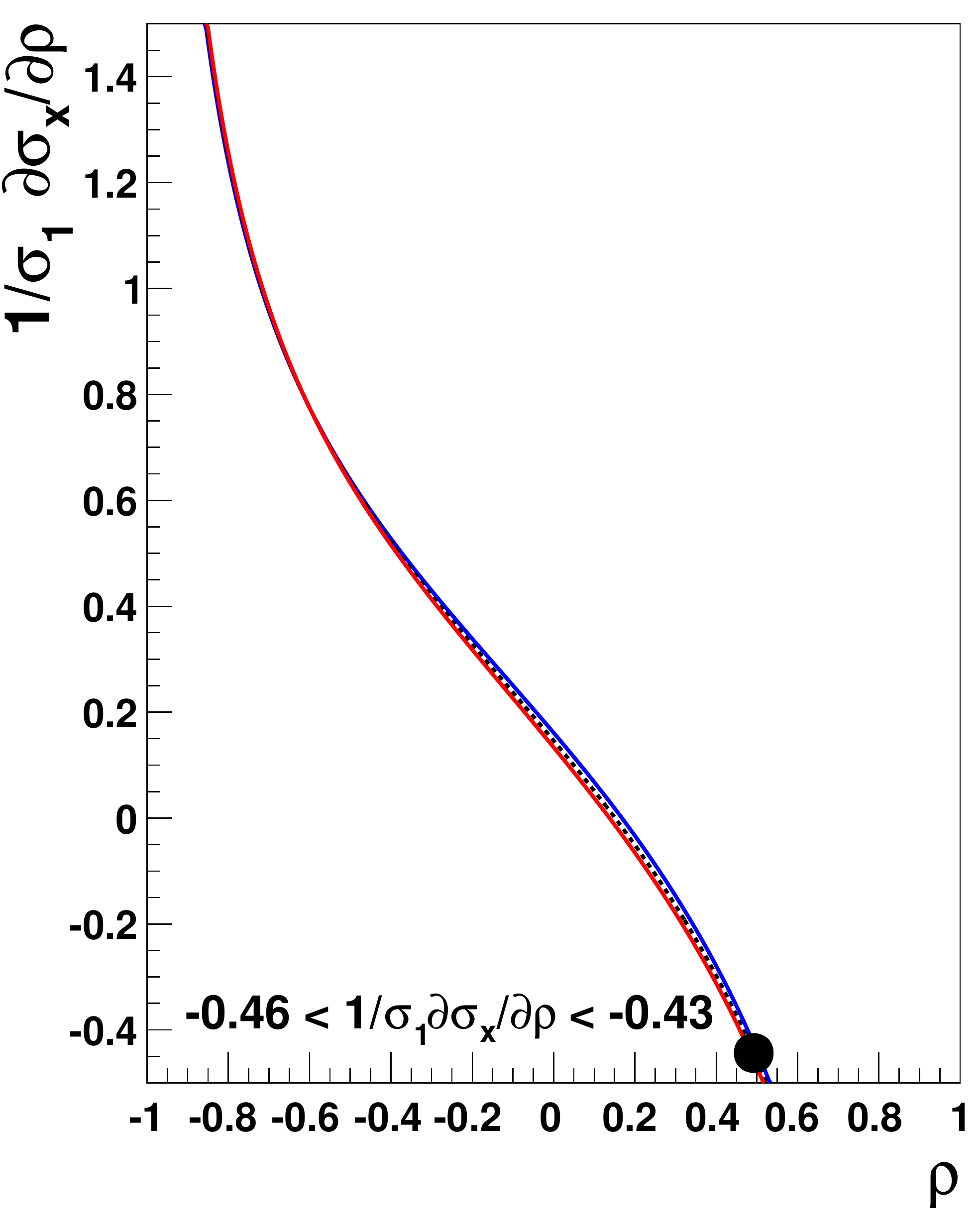}}
\subfigure[]{\label{fig:inspectmt:g}\includegraphics[width=0.245\textwidth]{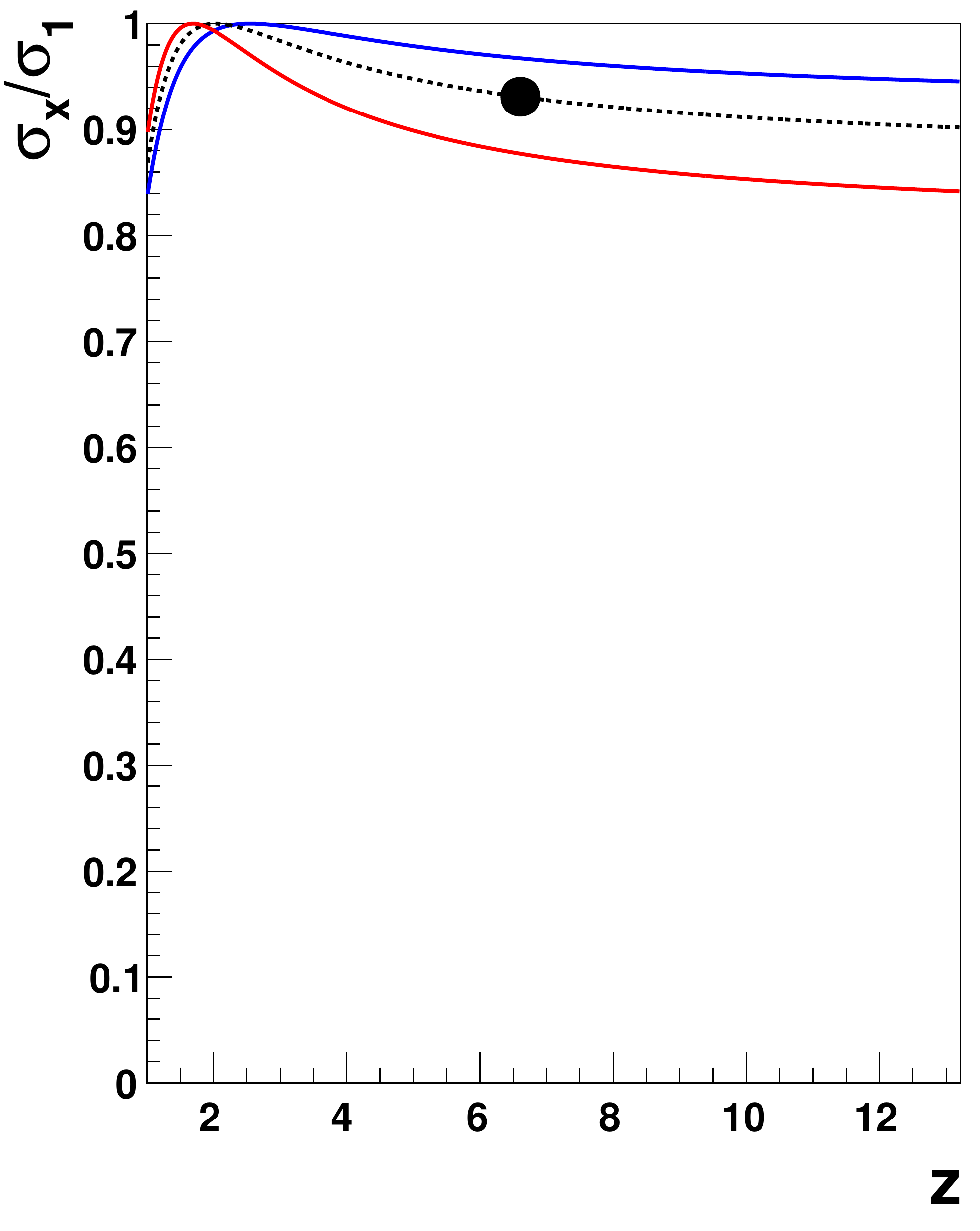}}
\subfigure[]{\label{fig:inspectmt:h}\includegraphics[width=0.245\textwidth]{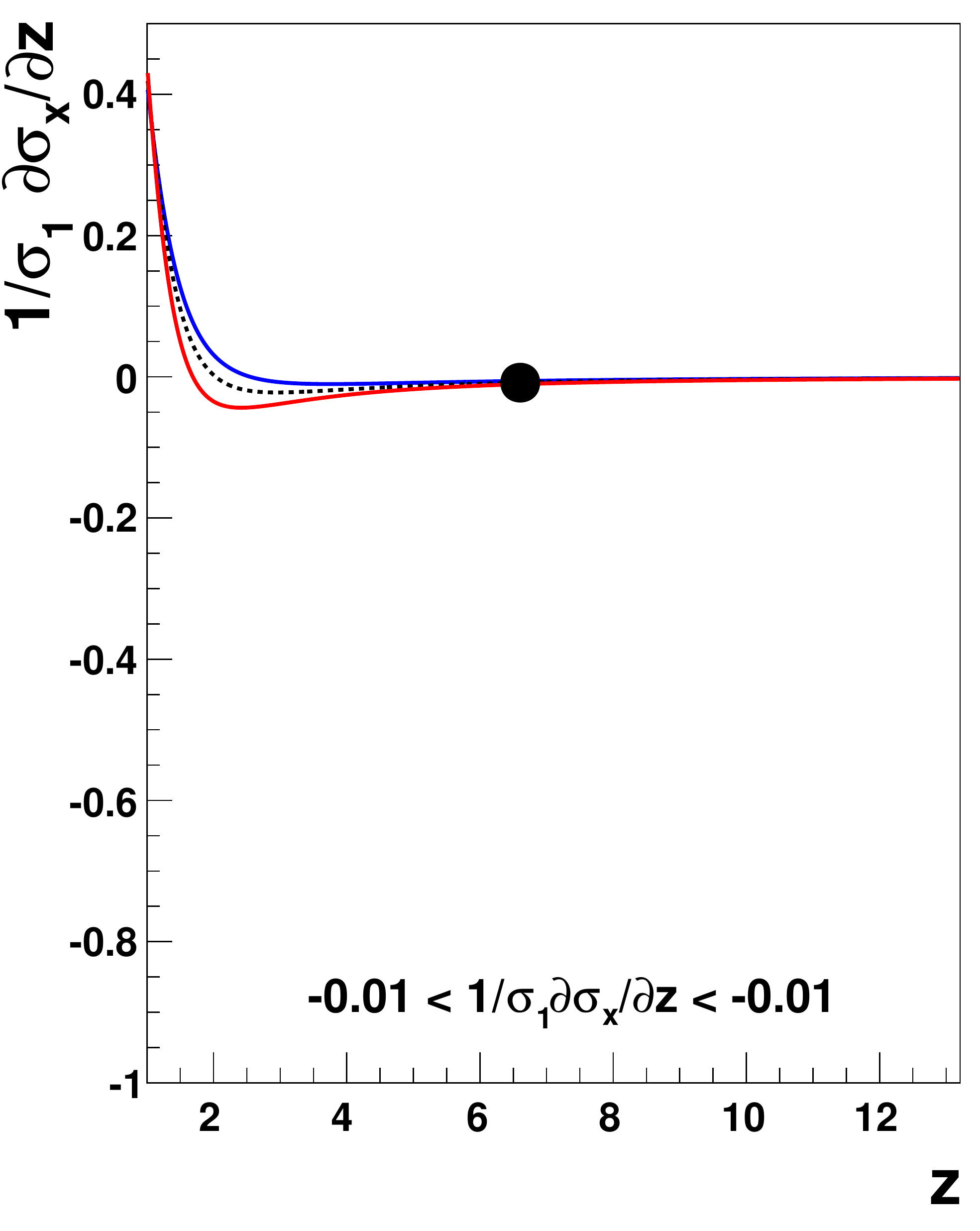}}
\caption{Same as Figure~\protect\ref{fig:inspect} but for the pair containing
  the most precise estimate, and the one with the negative \BLUE\ weight, for
  the \mt\ combination using input from~\protect\cite{TEV-1301}, see
  Table~\protect\ref{tab:BlueResTeV}.}
\label{fig:inspectmt}
\end{figure*}

 When using the proposed method, the corresponding result of \mt\ is shown in
 red. The \BLUE\ weights of the five estimates in the order they appear in
 Figure~\ref{fig:accimp} are: $0.61, 0.23, -0.06, 0.12, 0.10$.
 No combination below this line improves the total uncertainty by more than
 $1\%$.
 The situation for the pair containing the most precise estimate and the one
 with the negative \BLUE\ weight is shown in Figure~\ref{fig:inspectmt}.
 For all sub-figures and all coordinate axes Figures~\ref{fig:inspect}
 and~\ref{fig:inspectmt} are drawn using identical ranges.
 Compared to Figure~\ref{fig:inspect}, there is a very flat behaviour around the
 point representing this pair of estimates, but for \sx/\so\ and its derivative
 of \sx/\so\ with respect to \rhof, Figures~\ref{fig:inspectmt:e} and
 \ref{fig:inspectmt:f}.

 After performing the selection, the combination of all selected estimates is
 performed to determine the central value and the breakdown of uncertainties.
 The compatibility of the pairs of selected estimates is improved, only two
 \chiq\ values exceed one, and the smallest \Prob{\chiq}{1}\ value is about
 19$\%$.
 For the selected estimates the total \chiq\ amounts to $\chiq=2.5$ for four
 degrees of freedom yielding a very similar \chiq\ probability as for the full
 set of estimates of 0.65.
 By construction, the result of the combination is very close to the one based
 on all estimates. Only little information is lost, but it is much more clear
 which estimates contain the information, and the investigation of the stability
 of the result is more simple.
%
%%%%%%%%%%%%%%%%%%%%%%%%%%%%%%%%%%%%%%%%%%%%%%%%%%%%%%%%%%%%%%%%%%%%%%%%%%%%%%%
%%%%%%%%%%%%%%%%%%%%%%%%%%%%%%%%%%%%%%%%%%%%%%%%%%%%%%%%%%%%%%%%%%%%%%%%%%%%%

 As said above, the values of \rhootk\ and \zotk\ are only known with some
 uncertainties. The task is to evaluate the consequences of this for the
 combined value.
 Looking at the figures of pairwise combinations like
 Figure~\ref{fig:inspectmt}, or the values listed in Table~\ref{tab:BlueResTeV},
 the most critical pairs and parameters can easily be identified.
 To assess the stability of the combined result, individual uncertainty sources
 have to be investigated for possible variations of \rhoijk\ and \zijk. This
 should be done in view of the details of the procedures applied, and it should
 be decided whether a variation in \rhoijk\ or \zijk\ is the appropriate choice.

 To investigate III) independent variations per source $k$ are performed in
 which \rhoijk\ is varied within a range determined by analysing the procedures
 used for the estimates. 
 This is performed by multiplying the initially assigned correlation by a factor
 $r$, using the range $r = 1 \to \rmin$, and investigating the difference in the
 uncertainty of the combined result.
 If found appropriate, the observed differences in the combined values could be
 added quadratically to the uncertainty of the combined result to account for
 the uncertainties in the assigned correlations.

 Since the detailed information on reasonable variations of the initially
 assigned correlations is only available to the experiments that actually
 determined the estimates, for the example presented, the full range of $r = 1
 \to 0$ has been used for all sources that remain correlated after the selection
 of estimates, which likely is an overestimation of the effect.
 For this example all variations lead to an increase of the combined value $x$.
 The square root of the quadratic sum of the differences between the combined
 value of the default assignments and the ones obtained with the changed
 assumption on the correlation for all sources $k$ amounts to $0.26$~\GeV. This
 number is dominated by a single source that contributes with $0.23$~\GeV.
 Given this, a simultaneous variation of the correlation assumption of all
 sources would result in an only slightly larger value of $0.29$~\GeV. However,
 this evaluation is disfavoured, since it violates property 3) of the
 estimates. In addition, the individual variations also reveal which sources are
 the important ones for the stability.

 Given these variations, in principle the list of importance for the estimates
 may differ from the initial one. 
 If, as it is usually done, the above variations are only used as stability
 checks, and no additional uncertainty is assigned, this is of very little
 concern. This is the case for many combinations including the example
 presented, see
 e.g.~Refs.~\cite{ATLAS-CONF-2013-102,ATLAS-CONF-2013-098,TEV-1301}.
 If an additional uncertainty is assigned, one may want to perform the selection
 iteratively or refrain from selecting estimates.
 Again, for the example presented, this is of minor numerical importance. When
 using the recommended individual variations, the first six estimates of the
 list of importance are always the same for the full range of $r = 1 \to 0$.
 Only for two sources and for correlations below $0.6$ and $0.4$ respectively,
 one of the first five estimates is exchanged with the sixth one.

 To investigate IV) an indicative procedure is to assume identical values
 $\sik=\sjk$ for pairs of estimates and to repeat the combination.
 If this test results in large variations, it is advisable to understand whether
 the difference of \sik\ and \sjk\ is due to different sensitivities of the
 estimators, or caused by different procedures followed in determining the
 uncertainties. In the latter case one should try to harmonise the
 procedures. For a numerical example of such a situation see
 Table~\protect\ref{tab:BlueRes}. Investigating the procedures in detail, likely
 smaller variations of $\sik$ turn out to be appropriate.
 Since this information is only available to the experiments that actually
 determined the estimates, for the example presented, this has not been
 investigated here.
 Depending on the details of the situation this can easily be more important
 than variations of \rhof, as can been seen from the example of the
 hadronisation uncertainty for the LHC
 \mt\ combination~\cite{ATLAS-CONF-2013-102}.
%
%------------------------------------------------------------------------------
%
\section{Summary and conclusions}
\label{sec:summar}
 In this paper the combination of correlated estimates has been reviewed using
 the Best Linear Unbiased Estimate (\BLUE) method, mainly concentrating on the
 special case of two estimates of the same observable.

 It has been shown that the underlying conditional probability inevitably leads
 to the fact that for positively correlated estimators, for a given pair of
 estimates to be combined, in most of the cases the true value is not within the
 interval spanned by the estimates.
 This fact should be respected by any combination method. All combination
 methods deliberately constructed to force the combined value to always lie
 within the interval spanned by the estimates, violate this consequence of the
 conditional probability, and are wrong by construction. These methods will lead
 to worse results than the \BLUE\ method that achieves this predicted behaviour
 by means of negative weights, which occur if they reduce the variance of its
 unbiased result.
 This situation is realised if the mean of the conditional probability of the
 less precise estimator is further away from the true value than the more
 precise estimate. This is the case whenever the correlation of the estimates
 \rhof\ is larger than $1/z$, the ratio of the smaller and the larger
 uncertainty.

 For any pair of estimates, their combination is fully determined by the values
 of \rhof\ and $z$, which determine the main parameters of the combination,
 namely the weight of the less precise estimate \bet, and the ratio of the
 uncertainty of the combined results and the more precise estimate \sx/\so.
 However, \rhof\ and $z$ themselves are typically only known with some
 uncertainty. Therefore, for visualising the sensitivity of the central result
 to these uncertainties, derivatives of \bet\ and \sx/\so were derived with
 respect to \rhof\ and $z$. 
 The derivatives can be used to identify the sources of estimates and
 uncertainties for which the knowledge on \rhof\ and $z$ is most critical.

 The differences observed when using either relative or absolute uncertainties
 in the \BLUE\ combination have been investigated, including a simulation of
 Peelle's Pertinent Puzzle. It has been found that the apparent difference
 observed for the original formulation of the puzzle, i.e.~for a single pair of
 estimates, is mainly a reflection of the unlikeliness of this pair of
 estimates.
 When instead combining numerous pairs of estimates based on a number of
 hypothetical underlying probability distributions that cover the full range of
 combined results observed for the original version of the puzzle, and both
 uncertainty models, the differences of the two methods are insignificant. The
 same holds true for a number of specific examples of publicly available
 combinations.

 A critical assessment of methods proposed to deal with the uncertainty on the
 correlations has been given. Especially, it has been argued that reduced
 correlations mix \rhof\ and $z$ and act in an unphysical way.
 Other methods constructed to maximise the variance of the combined result are
 too general, do not respect all properties of the estimates, and do not reflect
 the different knowledge on the correlations that likely is available for
 estimates of the same experiment, or those obtained at the same collider
 compared to those from different experiments and/or colliders.
 For all other methods discussed, the uncertainty in the knowledge on the
 relative size of the uncertainties per source $k$, is ignored throughout,
 however, this can be numerically much more important.

 A detailed proposal for a procedure to combine a number of estimates and to
 evaluate the stability of the result has been made.
 It has been argued that the decision on including a given estimate into the
 combination should be based on its potential improvement with respect to the
 most precise estimate, i.e.~on the relative gain of uncertainty of the combined
 value with respect to the most precise one for hypothetical pairwise
 combinations, irrespectively of the existence of other estimates.
 The most precise estimate is chosen since it is special in the sense that, if
 no combination is performed, it represents the best knowledge of the
 observable, and the aim of any combination is to improve this information.
 It is proposed to only include other estimates if they significantly improve on
 the most precise one.
 By construction, this definition is a subjective and not unique choice, and
 other measures of importance could be taken, or no selection could be
 performed.

 In any case, the stability of the result should be assessed source by source in
 view of the uncertainty on the knowledge on \rhoijk\ and \zijk, while
 respecting the properties of the estimates.
 Given the different dependence of the two parameters \bet\ and \sx/\so\ of the
 pairwise combination on \rhof\ and $z$, it is advisable to assess the impact on
 a case by case basis performing appropriate changes in \rhoijk\ or \zijk.
 A freely available software package to perform these investigations has been
 written.

 Finally, all ways to assess the uncertainty on the combined result by
 variations of the \rhoijk\ and \zijk\ are only indicative of possible
 sensitivities. If large sensitivities occur, a better understanding and
 possibly harmonisation of the input, and ways to calculate, rather than
 postulate the correlations as is frequently done, are much preferred.
%
%------------------------------------------------------------------------------
%
\begin{acknowledgements}
 I like to thank Allen Caldwell and Giorgio Cortiana for fruitful discussions on
 the subject and useful suggestions to the manuscript.
\end{acknowledgements}
%
%------------------------------------------------------------------------------
%
%\bibliographystyle{spbasic}   % basic style, author-year citations
\bibliographystyle{spphys}     % APS-like style for physics
\bibliography{Blue}   % name your BibTeX data base
\end{document}